\documentclass[fleqn,usenatbib]{mnras}

\usepackage{newtxtext,newtxmath}
\usepackage{url}
\usepackage{color}
\usepackage{tabulary}
\usepackage{multirow}
\usepackage{amsfonts}
\usepackage{array}
\usepackage{ulem}
\usepackage[T1]{fontenc}
\usepackage{ae,aecompl}
\usepackage{hyperref}

\usepackage{graphicx}	
\usepackage{amsmath}	
\usepackage{amssymb}	
\usepackage{cases}

\graphicspath{ {pictures/} }

\newcommand{\beq}{\begin{equation}}
\newcommand{\eeq}{\end{equation}}
\newcommand{\simlt}{\mathrel{\hbox{\rlap{\hbox{\lower4pt\hbox{$\sim$}}}\hbox{$<$}}}}
\newcommand{\simgt}{\mathrel{\hbox{\rlap{\hbox{\lower4pt\hbox{$\sim$}}}\hbox{$>$}}}}

\newcommand{\s}{\;\mathrm{s}}
\newcommand{\dd}{\partial}
\newcommand{\Msol}{\;\mathrm{M}_{\odot}}

\newcommand{\Gyr}{\;\mathrm{Gyr}}
\newcommand{\Mpc}{\;\mathrm{Mpc}}

\newcommand{\km}{\;\mathrm{km}}

\newcommand{\pc}{\;\mathrm{pc}}
\newcommand{\yr}{\;\mathrm{yr}}

\def\apjl{ApJL}
\def\apj{ApJ}
\def\mnras{M.N.R.A.S.}
\def\aap{A\&A}
\def\nat{Nat.}

\def\araa{Ann. Rev. A\&A}

\def\aj{AJ}

\def\prd{prd}

\title[SMBHs]{Interactions between multiple supermassive black holes in galactic nuclei: a solution to the final parsec problem}

\author[T. Ryu et al.]{
Taeho Ryu$^{1}$\thanks{email: taeho.ryu@stonybrook.edu}, 
Rosalba Perna$^{1}$, 
Zolt\'an Haiman$^{2,3}$,
Jeremiah P. Ostriker$^{2}$,
\newauthor Nicholas C. Stone$^{2}$\\ 
$^{1}$Department of Physics and Astronomy, Stony Brook University, Stony Brook, NY 11794-3800, USA\\
$^{2}$Department of Astronomy, Columbia University, 550 W. 120th Street, New York, NY 10027, USA\\
$^{3}$Department of Physics, New York University, New York, NY 10003, USA}

\date{Accepted XXX. Received YYY; in original form ZZZ}

\pubyear{2017}

\begin{document}
\label{firstpage}
\pagerange{\pageref{firstpage}--\pageref{lastpage}}
\maketitle

\begin{abstract}
Using few-body simulations, we investigate the evolution of
supermassive black holes (SMBHs) in galaxies ($M_{\star}=10^{10}-10^{12}\Msol$ at $z=0$) at $0<z<4$. Following galaxy merger trees from the Millennium simulation, 
we model BH mergers with two extreme binary decay scenarios for the `hard binary' stage: a full or an empty loss cone.  These two models should bracket the true evolution, and
    allow us to separately explore the role of dynamical friction and
    that of multi-body BH interactions on BH mergers.  Using the
computed merger rates, we infer the stochastic gravitational wave
background (GWB).  Our dynamical approach is a first attempt to study the dynamical
evolution of multiple SMBHs in the host galaxies undergoing
mergers with various mass ratios ($10^{-4} < q_{\star} <
1$). Our main result demonstrates that SMBH binaries are able to
merge in both scenarios.  In the empty loss cone case, we find that
BHs merge via multi-body interactions, avoiding the `final parsec'
problem, and entering the PTA band with substantial orbital eccentricity. Our full loss cone treatment, albeit more approximate, suggests that the eccentricity becomes even higher
when GWs become dominant, leading to rapid coalescences (binary lifetime $\lesssim1\Gyr$). 
Despite the lower merger rates in the empty loss cone case, due to their higher mass
ratios and lower redshifts, the GWB in the full/empty loss cone models are comparable ($0.70\times10^{-15}$ and $0.53\times10^{-15}$ at a
frequency of $1\yr^{-1}$, respectively). Finally, we compute the
    effects of high eccentricities on the GWB spectrum.
\end{abstract}




\section{Introduction}
\label{sec:intro}

It is known that almost every nearby massive galaxy harbors a
supermassive black hole (SMBH) in its nucleus
\citep{KormendyHo2013}. In the $\Lambda$CDM cosmology, galaxies evolve
as they hierarchically merge. As a result, it is expected 
that more than two SMBHs could coexist in a galaxy.  If they
successfully get close to each other, they form a bound pair.
Recently, the presence of multiple SMBH systems has been
observationally confirmed, such as a SMBH binary system at $z=0.055$
with the projected separation of $\sim7\pc$
\citep{Rodriguez2006,Bansal+2017} and a triple SMBH at $z=0.39$ with
the closest pair separated by $\sim140\pc$ \citep{Deane+2014} .

However, it is still unknown whether the SMBH binary would further
decay and eventually merge. This is one of the fundamental questions
in astrophysics.  The coalescence of two SMBHs, possibly being the
loudest gravitational wave (GW) event in the Universe, has received
much attention recently as we enter a new era of GW astronomy. In
particular, Pulsar Timing Arrays (PTAs) are expected to be a powerful
tool to detect the GWs emitted during the inpiral and
  coalescence of SMBH binaries. Therefore, it is important to
study formation and evolution of SMBH binaries.

Generally speaking, the evolution of a SMBH binary involves three
stages from its formation to coalescence \citep{Begelman+1980}.  (i)
As galaxies merge, SMBHs spiral to the core regions of the merged
galaxies due to dynamical friction and form binaries. (ii) As the
orbit shrinks, dynamical friction becomes inefficient and three-body
interactions with surrounding stars or other orbiting BHs can cause
the orbit of the SMBH binary to further decay.  Viscous torques from a
surrounding circumbinary disk can also play a role at this stage in a
wet merger \citep[e.g][]{Mayer+2007,Lodato+2009,Mayer2013,Tang+2017}.
(iii) Finally, at small enough separations, GW emission takes over,
driving the SMBHs to merge.  Whether or not SMBHs can merge is mostly
determined by how smoothly and rapidly a transition from (i) to (iii)
takes place\footnote{A bottleneck can arise earlier, in phase (i), for
  very small mass ratios $q_\star \ll 1$, when the dynamical friction
  time exceeds the Hubble time \citep{Taffoni+2003}.}.

In order for such transition to occur in less than the Hubble time,
there must be a sufficient number of central stars to extract the
orbital energy of the SMBH binary until it enters the GW-dominated
regime. However, as the binary becomes more tightly bound, a
significant fraction of stars are ejected, leaving behind empty phase
space regions (the so-called empty ``loss cone") around the binary
with no stars remaining to interact with. The empty loss cone is
replenished by dynamical processes, the simplest of which is two-body
relaxation.  Given the long relaxation time in the nuclei of bright
elliptical galaxies ($\gtrsim10\Gyr$,
\citealt{MerrittWang2005,Merritt+2010}), once the loss cone is cleared
out, it is unlikely that it can be refilled fast enough - via two-body
relaxation - to merge within a Hubble time.  This may stall the SMBH
binary at parsec scales, and is famously known as the ``final parsec
problem'' \citep{MilosavljevicMerritt2003}.  However, alternative
dynamical mechanisms for sufficiently fast loss cone repopulation have
been proposed, such as enhanced stellar flux into the core regions in
non- spherical (triaxial or axisymmetric) nuclei
\citep[e.g][]{Yu2002,Gualandris+2017}.  Many studies have shown that
some level of triaxiality is a characteristic of galatic merger
remnants
\citep[e.g][]{Preto+2011,Khan+2011,Khan+2016,GualandrisMerritt2012}
and that triaxial potential-density configurations can be dynamically
stable over long timescales
\citep[e.g][]{PoonMerritt2002,PoonMerritt2004}.  This implies that
aspherical geometries may prevent SMBH stalling, and yield a quick
transition from phase (i) to (iii). In other words, SMBHs might
coalesce on a shorter timescale than estimated assuming an empty
loss cone. 

However, it is also possible that the loss cone is not replenished
efficiently.  Because of resolution limitations, most $N$-body
simulations of the final parsec problem in galaxy mergers are not
likely converged \citep{Vasiliev+2014}.  More approximate Monte Carlo
studies indicate that while axisymmetric potentials cannot solve the
final parsec problem, realistic levels of triaxiality {\it can}
\citep{Vasiliev+2015}.  However, triaxiality can erode over time due
to chaotic diffusion \citep{MerrittValluri1996}, and, particularly in
minor mergers, it is not clear that sufficient triaxiality is
generated on small scales to refill the loss cone in
time. Furthermore, the core/cusp dichotomy in the surface brightness
profiles of galactic nuclei suggests that the (large) galaxies of
greatest interest for pulsar timing efforts merge in a preferentially
gas-poor way \citep{Faber+1997, Lauer+2005}. Interestingly,   \citet{DvorkinBarausse2017} suggest that in their extreme ``nightmare scenario'' in which SMBH binaries are assumed to stall and never complete their mergers, such a population of stalled binaries would produce a stochastic GW background at lower frequencies that should be detectable with PTAs
.

The lack of theoretical consensus concerning solutions to the final
parsec problem motivates us to consider the outcomes of stalled SMBH
binaries {\it in a cosmological context}.  If a SMBH binary fails to
merge before another BH makes it to the nucleus as a result of a
subsequent galaxy merger, multi-body interactions between the binary
SMBH and the incoming BH will occur. Such triple BH interactions could be even more abundant at early times if more numerous SMBHs were assembled earlier, possibly promoting the formation of sufficiently compact binaries at high redshift which could merge by GW emission \citep{Volonteri+2003}.
The intrusion
of another BH into the SMBH binary system can enhance the loss cone
refilling rate by disturbing stellar orbits
\citep{Perets+2007,PeretsAlexander2008}.  Moreover, chaotic,
non-hierarchical three-body interactions tend to shrink the binary
semimajor axis and to increase the eccentricity of an initially
circular binary \citep{ValtonenMikkola1991}.  If they form a
hierarchical triple, the merger time of the inner binary can be
dramatically reduced due to eccentricity oscillations induced by the
Kozai-Lidov mechanism \citep{Blaes+2002}. All of these effects likely
accelerate the BH coalescence rate \citep[][]{Iwasawa+2006,Bonetti+2017} as well as the ejection rate of
(typically less massive) SMBHs \citep{HoffmanLoeb2007}. Ejection
events - which can also occur due to GW recoil following successful
SMBH mergers \citep{Bekenstein1973, Campanelli+2007} - are
observationally important for SMBH demographics \citep{Schnittman2007,
  KulkarniLoeb2012}. Understanding the outcomes of multiple SMBH
interactions is therefore of great importance not just for determining
merger rates, but also cosmological SMBH evolution.

In order to gain an in-depth understanding of SMBH binary evolution,
observations of GWs using PTAs are crucial.  There are currently three
ongoing PTA groups, the North-American Nanohertz Observatory for
Gravitational Waves (NANOGrav, \citealt{NANOGravCollaboration+2015}),
the European PTA (EPTA; \citealt{Desvignes+2016}), and the Parkes PTA
(PPTA; \citealt{Manchester+2013}). Their combined effort, the
International PTA (IPTA, \citealt{Hobbs+2010}), recently released its
first datasets \citep{Verbiest+2016}.  With the duration of the
observation $T\sim$ a few years to a few months and  the
  observing cadence of $\Delta t\sim$ a few weeks, the relevant
frequency band is beween $1/T$ and $1/2\Delta t$. This corresponds to
approximately $n{\rm Hz}-\mu {\rm Hz}$. This frequency range is
comparable to that of GWs from compact sub-parsec ($0.01-0.1\pc$) SMBH
binaries. This makes the SMBHs one of the most promising astrophysical
sources of GWs accessible to PTAs. A stochastic GW signal can be
described by its amplitude $h_{\rm c}$, also known as the
  characteristic strain. In particular, for each individual SMBH
binary in a circular orbit, it is easily shown that the strain scales
as $h_{\rm c}(f)\propto f^{-2/3}$, where $f$ is the observed frequency
\citep{Phinney2001}. The strain is usually quoted at
the frequency $f=1\yr^{-1}$, and then referred to as $A$
(\citealt{Jenet+2006}; Eq.~\ref{eq:hc} of this paper).  The stochastic
GWB from massive BH mergers has been extensively examined via
semi-analytical \citep[e.g][]{WyitheLoeb2003,Ravi+2014} or Monte Carlo
approaches
\citep[e.g][]{Sesana+2009,McWilliams+2014,Kulier+2015,Kelley+2017},
and it is typically estimated that $A\simeq (0.1 -
6)\times10^{-15}$. However, so far, most of the studies have relied on
the galaxy (or dark matter halo) merger history (merger rate and
merger mass ratio) and assumed that the SMBH coalescence rates track
the galaxy merger rates.

In this paper we adopt a {\it dynamical} approach to SMBH orbital
evolution following mergers, and we use it to estimate BH merger rates
for both the full and the empty loss cone scenarios.  Given the merger
histories of galaxy samples in a mass range
$M_{\star}=10^{10}-10^{12}\Msol$, for $0<z<4$, from the Millennium
simulation \citep{Springel+2005}, we follow the evolution of SMBH
binaries and their coalescences as the host galaxies go through
minor/major mergers.  Based on the inferred merger rates, we then
predict the stochastic GW background. Our work is a first attempt to
compute the global GWB by using few-body simulations to follow the
dynamical evolution of multiple SMBH systems as a consequence of
multiple galaxy mergers with a broad range of mass ratios
($10^{-4}<q_{\star}<1$, where $q_{\star}$ is the mass ratio of two
merging galaxies, defined to be smaller than 1).  We explore two
extreme scenarios for the last stage of the decay of a hard binary to
bracket the range of outcomes to the final parsec problem: the full
loss cone and the empty loss cone limits. In the empty loss cone case,
dynamical friction no longer affects the evolution of the orbits when
binaries become hard.  We treat the full loss cone case in a more
approximate way, assuming that dynamical friction always operates
efficiently to cause orbital decay down to the merger.  This is merely
an approximation to the more complex physics of stellar scattering in
the full loss cone regime (and furthermore neglects hydrodynamical
solutions to the final parsec problem), but as we argue later, it is
 a reasonable approximation for high mass-ratio systems.

In our suites of simulations, we find that SMBH binaries merge in both
scenarios, but with higher coalescence rates in the full loss cone
case than in the empty loss cone one.  In the full loss cone model,
when GW-driven evolution becomes more dominant, the binary
eccentricities are almost unity ($e>0.99$), confirming some past
predictions \citep{Quinlan1996, AntoniniMerritt2012}. Subsequently,
SMBH binaries coalesce rapidly (binary lifetimes $\lesssim1\Gyr$).  On
the other hand, in the empty loss cone model, multi-body interactions
of SMBHs play an important role in the decaying and coalescing of SMBH
binaries.  The binary lifetimes are longer ($\gtrsim1\Gyr$).  Using
the inferred BH coalescence rates, we estimate 
$A$ between the two models, $A=0.70\times10^{-15}$ and
$A=0.53\times10^{-15}$ for the full loss cone and the empty loss cone
case, respectively. They are comparable because 
(i) the higher coalescence rates of the full loss cone model come mostly from higher rates of low mass ratio mergers that contribute little to the GWB; high mass ratio systems merge in both models, (ii) more abundant and louder BH coalescence
    events at a later time (i.e. more massive mergers via mass growth
    and multi$-$BH interactions at small $z$), and (iii) the larger
mass ratios of merged binaries, which increase the contributions of
less massive binary mergers (in less massive galaxies) to the
stochastic background signal, relative to the full loss- cone regime.

This paper is organized as follows. In \S
\ref{sec:numericalsetup}, we explain our numerical setup including
galaxy sampling (\S \ref{subsec:galaxysampling}) and describe 
our model galaxies (\S \ref{sec:modeldescription}) and
prescriptions for BH mergers (\S \ref{subsec:SMBHseeds} -
\ref{subsec:BHmergers}). We present our results in \S
\ref{result}. In \S \ref{discussion}, we estimate the stochastic GWB
 and further discuss the effects of high eccentricity on GW
spectra. Finally, we conclude with a summary of our findings in
\S \ref{summary}.

\section{Numerical Setup}\label{sec:numericalsetup}

In this section we describe the main ingredients of our galaxy/SMBH
modelling.  In particular, we detail how we select galaxy samples and
how we treat galaxy mergers and the consequent rearrangement in the
background potentials of dark matter (DM) and stars. We also describe
how we take into account the formation of SMBH binaries and how we
define a BH merger.

\begin{figure}
	\centering
	\includegraphics[width=7.9cm]{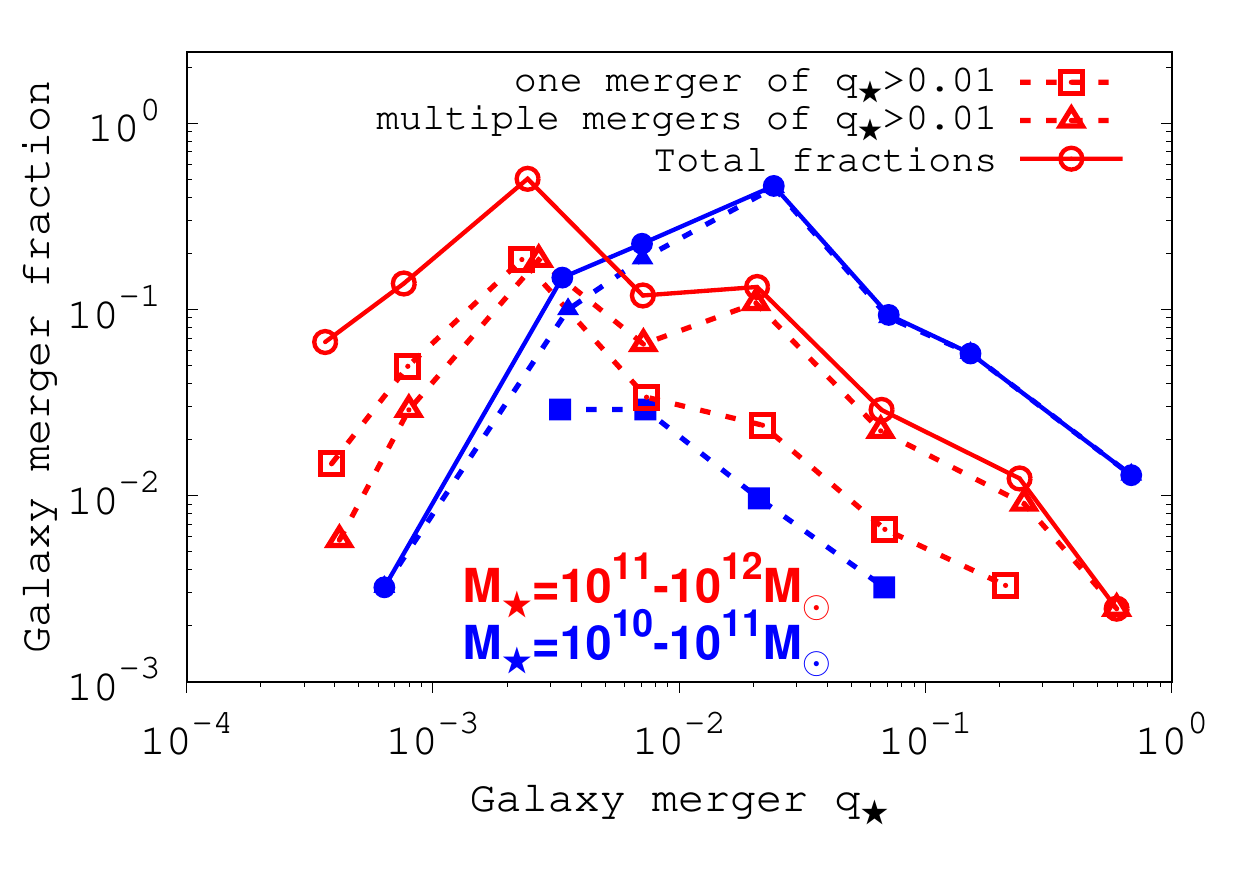}
	\caption{The fraction of galaxies that experience mergers 
		(solid line with circles) with a given galaxy merger mass ratio $q_{\star}$. We distinguish the host galaxies by the number of
          significant mergers ($q_{\star}>0.01$): the host galaxies
          with one significant merger (dotted line with squares) and
          those with multiple significant mergers (dotted line with
          triangles). Note that, while the lines for the
            galaxies experiencing no significant merger are not drawn,
            their contributions are included in the total fractions.}
	\label{fig:galaxymerger_q_count}
\end{figure}

\subsection{Sampling of dark matter (DM) and galaxy merger trees}
\label{subsec:galaxysampling}

We follow merger trees of DM subhalos sampled from the
Milli-Millennium simulation
\citep{Springel+2005}\footnote{http://gavo.mpa-garching.mpg.de/Millennium/}.
The Millennium simulation \footnote{There has been
  much progress in cosmological simulations since the
  Millennium: More advanced numerical techniques have been used in
  several simulations, such as `Illustris`
  \citep{Vogelsberger+2014}. Those simulations have successfully
  captured complicated effects induced by mutual interactions between
  gas, stars, DM and even BHs, which could not be achieved in DM-only
  simulations like the Millennium.  However, the general physical
  picture should be shared by all those simulations, in particular
  the treatment of DM/galaxy mergers
  \citep[e.g][]{RodriguezGomez+2015}, which is one of our main model
  ingredients. For the purpose of our study, the Millennium simulation allows us 
  more freedom to choose/implement different model ingredients
  under the same physical framework of galaxy formation.} is a large
$N$-body simulation of cosmological structure formation performed with
the GADGET-2 code assuming the standard $\Lambda$CDM cosmology with
the cosmological parameters of $\sigma_{m}=0.25, \sigma_{b}=0.045,
\sigma_{\lambda}=0.75, h=0.73$ and $\sigma_{8}=0.9$.  The simulation
follows the evolution of $N\approx10^{10}$ particles in a periodic box
of 500 $h^{-1}$Mpc on a side from $z=127$ to $z=0$. The simulation
provides a total of 64 snapshots at redshifts from $z\approx20$ to
$z=0$, equally spaced in $\log (1+z)$. Throughout this
paper, we assume that each DM halo hosts a galaxy whose mass is
proportional to that of the DM halo.

For galaxies at $z=0$ (denoted by ``host'' galaxy) in the Millennium
simulation, we follow the merger history of each host galaxy assuming
a SMBH seed located at the center, from the past ($z>0$) to the
present day ($z=0$). We will describe the detailed prescriptions for
seed SMBHs in \S \ref{subsec:SMBHseeds}. The total number of the
sampled host galaxies is 212. We consider galaxy mergers in each
tree up to 10 - 12 for each host galaxy. This amounts to a total of
1733 galaxy mergers. The stellar masses of the host galaxies (the scaling relation \ref{eq:scaling1} in \S \ref{subsec:DMstarBH_potential}) range within
$M_{\star}=10^{10}-10^{12}\Msol$\footnote{In this paper, the subscript
	$\star$ indicates physical quantities related to galaxies, while
	quantities with a subscript ``BH" or without one refer to the
	SMBHs. For example, we have galaxy masses $M_{\star}$, but BH masses
	$M_{\rm BH}$.  Similarly, galaxy merger mass ratios are indicated
	with $q_{\star}$, while BH binary mass ratios with $q$.}
(corresponding to virial masses of the host dark matter halos ranging from $M_{\rm DM, host}=10^{12}-10^{14}\Msol$ ). The earliest
galaxy merger occurs at redshift $z=3.58$ (or a cosmic time of
$t=1.76\Gyr$) and $z=4.18$ (cosmic time of $t=1.47\Gyr$) for host
galaxies in the mass ranges of $M_{\star}=10^{11}-10^{12}\Msol$ and
$M_{\star}=10^{10}-10^{11}\Msol$, respectively. In this paper, we
refer to (smaller) galaxies merging with the host galaxies as
``satellite'' galaxies of mass $M_{\star, \rm sat}$. The galaxy
merger ratio $q_{\star}$ between the satellite and host galaxy is
defined to be smaller than 1, namely $q_{\star}=M_{\star, \rm
	sat}/M_{\star, \rm host}$.

\begin{figure}
	\centering
		\includegraphics[width=7.9cm]{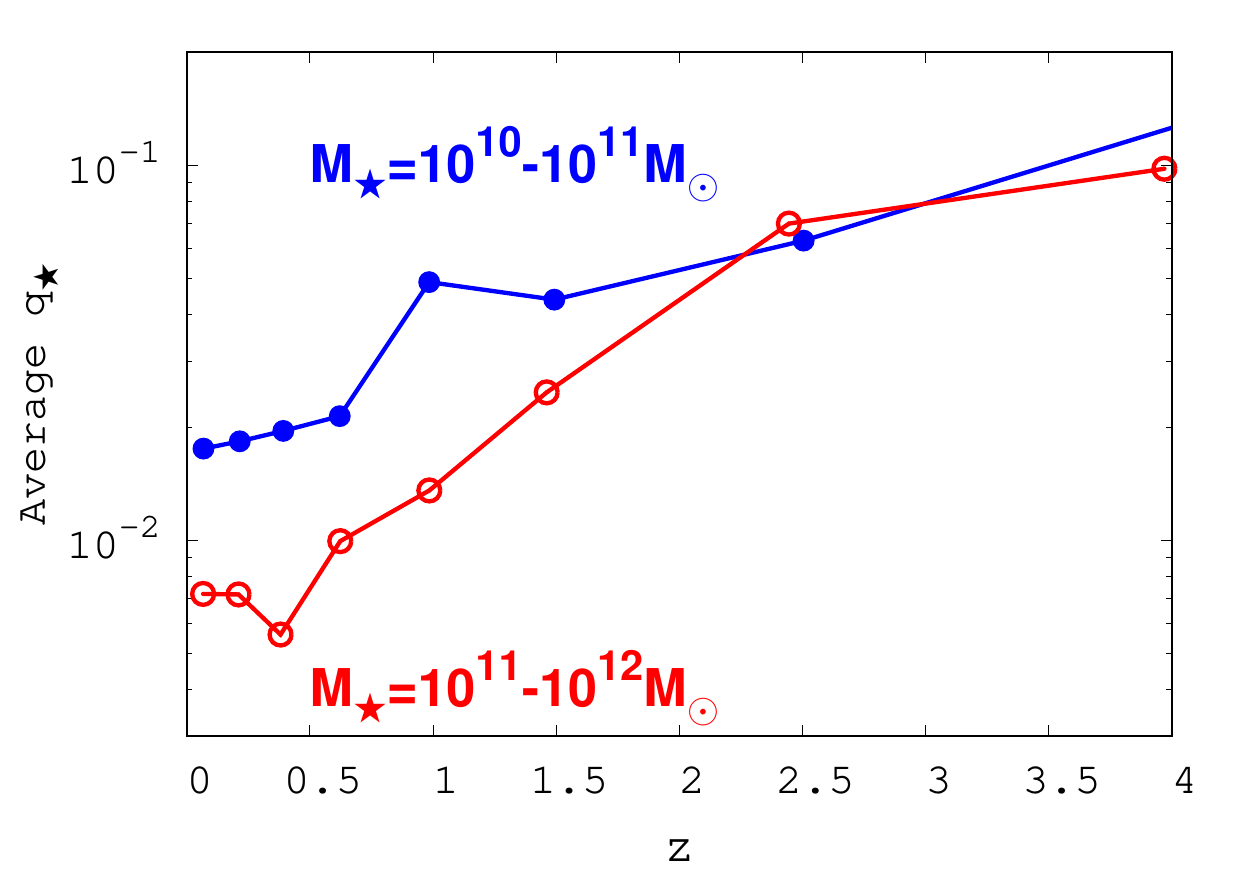}
	\includegraphics[width=7.9cm]{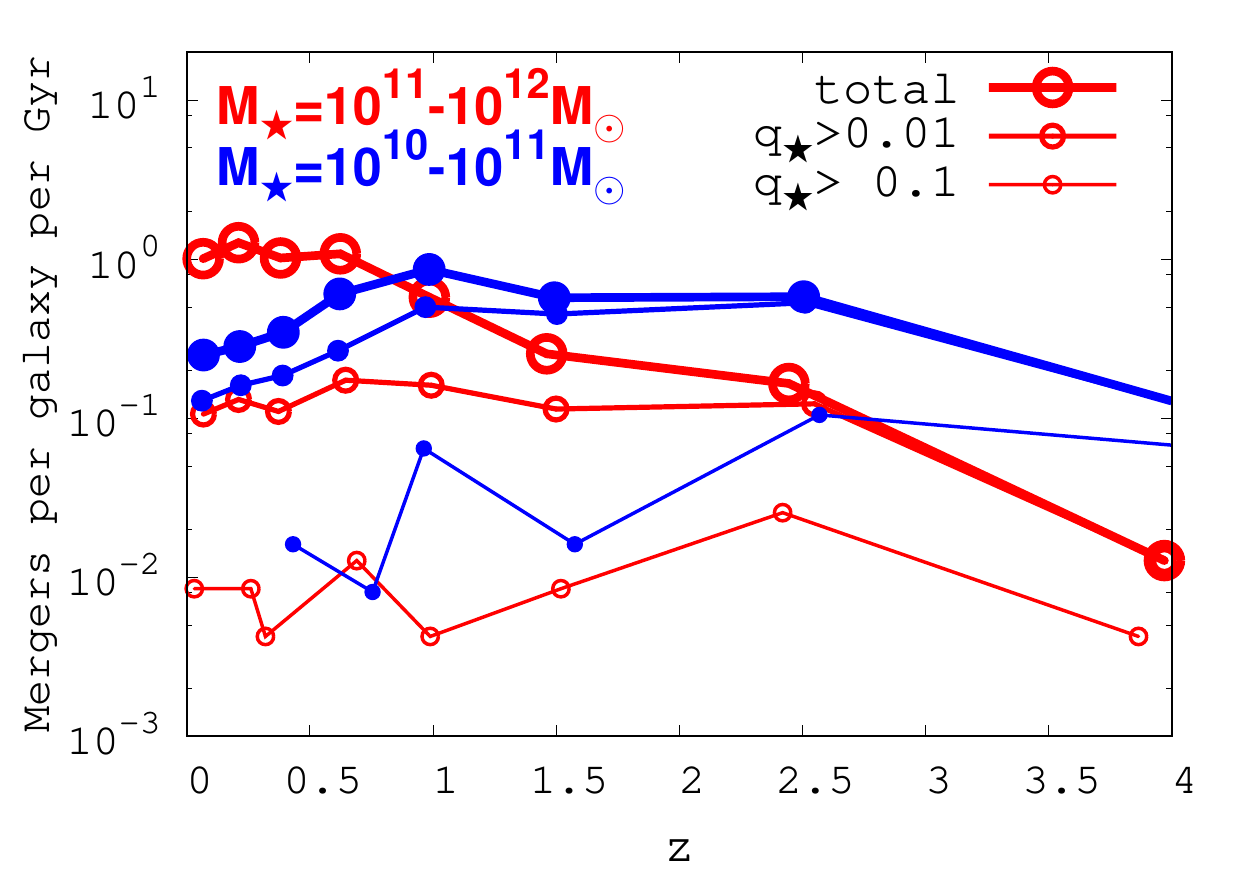}
	\includegraphics[width=7.9cm]{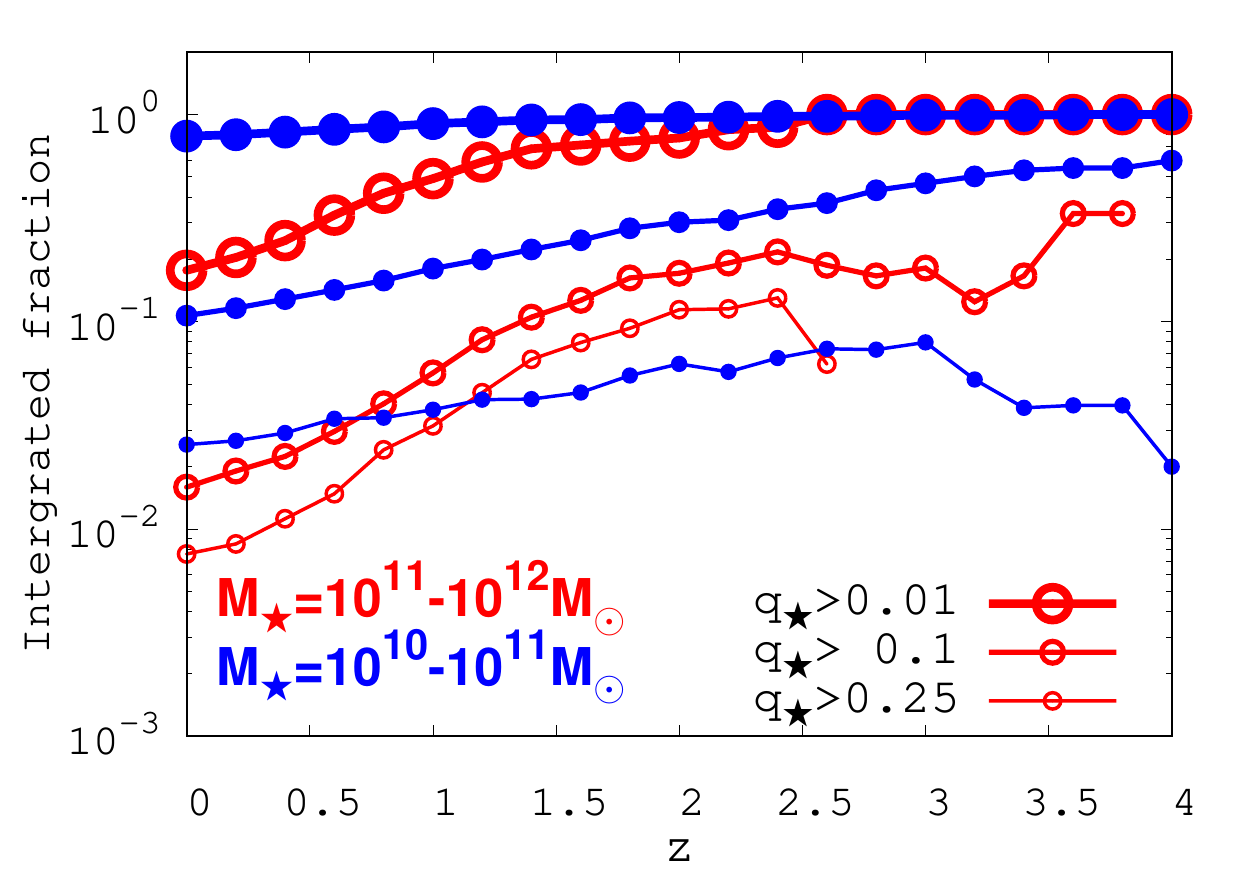}
	\caption{The evolution of the galaxy merger mass ratios (or the stellar mass ratio)
		$q_{\star}$, galaxy mergers and their number fractions as a function of redshift.  The \textit{upper}
		panel shows the merger mass ratios averaged per Gyr as a function
		of redshift. 
		Also, we present in the \textit{middle} panel
		the number of mergers per galaxy per Gyr. The
		line thickness indicates different, progressive cutoffs on the mass
		ratio: from the merger count without any cutoff (thickest
		line) to the mergers of $q_{\star}>0.1$ (thinnest line). In the \textit{bottom} panel, we present the number of significant galaxy mergers
		normalized by the total galaxy merger counts up to a given redshift. }
	\label{fig:galaxymerger_q_count2}
\end{figure}

In Figure \ref{fig:galaxymerger_q_count}, we show the fraction of
galaxies that go through mergers (solid line with circles) with a
given $q_{\star}$. We sub-categorize the host galaxies into
two bins depending on the number of significant mergers
($q_{\star}>0.01$)
\footnote{Throughout this paper, we only use the terms ``significant
  ($q_{\star}>0.01$)'', ``major ($q_{\star}>0.25$)'' and ``minor
  ($q_{\star}<0.25$)'' mergers to refer to galaxy mergers.}  they
experience: the host galaxies with one significant merger and those
with multiple significant mergers. We find that mergers with 
$q_{\star}<0.01$ are more common for galaxies of
$M_{\star}=10^{11}-10^{12}\Msol$ (the maximum count at
$q_{\star}\sim3\times 10^{-3}$). For galaxies of
$M_{\star}=10^{10}-10^{11}\Msol$, the most frequent are mergers of
$q_{\star}\sim3\times10^{-2}$ and the $q_{\star}$ distributions have
shorter low-$q_{\star}$ tails. In addition, we can see that the
majority of the host galaxies of $M_{\star}=10^{10}-10^{11}\Msol$
experience more than two major mergers.  Therefore, all of these
indicate that the merger mass ratio $q_{\star}$ is generally higher
for less massive host galaxies. This may imply that SMBHs merge with
relatively high rates in galaxies of $M_{\star}=10^{10}-10^{11}\Msol$
compared to more massive galaxies, which will be shown in \S
\ref{sub:BHmergerrates}. We also summarize those merger counts in
Table \ref{table:galaxymergercount}.

We present in Figure \ref{fig:galaxymerger_q_count2} the average mass
ratio $q_{\star}$, the merger rate per galaxy and the merger fraction,
as a function of redshift.  The \textit{upper} panel shows the merger
mass ratios averaged per Gyr as a function of redshift. It can be seen
that the merger mass ratio is generally higher for galaxies of
$M_{\star}=10^{10}-10^{11}\Msol$, independent of redshift.  In the
\textit{middle} panel we show the number of mergers per galaxy per
Gyr. The thickness of the line represents progressive, different
cutoffs on the mass ratio: from the merger count without any cutoff
(thickest line) to the mergers of $q_{\star}>0.1$ (thinnest line). In
the \textit{bottom} panel we present the cumulative distribution of
significant mergers in $z$, or the number fraction of significant
mergers integrated up to a given redshift.

\begin{table*}
	\centering
	\setlength\extrarowheight{4pt}
			\caption{Overview of our galaxy samples. From
                          top to bottom: The mass of sampled galaxies,
                          the total number of host galaxies, the total
                          number of galaxy mergers and BHs. In the
                          last four rows, we present the number
                          fraction of host galaxies experiencing
                          mergers with different mass ratios
                          $q_{\star}$ and frequencies $N_{\star}$:
                          (from the \textit{$4^{\rm th}$} to the last
                          row) one significant merger of
                          $q_{\star}>0.01$, more than one significant
                          mergers of $q_{\star}>0.01$, at least one
                          major merger of $q_{\star}>0.25$ and
                          $q_{\star}>0.5$.}
				\label{table:galaxymergercount}
	\begin{tabulary}{1\linewidth}{c c c c }
		\hline
		\multicolumn{2}{c}{Galaxy mass $M_{\star}$ }& $10^{10}-10^{11}\Msol$&   $10^{11}-10^{12}\Msol$  \\
		\hline
		\multicolumn{2}{c}{Total number of host galaxies (at $z=0$)} & 75& 137 \\
	\multicolumn{2}{c}{	Total number of galaxy mergers~/~BHs} & 462~/~457& 1271 ~/~1256\\
		Number fraction of host galaxies with one significant merger &$(N_{\star}=1, q_{\star}\geq0.01)$ & 7\%& 32\% \\
		Number fraction of host galaxies with more than one significant mergers &$(N_{\star}\geq2, q_{\star}\geq0.01)$ & 93\%& 42\% \\
		Number fraction of host galaxies with at least one major merger &$(N_{\star}\geq1, q_{\star}\geq0.25)$& 12\% & 7\% \\
		Number fraction of host galaxies with at least one major merger &$(N_{\star}\geq1, q_{\star}\geq0.5)$ & 8\% & 2\%\\
		
		\hline
	\end{tabulary}

\end{table*}

\subsection{Model description}
\label{sec:modeldescription}

In this section we describe our modelling of DM and
galaxy potentials, as well as the treatment of SMBHs, and in
particular their seed masses, their orbital parameters at galaxy
mergers, and their mass growth. Furthermore, we describe our treatment
of dynamical friction and the BH merger conditions using two different
prescriptions.

\subsubsection{Dark matter and stellar distribution and seed BH mass}
\label{subsec:DMstarBH_potential}

We model gas-poor galaxies with three components: DM, 
stars and SMBHs.
As we follow the merger histories of the host galaxies in the
Millennium simulation, at every galaxy merger DM and stellar
potentials are re-established based on the new mass of the galaxy
after the merger.

We adopt the NFW profile for the DM density distribution $\rho_{\rm DM}$ with 
concentration parameter $C=3$ \citep[e.g][]{VanWassenhove+2014}. For numerical 
convenience, we slightly modify the inner region of 
the NFW profile ($\rho_{\rm DM}\sim r^{-1}$) so that the DM density
does not exceed the stellar density at the very center of the galaxy core.
This only affects the region inside $\sim (10^{-3}-10^{-4})r_{\rm
	c}$, and does not appreciably affect our results. 

We consider the stellar density distribution for merged galaxies explored in \citet{StoneOstriker2015}:
\begin{equation}
\rho_{\star}=\frac{\rho_{\rm c}}{(1+r^{2}/r_{\rm c}^{2})(1+r^{2}/r_{\rm h}^{2})}\,,
\label{eq:stellardensity}
\end{equation}
where $\rho_{\rm c}$ is the central density, $r_{\rm c}$ is the core radius and 
$r_{\rm h}$ is the outer halo radius (or half-mass radius). The profile has a flat central core in the innermost region ($r<r_{\rm c}$), smoothly extending outward with $\rho_\star \propto r^{-2}$ 
for $r_{\rm c}\leq r<r_{\rm h}$ and $\rho_{\star} \propto r^{-4}$ for $r_{\rm h}\leq r$.

The post-merger stellar density profile $\rho_\star(r)$ of a merged galaxy is more complex than this idealized model, but our choice of $\rho_\star(r)$ is motivated by observations of large elliptical galaxies that are likely the primary hosts of PTA sources.  Specifically, {\it Hubble Space Telescope} ({\it HST}) observations of the nuclear regions of nearby early type galaxies find a bimodality in surface brightness profiles $I(R)$ (here $R$ is a projected 2D radius, as opposed to the 3D radial coordinate $r$).  When power law profiles are fit to the inner isophotes of {\it HST} data, i.e. $I(R) \propto R^{-\Gamma}$, the resulting $\Gamma$ distribution is strongly bimodal, with most galaxies having either $0<\Gamma<0.3$ or $0.5<\Gamma<0.9$ \citep{Lauer+2005}.  The former type of galactic nucleus, known as a ``core'' profile, is dominant among galaxies brighter than $M_V \approx -20$ \citep{Graham+2003,Graham+2003b}, and is roughly consistent with the flat inner slope one obtains by projecting Equation \ref{eq:stellardensity}.

Flat cores in surface brightness profiles could be created by the
dynamical effects of SMBH binaries. In the aftermath of a galaxy
merger, hosted SMBHs are effectively dragged towards the centre of the
merged galaxies by dynamical friction, and eventually form a
binary. The binary acts as a heating source as its orbit shrinks,
pumping the lost energy to the background stellar populations. The
deposition of the binary's orbital energy can scour out a flat core of
stars in the inner region, creating a mass deficit relative to the
initially steeper density profile \citep[e.g. see Chapter
    7 in ][and references therein]{Merritt2013}. The creation of flat
cores by SMBH binaries has been confirmed in numerical simulations
\citep[e.g.][]{Merritt2006,Gualandris+2012,KulkarniLoeb2012,Bortolas+2016}.
Furthermore, stellar scouring has been inferred in a
  number of core elliptical galaxies from observations
  \citep[e.g.][]{Thomas+2014}, and is widely predicted in numerical studies
  \citep[e.g.][]{Milosavljevic2001,KormendyHo2013}.  Although both
dynamical friction and three-body stellar scatterings contribute to
core creation in the vicinity of an SMBH binary, we only include the
former \citep{Ebisuzaki+1991} in our model\footnote{The anisotropic
  emission of GWs (or ``gravitational rocket effect'') during the
  final coalescence of two SMBHs may also produce a mass deficit in
  galactic nuclei \citep{Merritt+2004,Gualandris2008} following recoil
  of the merged SMBH, but we neglect this in our model.}.  We discuss
limitations of our simple treatment of phase (ii) later in this
section.

\begin{figure*}
	\centering
	\includegraphics[width=8.4cm]{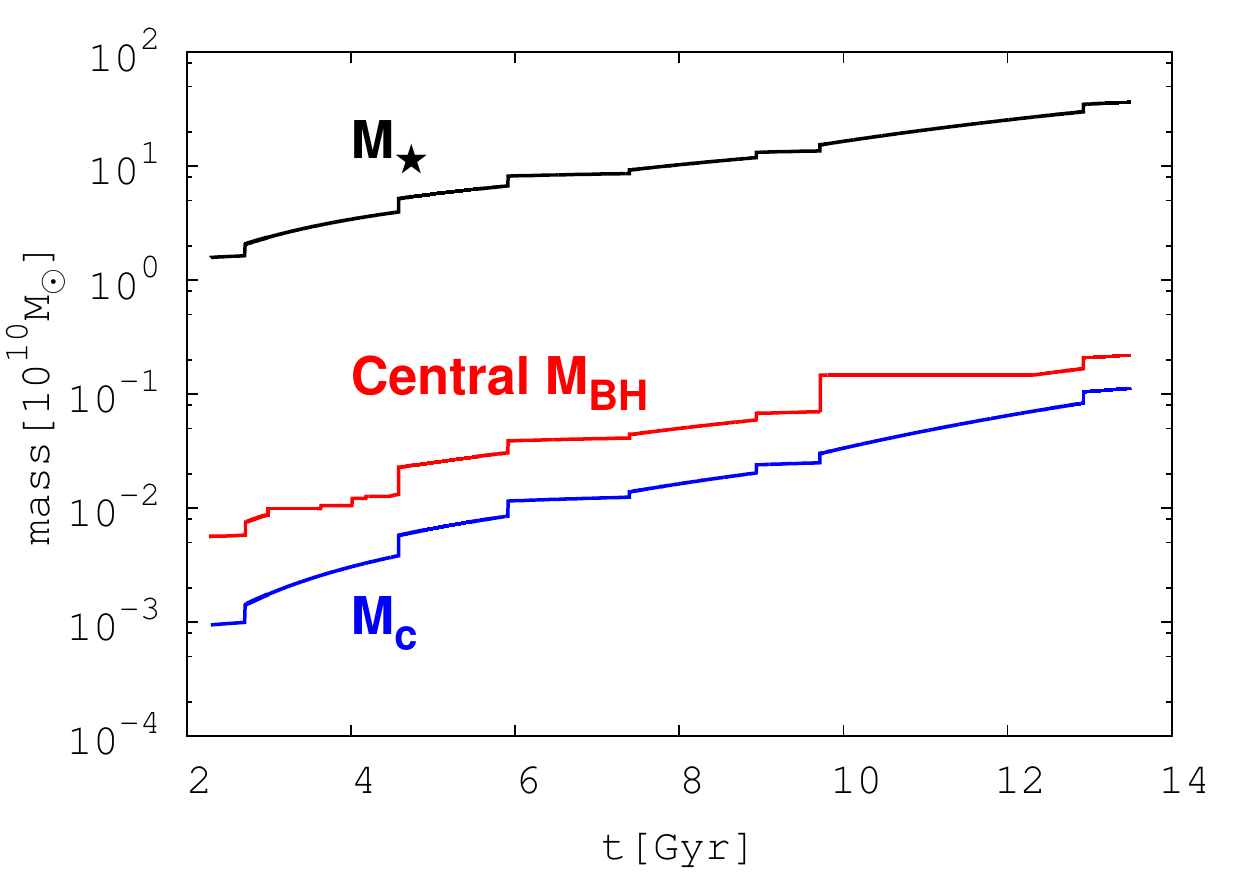}
	\includegraphics[width=8.4cm]{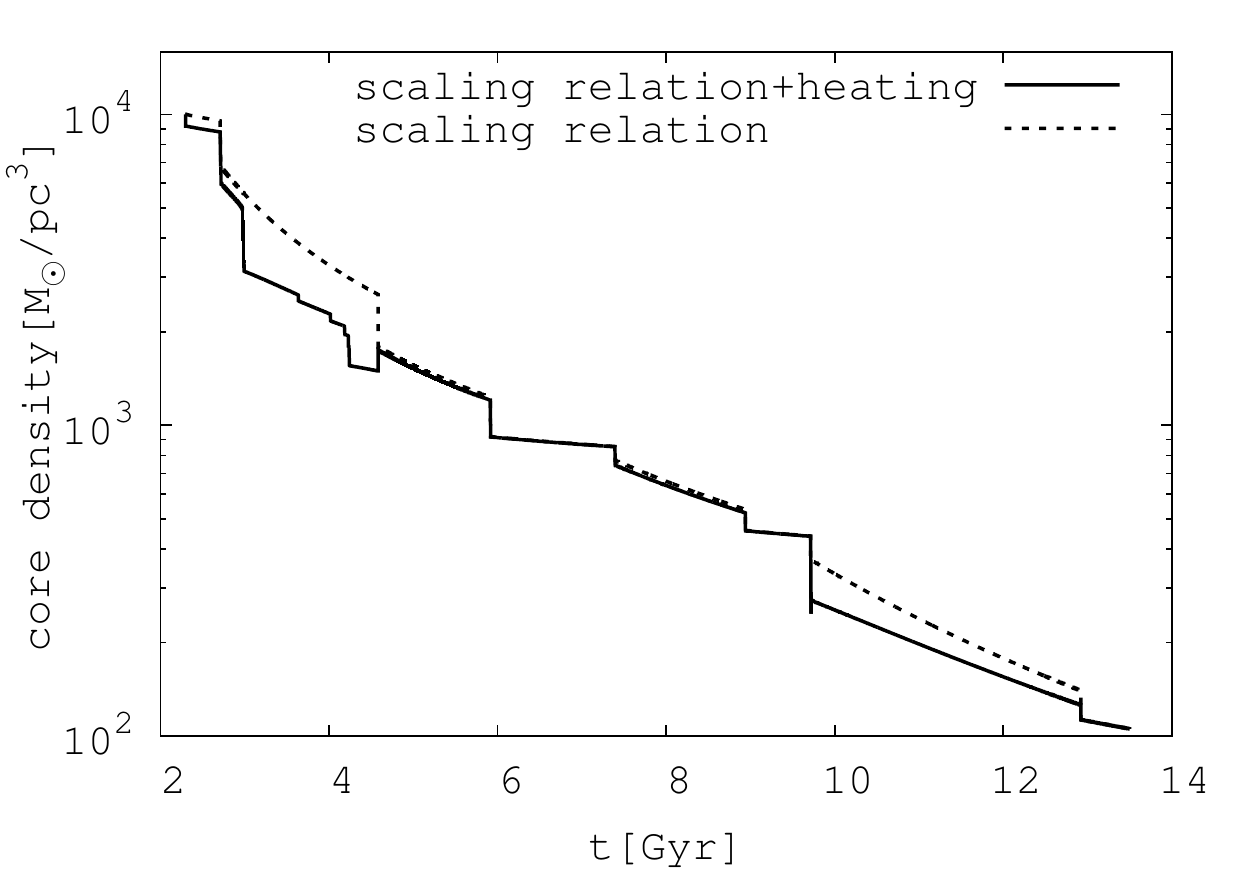}
	\includegraphics[width=8.4cm]{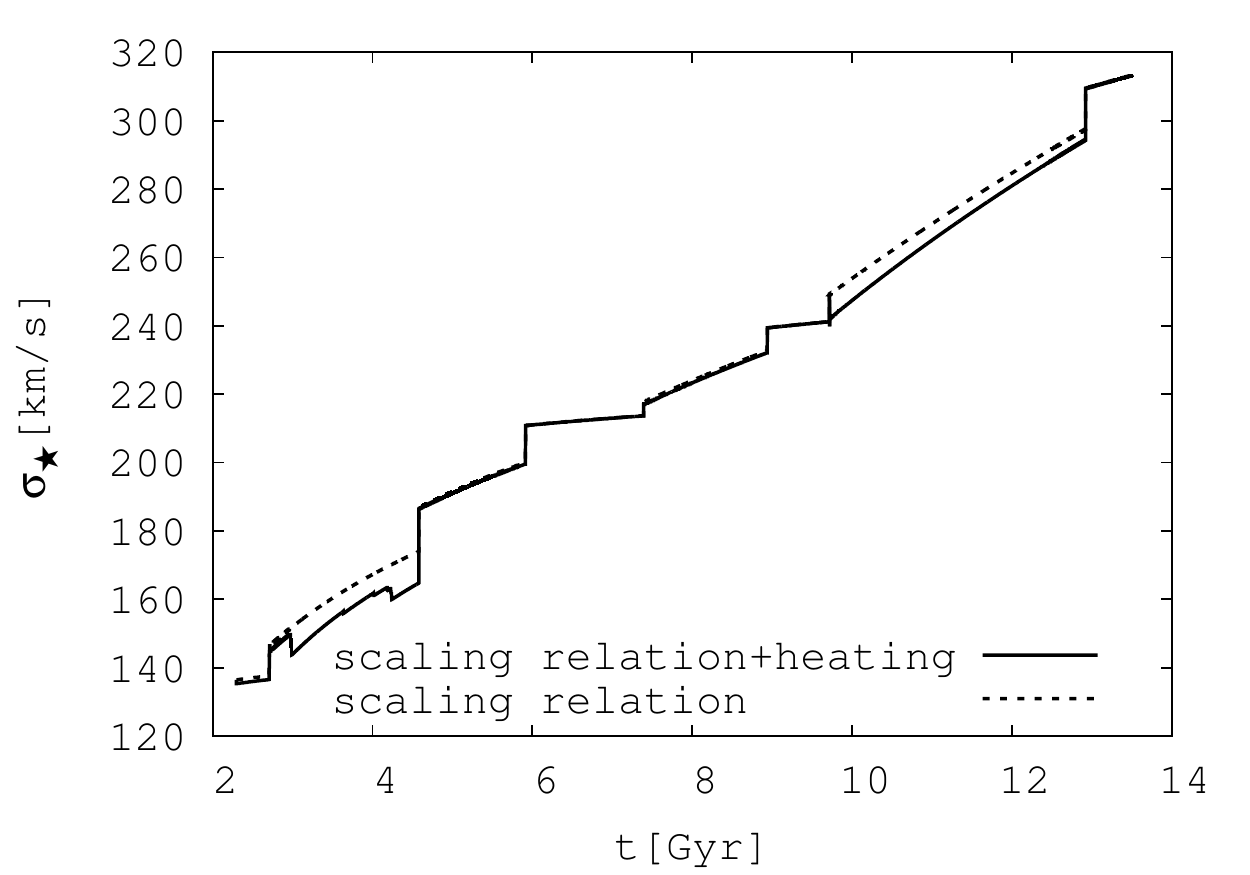}
	\includegraphics[width=8.4cm]{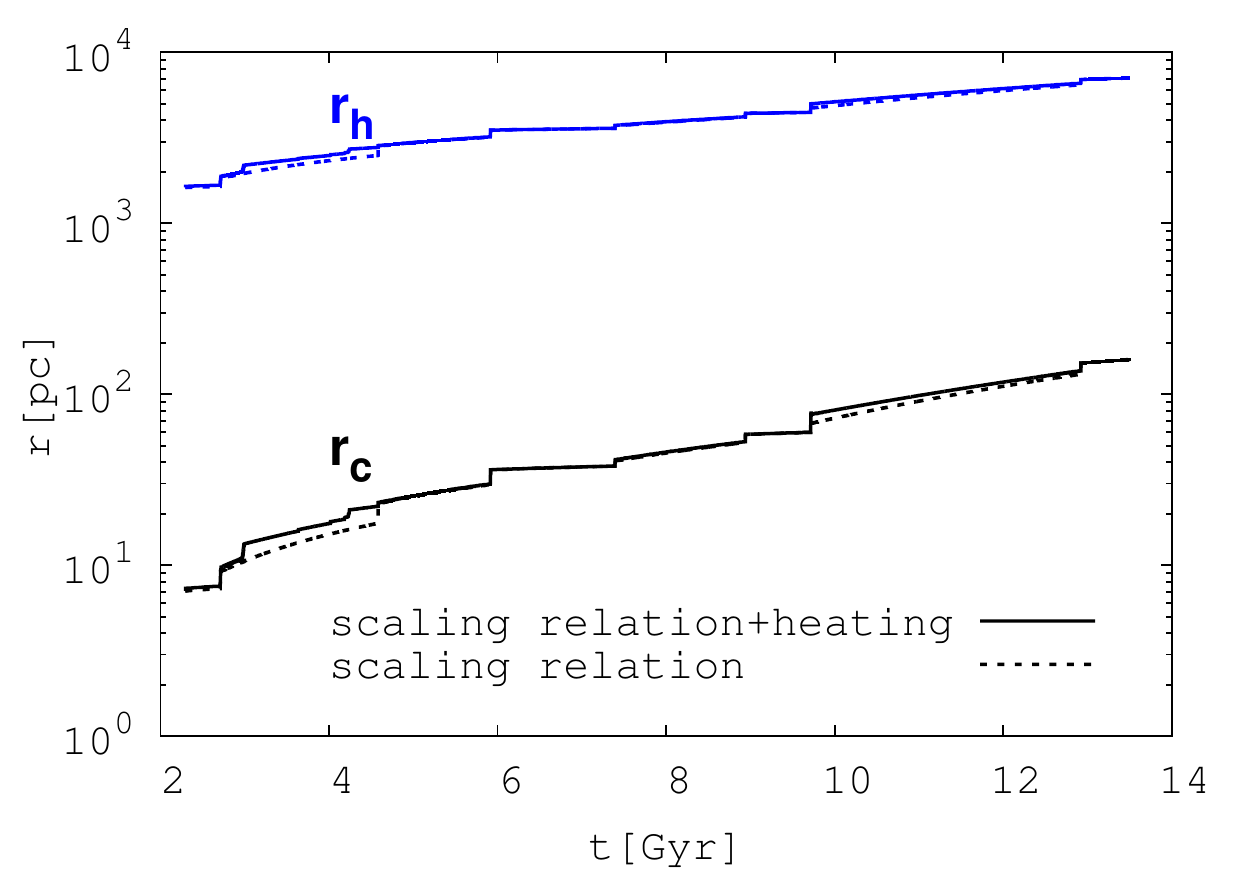}
	\caption{The evolution of stellar mass ($M_{\star}$), central BH mass (central $M_{\rm BH}$) and core mass ($M_{\rm c}$)
		(\textit{top-left}), core density $\rho_{\rm c}$ (\textit{top-right}), $\sigma_{\star}$ 
		(\textit{bottom-left}) and two characteristic radii $r_{\rm h}$ and $r_{\rm c}$ (\textit{bottom-right}) of 
		one massive galaxy among the sampled galaxies. We show those 
		variables as determined only by the scaling relations (dotted lines) as well as when 
		the heating effect due to dynamical friction is additionally taken into account 
		(solid lines). The host galaxy grows via 9 mergers from 
		$M_{\star}\simeq10^{10}\Msol$ at $z=2.8$ ($t=2\Gyr$) to 
		$M_{\star}\simeq 3\times10^{11}\Msol$ at $z=0$ ($t=13.8\Gyr$). 
		The mass of the central (most massive) BH has reached 
		$M_{\rm BH}\simeq10^{9}\Msol$ at $z=0$. Overall, 
		the core swells ($r_{\rm c}$ and $M_{\rm c}$) as 
		$M_{\star}$ increases, but $\rho_{c}$ declines. }
	\label{fig:scaling_heating}
\end{figure*}

For a given DM halo mass $M_{\rm DM,host}$ at redshift $z$, the DM density 
distribution is completely determined. However, we have three free parameters 
for the stellar potential to be fixed, namely, $r_{\rm c}$, $r_{\rm h}$ and $\rho_{\rm c}$. 
In order to fix those parameters as well as the seed SMBH mass, we solely depend 
on four observational scaling relations:

\begin{enumerate}
	\item \label{eq:scaling1}$M_{\rm DM}-M_{\star}$ relation  : $\left(\frac{M_{\rm DM}}{10^{13}\Msol}\right)=0.50\left(\frac{M_{\star}}{10^{11}\Msol}\right)$ \\ \citep{Lin+2012,Kulier+2015}
	\item \label{eq:scaling2}$M_{\rm BH}-M_{\star}$ relation : $\left(\frac{M_{\rm BH}}{10^{9}\Msol}\right)=0.49 \left(\frac{M_{\star}}{10^{11}\Msol}\right)^{1.16}$ \citep{KormendyHo2013}
	\item\label{eq:scaling3} $M_{\rm BH}-\sigma$ relation    : $\left(\frac{M_{\rm BH}}{10^{9}\Msol}\right)=0.309 \left(\frac{\sigma}{200\km \s^{-1}}\right)^{4.38}$ \citep{KormendyHo2013}
	\item \label{eq:scaling4}$M_{\rm BH}-r_{\rm c}$ relation  : $\left(\frac{r_{\rm c}}{\rm kpc}\right)=0.0821\left(\frac{M_{\rm BH}}{10^{9}\Msol}\right)^{0.855}$ \citep{Thomas+2016},
\end{enumerate}
We note that we ignore the scatter in the above relations,
  assuming them to be exact. \footnote{Taking
    into account scatter and the impact of different choices of
    the SMBH-galaxy relation to populate SMBHs may overly introduce
    complexicities to our analysis. For simplicity, we neglect scatter in galaxy scaling relations used in our model.  However, this issue has been addressed before, for example in 
    \citet{Shankar+2016,Sesana+2016,RasskazovMerritt2017}. } While
the general $M_{\rm DM}-M_\star$ relation is more complicated than our
prescription \citep[e.g][]{Moster+2013}, a single power law is a
reasonable approximation to the high mass end of this relation that we
focus on. With these relations, the DM, the stellar distributions, and
the seed SMBH mass are determined given the DM halo mass. In
particular, we assume that central SMBHs are missing in galaxies of
$M_{\star}<10^{8}\Msol$, which correspond to the minimum BH mass
$M_{\rm BH}=10^{5.5}\Msol$ in our simulations. For mergers with such
small galaxies, we simply add the masses of the small galaxies to the
host galaxy masses without placing the seed SMBHs. Those small
galaxies occupy around 1\% of the total number of satellite galaxies
for all galaxy mass ranges.

In order to see how the stellar potential evolves as the total stellar mass increases, 
we express $r_{\rm c}$, the core stellar mass $M_{\rm c}$ and $\rho_{\rm c}$ in 
terms of $M_{\star}$:
\begin{align}
r_{\rm c}&\propto M_{\star}^{0.99}\,, \label{eq:relation1}\\
M_{\rm c}&\propto M_{\star}^{1.52}\,,\label{eq:relation2}\\
\rho_{\rm c}&\propto M_{\star}^{-1.46}\,. \label{eq:relation3}
\end{align}
The relations are derived in Appendix \ref{appendix:scalingrelation} and imply that as galaxies (DM subhalos) grow in mass, the
core regions expands in size and mass whereas the core density
declines \citep{DulloGraham2014}. Even though the dependencies on $M_{\star}$ differ, those
trends are consistent with those of \citet{Faber+1997}, i.e. $r_{\rm
  c}\sim M_{\star}^{0.92}$, $M_{\rm c}\sim M_{\star}^{1.24}$ and
$\rho_{\rm c}\sim M_{\star}^{-1.52}$.  As an example, in Figure
\ref{fig:scaling_heating} we show the evolution of $M_{\star}$, the
central BH mass and $M_{\rm c}$, $\rho_{\rm c}$, $\sigma_{\star}$ and
$r_{\rm h}$ and $r_{\rm c}$ of one of the more massive galaxies in our sample. In the plots, we show
those variables as determined only by the scaling relations as well as when the heating effect due to dynamical friction
(see below) is additionally taken into account.  The host
galaxy grows via 9 mergers from $M_{\star}\simeq10^{10}\Msol$ at
$z=2.8$ ($t=2\Gyr$) to $M_{\star}\simeq 3\times10^{11}\Msol$ at $z=0$
($t=13.8\Gyr$). The mass of the central (most massive) BH has reached
$M_{\rm BH}\simeq10^{9}\Msol$ at $z=0$. As expected, the core swells
($r_{\rm c}$ and $M_{\rm c}$) as $M_{\star}$ increases, but $\rho_{c}$
declines.  Notice that $r_{\rm h}$ has a relatively weak dependence on
$M_{\rm c}$ compared to $r_{\rm c}$, i.e., $r_{\rm h}\sim
M_{\star}^{0.47}$. We go into greater detail on evolutionary deviations due to heating effects and the prescription 
for BH mass growth in the following section (\S~\ref{subsubsec:DM_stars_BHmass}).

We provide in Table \ref{table:scalingrelation} in Appendix \ref{appendix:scalingrelation} 
the scaling relations between the relevant variables in our model in terms of $M_{\rm BH}$ 
(as well as $M_{\rm DM}$), derived with the four scaling relations \ref{eq:scaling1}-\ref{eq:scaling4}.

\subsubsection{Evolution of DM and galaxy potential and BH mass growth}
\label{subsubsec:DM_stars_BHmass}
As galaxies merge, DM and stellar potentials evolve in time. For the DM
potential, we interpolate the DM halo masses between two adjacent
galaxy mergers (or two different redshifts or snapshots at mergers) in
the Millennium simulation.  In particular, we use a fitting formula
derived by \citet{Wechsler+2002}, which can be written as follows,

\begin{align}
M_{\rm DM}(z)=M_{\rm DM,0}\exp\left[-\Delta\left(\frac{z+1}{z_{0}+1}-1\right)\right]\,,
\end{align}
where $z_{\rm 0}$ is the redshift when a halo is observed. Here, we
assume $z_{\rm 0}$ to be the same as the redshift at which two halos
merge in the Millennium simulation.  Therefore, given the mass of a
merged halo (or merged galaxy) at $z=z_{0}$ and the subsequent merger
at $z=z_{1}$, we determine $\Delta$\footnote{More explicitly, $\Delta$
  is expressed in \citet{Wechsler+2002} as $S/(1+z_{\rm c})$, where
  $z_{\rm c}$ is the redshift at which the halo collapses and $S$ is a
  characteristic factor which relates the accretion rates of
  halos.}. In the Millennium simulation, DM halos typically grow in
mass from one merger to the following merger.  However, there are also
cases where the DM halo masses at subsequent mergers are found to be
smaller. On average, sampled host galaxies experience such decreases
in mass once in their merger histories.  This could be caused by
several mechanisms, and in particular tidal stripping. But the precise
cause cannot be determined from the information provided in the
snapshots alone. In this paper, for such cases, we conservatively
assume that DM halo masses do not change between the two mergers, but
we update the halo masses accordingly at the later merger.  In
addition to the growth of DM halos, we also take into account the
widening of the stellar potential due to the ``scouring effect''
\citep{Milosavljevic+2002,Merritt2006} of SMBH binaries as a result of
dynamical friction. As the orbits of SMBH binaries shrink, they lose
their energy to background stars, which will clear out some stars onto
wider orbits. To quantify this effect {on the stellar
  potential \footnote{Our treatment of scouring
    only alters the stellar, not the DM, density profile. However, our
    results are not significantly affected by whether or not the DM
    density profile is influenced by scouring effects since the
    stellar potential is dominant near the core regions where binaries
    and multiple BHs interact.} we compute, at every time step
$\Delta t$, the dissipative energy $E_{\rm dis,i}$ due to the
dynamical friction force $\textbf{\textit{f}}_{\rm df,i}$ (see
Equation \ref{eq:dynamicalfriction}) for the $i_{\rm th}$ BH moving at
velocity $\textit{\textbf{v}}_{i}$,
\begin{equation}
E_{\rm dis}=\sum_{i}|\textbf{\textit{f}}_{{\rm df},i}\cdot\textbf{\textit{v}}_{i}|\Delta t.
\end{equation}
Hence we deposit $E_{\rm dis}$ into the virialized stellar potential assuming the 
total mass of stars $M_{\star}$ is fixed and the three-parameter structure of the density distribution 
is maintained. By the virial theorem, the total potential energy of stars $W_{\star}$ 
can be expressed in terms of the total binding energy of stars $E_{\star}$, the 
dissipative energy $E_{\rm dis}$ and the virially averaged dispersion $\sigma_{\star}$ as follows,
\begin{align}
-W_{\star}=-2(E_{\star}+E_{\rm dis})=M_{\star}\sigma_{\star}^{2}.
\label{eq:scouring}
\end{align}
\citet{StoneOstriker2015} provide the explicit expressions for the
total potential energy $W_{\star}$ (Equation 8) and $M_{\star}$
(Equation 5) in terms of $\rho_{\rm c}$, $r_{\rm c}$, $r_{\rm
	h}$ and $\sigma_{\star}$. With the scaling relations \ref{eq:scaling1}-\ref{eq:scaling4},
we can then estimate the adjusted values of $\rho_{\rm c}$, $r_{\rm
	c}$ and $r_{\rm h}$, and update them accordingly at every time step.
Given $W_{\star},~E_{\star}<0$, the scouring effect produces an expansion
of the chracteristic size of the potential ($r_{\rm c}$ and $r_{\rm h}$), while 
lowering the core density $\rho_{\rm c}$, as shown in Figure
\ref{fig:scaling_heating}. However, note that the decrease in the core
density is accompanied by mass growth of the galaxies.

In our simulations, the masses of the central BHs increase such that
$M_{\rm BH}-\sigma$ (scaling relation iii) is always satisfied. The
central BHs are defined in this paper as BHs whose entire orbits
(either with respect to galaxy potential or in binaries with other
BHs) are confined to the core. If BHs only temporarily stay in the
core region at their closest approach (pericentre) to the origin, they
are not identified as central BHs. In our simulations, we find that
the central BHs typically include the most massive BHs (denoted by
BH$_{1}$ and their masses $M_{\rm BH,1}$) and the BHs forming bound
pairs with BH$_{1}$. The total mass of the central BHs (denoted by
$M_{\rm cBH}$) is mostly dominated by $M_{\rm BH,1}$. If the BH mass
required by $M_{\rm BH}-\sigma$ (denoted by $M_{\rm
  BH,\sigma_{\star}}$ where $\sigma_{\star}$ is the virially averaged
dispersion defined in Equation \ref{eq:scouring} \footnote{
  Note that the variable $\sigma$ used in the $M_{\rm BH}-\sigma$
  relation may not have exactly the same meaning as $\sigma_{\star}$
  of the virially averaged dispersion. However, considering systematic
  uncertainties in the dispersion measure \citep{Tremaine2002}, we
  conservatively assume the virially averaged value of
  $\sigma_{\star}$ as a representative for the dispersion of the host
  galaxy. \citet{StoneOstriker2015} show that the virially averaged
  dispersion is comparable to the central dispersion for $r_{\rm
    h}\gg r_{\rm c}$.}) is already smaller than $M_{\rm cBH}$, the mass
of each central BH stays the same. On the other hand, for $M_{\rm
  BH,\sigma_{\star}}>M_{\rm cBH}$, given the mass $M_{\rm BH,i}$ of
each central BH$_{i}$ at a certain time step, the mass of the BH$_{i}$
at the following time step $M_{\rm BH,i}'$ increases by a factor of
$M_{\rm BH,\sigma_{\star}}/M_{\rm cBH}$, or simply,
\begin{align}
\label{eq:massgrowt}
M_{\rm BH,i}'=M_{\rm BH,i}\frac{M_{\rm BH,\sigma_{\star}}}{M_{\rm cBH}}.
\end{align}
With this crude approximation for mass growth through gas accretion
we ensure that the masses of the central BHs are
maintained at realistic values.  On the other hand, whenever the
central massive BHs are missing in the core regions, given the mass
reservoir in these regions, the masses of other small BHs, which
fall into the core later or have already existed, could grow rapidly
up to masses comparable to the missing central BHs. In particular, in
our models assuming instantaneous formations of post-merger galaxies,
this can take place while the central BHs are dislocated off center at
galaxy mergers. Only two such cases occurred in our simulation suite,
and to be conservative, we exclude the contribution of these to the
GWB (see \S \ref{discussion}).

\subsection{Initial orbital parameters of SMBHs at galaxy mergers}

\label{subsec:SMBHseeds}

In hierarchical models of structure formation in cosmology, DM halos
grow via mergers as well as accretion of DM. 
 During the process of merging, the orbital
properties of infalling satellite halos have been investigated in many
studies. Recent cosmological $N$-body simulations show that two halos typically merge on almost parabolic orbits with large
eccentricity for various ranges of the halo mass, mass ratio and
redshift
\citep{Benson2005,KhochfarBurkert2006,Wetzel2011,Jiang+2015}. 
Additionally, studies of SMBH binary formation in merging
galaxies generally assume such radial orbits for infalling SMBHs as
initial conditions \citep{KulkarniLoeb2012,
  VanWassenhove+2014,Capelo2015}. Motivated by those studies, we also
assume the initial orbits of incoming BHs with respect to the merged
galaxy potential to be highly eccentric. In particular, we adopt a
fitting formula for the eccentricity given in
\citet{Wetzel2011}. Using cosmological $N$-body simulations,
\citet{Wetzel2011} investigated the orbital parameters of infalling
satellite halos and their dependences on the halo mass and redshift.
The author provides a simple functional form of the orbital
distribution of the satellite circularity $\eta$ for $z=0-5$ and $M_{\rm
  DM,host}=(10^{10}-10^{15})h^{-1}\Msol$. The distribution of the circularity 
  		$df/d\eta$ adopted in this study is expressed as follows,
\begin{align}
\frac{df}{d\eta}&= 3.38\left( 1+0.567\left[\frac{M_{\rm DM,host}}{M_{0}}\right]^{0.152}\right)\nonumber\\
&\times\eta^{1.05} (1-\eta)^{0.242\left(1+2.36\left[\frac{M_{\rm DM,host}}{M_{0}}\right]^{0.108}\right)}\,,
\label{eq:initialeccentricity}
\end{align}
where $\eta=\sqrt{1-e^{2}}$ and
$\log[M_{0}/h^{-1}\Msol]=12.42-1.56z+0.038z^{2}$. We estimate the
eccentricity using $e=(r_{\rm a}-r_{\rm p})/(r_{\rm a}+r_{\rm p})$,
where $r_{\rm a}$ and $r_{\rm p}$ are the apocentric and pericentric
distances of BH orbits, respectively, with respect to the galactic
potential.  For simplicity, the initial eccentricity is given in the
simulations as the peak value of the fitting formula, $e\simeq0.8-0.9$
at mergers.

For a merger between a host galaxy already hosting several BHs 
(BH$_{i}$ with $i\geq1$) and an incoming $j_{\rm th}$ satellite galaxy, we assume 
only one BH per satellite galaxy but we allow multiple mergers at the same redshift (i.e., $j\geq1$).
In the Millennium simulation, when more than two galaxies disappear
from one snapshot to the next one, we assume that they merge with the
host galaxy at the same time. For galaxy mergers at a given redshift, 
we find a post-merger galaxy system of a pre-existing BH cluster in a host galaxy 
and incoming BHs in satellite galaxies; these are found at the apocenter of their instantaneous orbits in the new spherical post-merger potential, re-established around the center of mass (CoM) of all BHs. The CoM of the pre-existing BH cluster and each of the incoming
    BHs are separated by $r\sim r_{\rm h}$   \footnote{Note that for multiple mergers ($j\geq2$), the separation between the two BH systems is not exactly $r_{\rm h}$ since they are not aligned on a line, but rather spread on a $2-$dimensional plane ($j=2$) or in a $3-$dimensional space ($j\geq3$).}.
For the given initial
positions (i.e., the apocenters of the initial orbits), the initial
velocities are assigned to give highly eccentric orbits as above. 
Finally, 
the positions and velocities of BHs in host galaxies
($\textbf{x}_{{\rm host},i}^{\rm BH'}$,$\textbf{v}_{{\rm host},i}^{\rm
  BH'}$) and in the $j_{\rm th}$ satellite galaxy ($\textbf{x}_{{\rm
    sat},j}^{\rm BH'}$,$\textbf{v}_{{\rm sat},j}^{\rm BH'}$) are
expressed for any number of mergers $(j\geq1)$ at a given redshift as follows,

\begin{align}
\textbf{x}_{{\rm host},i}^{\rm BH'}&=\textbf{x}_{{\rm host},i}^{\rm BH} + \frac{\sum_{j}M_{\star,j}}{M_{\star,\rm host}+\sum_{j}M_{\star,j}}\times r_{\rm h}\hat{\textbf{x}}_{{\rm host},i}^{\rm BH} \nonumber\\
\textbf{x}_{\rm sat,j}^{\rm BH'}&=\frac{M_{\star,\rm host}}{M_{\star,\rm host}+\sum_{j}M_{\star,j}}\times r_{\rm h}\hat{\textbf{x}}_{{\rm sat},j}^{\rm BH}\nonumber\\
\textbf{v}_{\rm host,i}^{\rm BH'}&=\textbf{v}_{{\rm host},i}^{\rm BH}\times \xi(q,n) +\sqrt{\frac{G M_{\rm en}(r<x_{\rm host,i}^{\rm BH'})}{x_{\rm host,i}^{\rm BH'}}(1-e)\alpha}\times\hat{\textbf{v}}_{{\rm host},i}^{\rm BH}\hspace{0.2in}\nonumber\\
\textbf{v}_{\rm sat,j}^{\rm BH'}&=\sqrt{\frac{G M_{\rm en}(r<x_{{\rm sat},j}^{\rm BH'})}{x_{{\rm sat},j}^{\rm BH'}}(1-e)\alpha}\times\hat{\textbf{v}}_{{\rm sat},j}^{\rm BH}\,,\nonumber
\end{align}	
where $\textit{\textbf{x}}_{i}$ and $\textit{\textbf{v}}_{i}$ (without
prime symbol) are the position and velocity vectors of BH$_{i}$ just
before mergers, and $\hat{\textbf{x}}_{i}$ and $\hat{\textbf{v}}_{i}$
are the randomly-generated unit vectors, satisfying
$\hat{\textbf{x}}_{i}\perp\hat{\textbf{v}}_{i}$ (same for $j$ as
well). $M_{\rm en}(r<x')$ is the enclosed mass inside of $r=x'$ and
$\alpha$ is a factor used to assign the eccentricity for the first
orbit in a non-Keplerian potential (See equation
\ref{eq:stellardensity}).  We conservatively use
  $\alpha\simeq1/5$ for the eccentricity ranges given by Equation
  \ref{eq:initialeccentricity}, i.e., $e>0.8$ \footnote{For
    the same velocity (not the circular velocity) at the same
    apocenter, the first pericenter distances are different in the
    Keplerian and Non-Keplerian potentials (i.e., different
    eccentricities). Therefore, some extra factor should be
    taken into account in the expression for $v$ at apocenter in the
    Keperian potential.  The value of $\alpha$ taken in this paper is
    comparable to that for the logarithmic potential ($\rho\sim
    r^{-2}$) \citep{Innanen+1982}. Recall that our stellar density
    approximately follows $\rho\sim r^{-2}$ at $r<r_{\rm h}$.}. Here,
we introduce a function $\xi(q,n)$ to quantify the extent by which a
host galaxy is disrupted by a merger. We define the function $\xi(q,n)$ 
		as a degree of memory for the orbits of existing BHs in the 
host galaxies at given mergers, scaling from 0 (complete loss of memory) 
to 1 (complete retention of memory). Motivated by the considerations
below, we define $\xi(q,n)$ assuming the following functional form,
\begin{align}
\xi(q,n)\equiv\left|\frac{q^{n}-1}{q^{n}+1}\right|\,,
\end{align}
where $q=\sum_{j}M_{\star,j}/M_{\star,\rm host}$. During the process of merger, 
it is more likely that the system of host galaxies is 
disrupted by mergers of high $q$. In other words, as they go through major mergers,
 the host galaxies lose memory of the dynamics before the mergers ($\xi\simeq0$ for 
 $q\rightarrow1$). BHs in the host galaxies, however, are less influenced by
  minor mergers, possibly keeping more memory of the dynamics ($\xi\simeq1$ for 
  $q\rightarrow0$). $n$ is meant to inform how much the dynamics of BHs in the host halo is affected by a given galaxy merger. For this study, we conservatively take $n=1$. We hope that a more 
  precise functional form will be found in future studies.

We note that with the prescriptions for $\textit{\textbf{v}}$ and the
assumption of instantaneous formation of post-merger galaxies, the
orbits of pre-existing BHs become possibly either more radial or more
circularized at mergers. We further note that it is possible that BHs
could escape from the potential or their apocenters could become
significantly larger than $r_{\rm h}$ if they happened to gain
sufficient kinetic energies at mergers. However, in our simulations,
we could not find such cases.

\subsection{BH mergers and prescriptions for BH merger remnants}
\label{subsec:BHmergers}
\subsubsection{BH merger conditions}
The fate of the SMBHs after galaxy mergers is still not fully
understood, with uncertainties remaining on whether SMBH mergers do
occur, and on which timescale.  However, under the assumption that
SMBHs do eventually merge, it is important to estimate how frequently
they do so given the merger histories of the host galaxies. At large
separations, dynamical friction plays a dominant role in bringing two
massive BHs together to form a bound binary. As they become more
tightly bound, a significant amount of stars may be ejected, leaving
behind an empty loss cone. Given the long relaxation time in the
nuclei of early-type galaxies ($\sim10\Gyr$, \citealt{Merritt2006}),
once the stars are cleared out, it is unlikely that collisional
processes can refill the loss cone before $z=0$.  Many alternative
mechanisms to solve the final parsec problem exist, from nuclear
triaxiality to circumbinary disks (see \S \ref{sec:intro}).  Treating
all of these mechanisms in a self-consistent way is far beyond the
scope of this paper, which primarily aims at studying the role of
multi-SMBH interactions in the solution of the final parsec problem.
We therefore focus only on dynamical friction and multi-SMBH
encounters as drivers of orbital evolution.

We consider two extreme scenarios for dynamical friction. In our
fiducial model, we assume dynamical friction stops affecting SMBH
orbits once binaries become sufficiently tight. We refer to this as
the ``empty loss cone model'', or ``ELC-model'' for short. In the
ELC-model, if binaries satisfy any of the following conditions,
dynamical friction is deactivated:

\begin{enumerate}
	\item 	Hard binary: when the semimajor axis of the BH binary is smaller than 
	the hard semimajor axis $a_{\rm h}$, or  $a<a_{\rm h}=G\mu/4\sigma_{\star}^{2}$ 
	($\mu$ is the reduced mass of the binary);
	
	\item Fast-moving stars: when the speeds of the BHs are slower than the local 
	circular velocity, or $v<\sqrt{G[M_{\rm en}(r)+M_{\rm BH}(r)]/r}$;
	
	\item Inside the influence radius $r_{\rm in}=2G M_{\rm
          BH,1}/\sigma_{\star}^{2}$ (where $M_{\rm BH,1}$ is the
          primary BH mass): when a less massive BH in a binary is
          inside the influence radius of a more massive BH but no
          $3^{\rm rd}$ BH is inside $r_{\rm in}$.
\end{enumerate}
The ELC-model is meant to investigate multi-SMBH interactions as a
``mechanism of last resort'' for solving the final parsec problem in
massive galaxies where alternative solutions are likely to be less
reliable.

In our alternative scenario, we assume that dynamical friction always
play a role until binaries merge. We refer to this case as ``full
loss cone model'' or simply ``FLC-model''  \footnote{This
  model name, as well as the assumptions behind this model, may be overly
  idealized. However, our strategy here is to anchor our two models as
  extreme, but physically possible end limits for BH merger
  scenarios. }. We emphasize that 
our FLC-model assumes full loss cones and the standard Chandrasekhar formula (see Equation \ref{eq:dynamicalfriction}) as a valid way to evaluate dynamical friction for hard binaries in the full loss cone regime. The standard Chandrasekhar formula was derived under the assumption of non-accelerated/linear motion in a uniform density distribution. When a binary enters the hard-binary regime, as the gravity from the second binary becomes more important, those assumptions of the dynamical friction formula may not be valid any more. 
However, by continuing to use the usual dynamical friction formula in the FLC-model down to the GW-driven regime, we ignore these corrections. We discuss the analytic validity, as well as
the limits and caveats of full loss cone assumption in more detail later.
In spite of our approximated treatments, 
it captures one very important, and unexplored
effect: the stochastic GWB from a cosmologically motivated population
of {\it high-eccentricity SMBH inspirals.}  When dynamical friction
acts on a satellite SMBH with $q \ll 1$ in a Keplerian potential and a
relatively flat density profile, the orbit becomes increasingly
eccentric \citep{AntoniniMerritt2012}. In some portions of the
parameter space, the final parsec problem can be self-consistently
bypassed by eccentric dynamical friction effects.  Specifically, for
$a_{\rm h} \ll r_{\rm in}$ and sufficiently small $q$, the secondary's
pericenter will decrease much more rapidly than its apocenter,
allowing it to bypass the final parsec problem altogether by using
apocentric interactions as a sink for angular momentum at roughly
fixed energy.  We analyze this effect in greater detail in later
sections.

Together, these two models allow us to separately explore the role of
dynamical friction (FLC-model) and that of possible three-body
interactions (ELC-model) on BH mergers, especially merger rates and
stochastic GWB.  To proceed further, it is very important to establish
a proper criterion for BH mergers. Given our two limiting treatments for
dynamical friction, we adopt two physically motivated, but distinct
merger conditions for BH mergers.  We assume that BHs merge under the
following conditions:

\begin{enumerate}
	\item \label{item:mergercondition1}When dynamical friction is not zero ($f_{\rm df}\neq0$): \\
	If gravitational wave (GW) emission becomes efficient ($P_{\rm GW}>P_{\rm df}$) over multiple orbits, the binary is declared as a merged BH when
	the decay time due to GW emissions is shorter than the dynamical time scale $t_{\rm dyn}$.
	
	\item \label{item:mergercondition2}When dynamical friction is zero ($f_{\rm df}=0$): \\
	If the decay time due to GW is shorter than the time left 
	until the next galaxy merger and $t_{\rm dyn}$, the binary is declared as a merged BH.

	\item \label{item:mergercondition3}For either $f_{\rm df}=0$ or $f_{\rm df}\neq0$:\\
	If the Schwarzschild radii of two BHs overlap, the binary immediately merges. Simply: 
	$r<r_{\rm sch,1}+r_{\rm sch,2}$, where $r$ is the separation of two BHs 
	and $r_{\rm sch}$ is the BH Schwarzschild radius.

\end{enumerate}
The decay time due to GW emissions is evaluated as $|a/\dot{a}_{\rm GW}|$ using Equation (5.6) in \citet{Peters1964}. The code computes, and updates at every time step, the decay time until
merger. 
	In condition \ref{item:mergercondition1}, $P$ represents the
        dimensionless dissipative power and time scale for each force,
        defined as $P_{\rm GW,df}=\textbf{\textit{f}}_{\rm GW,df}
        \cdot\textbf{\textit{v}}(E_{\rm b}/t_{\rm dyn})^{-1}$, where
        $E_{\rm b}$ is the orbital binding energy. In the
        simulations, whether BHs would merge in the FLC-model is
        mostly decided by condition \ref{item:mergercondition1}, while
        in the ELC-model, by condition
        \ref{item:mergercondition2}. Condition
        \ref{item:mergercondition3} may not even be relevant when two
        BHs form binaries and merge without the help of other BHs
        (likely in the FLC-model), but we include it to account for
        possible collision events in chaotic multi$-$BH interactions
        (the ELC-model).

\subsubsection{Gravitational wave recoils and remnant masses}
When two SMBHs merge, the remnant BH gets a kick due to anisotropic 
emission of gravitational waves \citep{Bekenstein1973, FitchettDetweiler1984, Favata+2004}. 
Recent numerical simulations of general relativity have confirmed that the 
recoil velocities could be as large as galactic escape velocities depending 
on progenitor spins and mass ratios \citep{Campanelli+2007,Campanelli+2007b,
	Lousto+2010,LoustoZlochower2011}. For such large kicks (up to $\sim 5000~{\rm km~s}^{-1}$), the remnant
BH could escape to infinity or end up orbiting in the outskirts 
of the halo. If the kicks are not large enough to completely eject the remnant 
BH, the BH may return to the core regions after temporarily being 
ejected, taking part in interactions again with other BHs.

We implement the effects of the recoil kick in the simulations and
take into account the mass loss to gravitational radiation for the remnant BH using the analytic
formulae with the best-fit values given in \citet{Lousto+2010}, with random spin orientations and dimensionless spin magnitudes
		chosen randomly between 0 and 0.9. 
We provide the detailed expressions and
    prescriptions used in this study in Appendix
    \ref{appendix:GWkicks}.

\subsection{The equations of motion}
\label{subsec:theequationofmotion}

Using a few-body code (see \citealt{Ryu+2016a,Ryu+2017} for code details), 
the equations of motion and mass
growth for each SMBH embedded in the evolving galaxies are integrated. 
The motion of
the BHs is determined by the following forces: (i) $\textit{\textbf{a}}_{\rm N}+\textit{\textbf{a}}_{\rm PN}$: their mutual
gravitational attraction 
including post-Newtonian terms up to 2.5th
order, (ii) $\textbf{\textit{a}}_{\rm df}$: dynamical friction from the 
surrounding medium (stars+ DM), (iii) $\textbf{\textit{a}}_{\rm bg}$: 
the gravitational pull of the background matter (stars+ DM) 
and (iv) $\textbf{\textit{a}}_{\rm mg}$: the deceleration due to BH mass increase with momentum conserved. 
The resulting equation of motion for the $i^{\rm th}$ BH includes the
sum of the five forces:
\beq
\textbf{\textit{a}}_{i}=\textbf{\textit{a}}_{{\rm N},i}+\textbf{\textit{a}}_{{\rm PN},i}+\textbf{\textit{a}}_{{\rm df},i}+\textbf{\textit{a}}_{{\rm bg},i}+\textbf{\textit{a}}_{{\rm mg},i}
\label{eq:eom}
\eeq
Given the solutions of the equation of motion at every time step, 
we update the positions and velocities for each BH and 
the evolution of galaxy potentials. We next describe each
contribution in detail.

\begin{enumerate}
	
	\item \noindent\textit{Mutual gravitational forces between BHs}\\
	We calculate the standard Newtonian gravitational force 
	$\textbf{\textit{a}}_{\rm N}$ as well as post-Newtonian terms $\textbf{\textit{a}}_{\rm PN}$, 
	
	\begin{align}
	\label{eq:Newton-pairs}
	\textbf{\textit{a}}_{\rm gr}&=\textbf{\textit{a}}_{{\rm N}, i}+\textbf{\textit{a}}_{{\rm PN}, i}  \nonumber\\
	&= -\sum_{j\neq i} G~M_{{\rm BH},j}~
	\frac {\dd ~\Phi(r_{ij})}{\dd ~r_{ij}}~
	\frac{\textbf{\textit{r}}_{i}-\textbf{r}_{j}}{r_{ij}}\nonumber\\
	&+\textbf{\textit{a}}_{{\rm 1PN}, i} +\textbf{\textit{a}}_{{\rm 2PN}, i}+\textbf{\textit{a}}_{{\rm 2.5PN}, i},
	\end{align}
	where $G$ is the gravitational constant, $\Phi$ is the pairwise
	gravitational potential, $\textbf{r}_{i}$ is the displacement of the
	$i^{{\rm th}}$ BH from the center of the host galaxy, and
	$r_{ij}\equiv |\textbf{r}_{i}-\textbf{r}_{j}|$.
	In our numerical implementation, we adopt the Plummer
	softening kernel \citep[e.g.][]{Galacticdynamics} with
	softening length equivalent to the Schwarzschild radius for a $100~\Msol$ BH.
	
	We include post-Newtonian terms ${\textbf{\textit{a}}}_{\rm PN}$ 
	up to order 2.5, which account for the loss
	of orbital energy and angular momentum via gravitational waves, 
	but do not account for spin-orbit or spin-spin coupling.
	The full expressions for these terms can be found in, e.g.,
	\citet[]{Kupi+06}.  
	
	\item \noindent\textit {Dynamical friction from background matter}\\
	When an object moves through a medium, it induces an overdensity of the medium, or wake,
	behind it. The gravitational force due to the overdense region acts 
	as a dissipative drag on the object's motion. 
	In this study, we consider dynamical friction due to
	both DM and stars.
	
	For the DM contribution, we adopt the standard Chandrasekhar
	formula \citep{Galacticdynamics},
	\begin{equation}\label{eq:dynamicalfriction}
	\textbf{\textit{a}}_{{\rm df},i}=-4\uppi ~\ln\Lambda ~f(X_{i})
	~ \frac{G^{2} M_{{\rm BH},i}}{v_{i}^{3}}~\rho(r_{i})~\textbf{\textit{v}}_{i},
	\end{equation}
	with
	\begin{equation}
	f(X_{i})\equiv {\rm erf}(X_{i})-\frac{2}{\sqrt{\uppi}} ~X_{i}
	~\exp\left(-X_{i}^{2}\right),
	\end{equation}
	where $X_{i}\equiv v_{i}/(\sqrt{2} \sigma_{v})$. We use the circular 
	velocity, defined as $\sqrt{G[M_{\rm en}(r\leq r_{i})+M_{\rm BH}(r\leq r_{i})]/r_{i}}$ 
	for $\sigma_{v}$. We do not include the contribution of stars bound to infalling BHs in estimating dynamical friction.
	Again, $M_{\rm en}(r\leq r_{i})$ is the enclosed mass 
	(DM+stars) and $M_{\rm BH}(r\leq r_{i})$ is the total mass of BHs 
	(including the $i$-th BH itself) inside $r=r_{i}$. The expression for 
	the enclosed mass of stars is given in \citet{StoneOstriker2015}. 
	We use
	$\ln\Lambda=5$ \citep{Spinnato+2003,Merritt2006} and we take the sum of local densities of stars 
	and DM for $\rho$, namely, $\rho=\rho_{\star}+\rho_{\rm DM}$, 
	at the location of the $i$-th BH.

	\item\noindent\textit{Gravitational force of the background matter}\\
	The background stars and DM exert an additional gravitational 
	force on the BHs. Because we assume a spherically symmetric 
	density profile, this force
	points toward the centre of the potential. It can be expressed as
	\begin{equation}
	\label{eq:bg}
	\textbf{\textit{a}}_{{\rm bg}, i}=-\frac{G~M_{{\rm en},i}(r\leq r_{i})}{r_{i}^{3}} \textbf{r}_{i},
	\end{equation}
	where $\textbf{r}_{i}$ is a vector pointing from the 
	centre of the galaxy to the $i$-th BH.
	
	\item \noindent\textit{Deceleration due to mass growth}\\ 
	We take into account the decrease in velocity due to mass 
	growth (see \S \ref{subsubsec:DM_stars_BHmass}). 
	Assuming BHs grow in mass in a spherically symmetric fashion, 
	the $i-$th BH decelerates 
	through conservation of linear momentum,  
	\beq
	\textbf{\textit{a}}_{{\rm mg},i}=-\frac{M_{\rm BH, i}'-M_{\rm BH, i}}{M_{\rm BH, i}'\Delta t}\textbf{\textit{v}}_i\,,
	\label{eq:a_acc}
	\eeq
	where $M_{i}'$ is the increased mass estimated using Equation 
	\ref{eq:massgrowt}, and $\Delta t$ is the time step.
\end{enumerate}

In summary, our simulations display several noticeable
features: (i) we follow the merger history of SMBH host galaxies as
extracted from cosmological $N$-body simulations for $0<z<4$, across a
wide range of merger mass ratios, $10^{-4}<q_{\star}<1$; (ii) we take
into account the evolution of the galactic potential (star+DM) in both
physical size and depth as a result of both galactic mass growth and
core scouring from SMBH binaries; (iii) we explore two different
models, the empty and full loss cones. These two extreme assumptions
plausibly bracket the true evolution of binary BHs close to their
merger. In addition, they allow us to clearly isolate the importance
of multi-body BH interactions between BHs at
coalescence. We show this by estimating the BH merger rates
  and the GWB independently for the two models.

\section{Results}
\label{result}
In this section we present the binary lifetimes of merged BHs and
their merger rates for the two models (FLC and ELC). Additionally,
given the merger rates, we infer the characteristic GW amplitude
$h_{\rm c}$. Given the eccentricities found in our simulations, we
show how $h_{\rm c}$ for eccentric orbits deviates from that for
circular orbits.

\begin{figure}
	\centering
	\includegraphics[width=8.4cm]{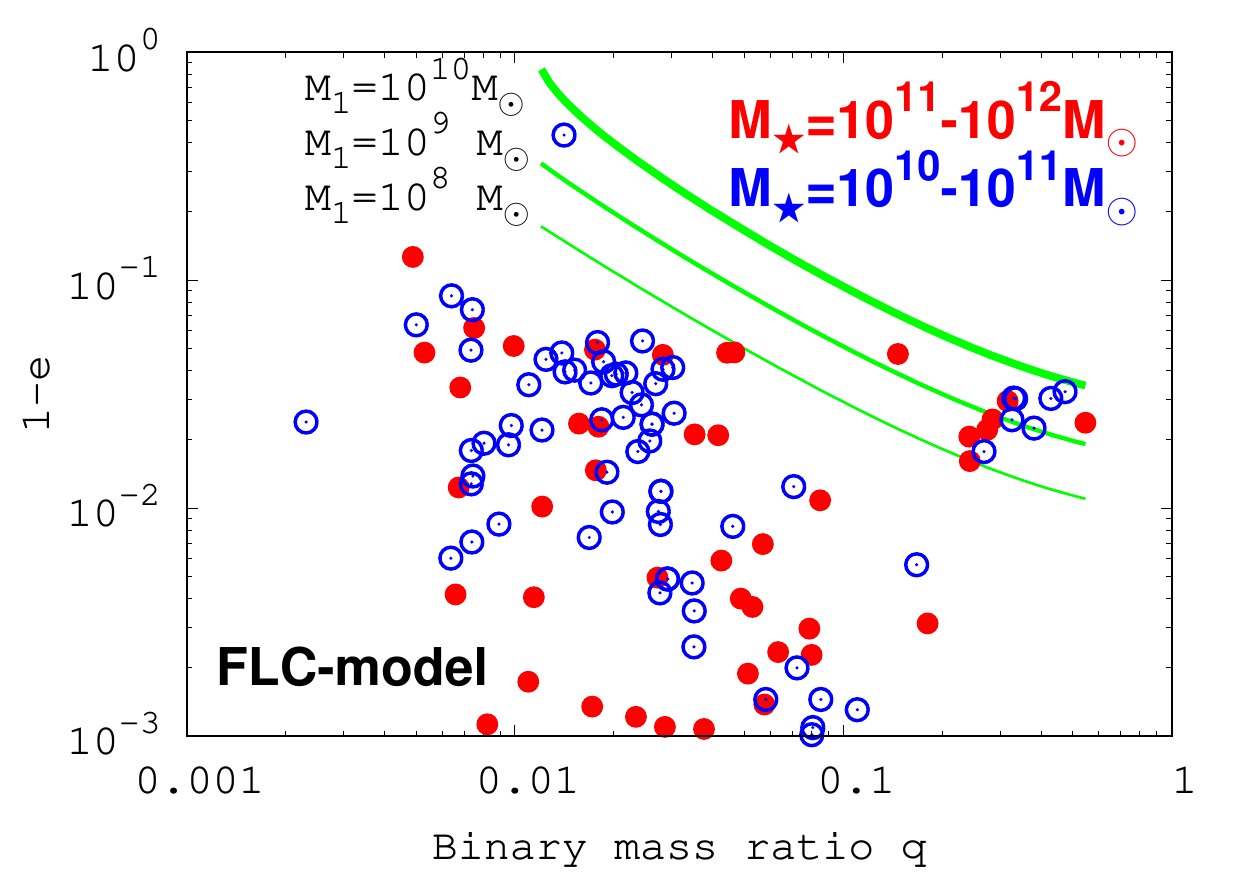}
	\caption{The distribution of the eccentricities of merged
          binaries in the FLC-model as a function of the binary mass
          ratio $q$ when energy loss by GWs becomes dominant. We use
          different colors to distinguish between merged binaries in
          the galaxies of $M_{\star}=10^{11}-10^{12}\Msol$ (red) and
          in $M_{\star}=10^{10}-10^{11}\Msol$ (blue). The
          eccentricities are quite high ($e>0.9$). Green curves show
          analytic limits for the validity of our approach
          (Equation \ref{eq:e_rGW}), which estimates
          dynamical friction in the absence of stellar scattering.
          High-$q$ mergers above these curves are not treated
          self-consistently by our FLC-model, but the majority of
          (low-$q$) mergers, which lie below these curves, are.}
	\label{fig:lifetime_eccentricity_q}
\end{figure}

\begin{table*}
	\centering
	\setlength\extrarowheight{4pt}
	\caption{Overview of BH coalescence events for the FLC- and
          ELC- model for the host galaxies of
          $M_{\star}=10^{10}-10^{11}\Msol$ and
          $M_{\star}=10^{11}-10^{12}\Msol$.  From top to bottom: The
          mass of sampled galaxies, the total number of BH
          coalescences, the average mass of merged binary BHs in unit
          of $\Msol$ and $M_{\star}$, the number fraction of BH
          coalescences in host galaxies with one significant merger
          ($N_{\star}=1$) with $q_{\star}\geq0.01$ and that in host
          galaxies with multiple significant mergers
          ($N_{\star}\geq2$) with
          $q_{\star}\geq0.01$. Notice that only 1\% of BH
            coalescences occur in host galaxies with
            $M_{\star}=10^{10}-10^{11}\Msol$ experiencing only one
            significant merger ($N_{\star}=1, q_{\star}\geq0.01$). This
            is because the number of such galaxies is small (See
            Figure \ref{fig:galaxymerger_q_count} and Table
            \ref{table:galaxymergercount}) and binary formation and
            BH mergers are less likely to happen.}
	\label{table:BHmergercount}
	\begin{tabulary}{1\linewidth}{c c c c c }
		\hline
		Galaxy mass $M_{\star}$ &             \multicolumn{2}{c}{$10^{10}-10^{11}\Msol$ }&         \multicolumn{2}{c}{$10^{11}-10^{12}\Msol$}   \\ loss cone models&
                FLC-model & ELC-model & FLC-model & ELC-model\\ \hline
                Total number of BH coalescences&     140&77& 76 &35\\ Average total mass of merged BHs
                [$10^{8}\Msol$]& 2.2&3.2& 15 &17\\ Average total mass
                of merged BHs [in unit of $10^{-3}M_{\star}$]& 4.4&5.0& 4.5
                &5.1\\ Number fraction of BH coalescences in
                host galaxies ($N_{\star}=1$, $q_{\star}\geq0.01$) & 1\%&1\%&
                17\%&8\% \\ Number fraction of BH
                coalescences in host galaxies ($N_{\star}\geq2$,
                $q_{\star}\geq0.01$) & 99\% &99\%& 83\%&92\% 
                \\ \hline
	\end{tabulary}
	
\end{table*}

\subsection{Overview of results}
\label{sub:binarylifetime}
We consider two different evolutionary paths of SMBH binaries in the
two models (FLC and ELC).  In the FLC-model, the orbits of the
binaries shrink only via dynamical friction until energy loss to GWs becomes more
efficient. In the ELC-model, three-body interactions play a role in
addition to dynamical friction.  Overall, we find coalescences of BH
binaries from both models but with different merger rates, which will
be described in more details in \S \ref{sub:BHmergerrates}. 
In this section, we focus on average properties of mergers in our two limiting regimes.

\subsubsection{Dynamical features}
\label{sub:dynamicalfeatures}
\begin{enumerate}
	\item \textit{FLC-model}
	
	In the FLC-model, the birth eccentricities of the binaries are moderate ($e\gtrsim 0.4$).  These are lower than the
	eccentricities assigned to BHs as initial conditions (see Equation
	\ref{eq:initialeccentricity}). This is because inspiralling 
	BHs experience the strongest dynamical friction forces (prior to binary formation) at pericenter (near denser core region), leading to orbital circularization. Given that the
	density profile adopted in this study approximately follows
	$\rho\sim r^{-2}$ at $r_{\rm c}<r<r_{\rm h}$, this is consistent
	with the eccentricity evolution of BHs in an isothermal density profile decaying towards the core shown in \citealt{Ryu+2016} (see their Figure 7). Once a binary forms, however, orbital eccentricities increase rapidly due to dynamical
	friction.  At the point when GW emission becomes the dominant driver of orbital decay, eccentricities can reach up to $e>0.99$ and
	semimajor axes down to $a\sim 0.01-1\pc$.  
	
We emphasize here that the evolution of the
          eccentricities in the FLC model likely represents the most extreme scenario of eccentricity evolution. Accounting
          for stellar 3-body scatterings would likely moderate the increase in eccentricity we observe. Hence, the eccentricity at
          which GW emission takes over may not be as high as that
          found in this study. Indeed, the eccentricities of compact
          binaries in our simulations tend to be higher than those found in
          some previous numerical works with large $N$-body simulations
          \citep[e.g.][]{Berentzen+2009,Khan+2011,Preto+2011}, even
          though a qualitatively similar increase in eccentricity has been
          seen in those studies. Since a long-term ``full'' loss
          cone in large $N$-body simulations cannot be easily
          achieved, the binary evolutions found in their studies may
          correspond to intermediate regimes bracketed by our two
          models. For example, \citet{Berentzen+2009} studied the
          evolution of SMBH binaries, focusing on the interactions
          with surrounding stars. In the eccentricity evolutions shown
          in their examples, we can see a rapid increase right after
          binary formation, followed by a relatively gradual
          rise. This may be due to quick depletion of the initially
          full loss cone reported in their paper, as noted above,
          possibly corresponding to a regime in between our two
          models.

	We show in Figure \ref{fig:lifetime_eccentricity_q} the distribution
	of the eccentricities of binaries in the FLC-model which will
	eventually merge, as a function of the mass ratio $q$.  The
	eccentricities are evaluated at the time when GWs become more
	efficient. For such eccentric binaries, the decay time
	\citep{Peters1964} is short (typically, $t_{\rm
		decay}<10^{8}\yr$). Considering the galaxy merger time scale of
	$\sim1\Gyr$ and the long infall times for BHs to reach the core, this
	means coalescences of BHs may occur even before a $3^{\rm rd}$ BH can
	arrive. Indeed, in almost all of our FLC-model simulations, incoming
	BHs which can reach the core form binaries with the central BH, and
	subsequently merge on a short time scale.  Of course, our FLC-model
	orbital evolution is quite approximate in that it neglects hardening
	via three-body interactions with surrounding stars.  This
	approximation is only justified in the subset of parameter space where
	a radializing binary orbit (inside the primary influence radius) can
	keep its apocenter outside the hard radius $a_h$.  In other words, the
	final parsec problem can only be bypassed when $r_{\rm p,GW} > a(1-e)$
	and, simultaneously, $a>a_{\rm h}$.  Here $r_{\rm p, GW}$ is the
	maximum pericenter for which a SMBH binary will merge in a Hubble time
	$t_{\rm H}$.  Combining these two inequalities gives a necessary
	condition for this bypass to occur, which is
	\begin{equation}
	\label{eq:e_rGW}
	\frac{q^{3/4}}{(1+q)^{5/4}} < \frac{4\sigma_\star^2}{c}\left(\frac{85 t_{\rm H}}{3GM_{\rm BH,1}c} \right)^{1/4} (1-e^2)^{-7/8}.
	\end{equation}
	Green curves illustrating this inequality are shown for different
	primary SMBH masses in Fig. \ref{fig:lifetime_eccentricity_q}.  Most
	of the mergers we simulate are at sufficiently low mass ratio that our
	simulations of high-eccentricity coalescence are
	self-consistent. However, we caution that Equation~\ref{eq:e_rGW} is a
	necessary, not a sufficient, criterion for an eccentric bypass of the
	final parsec problem (see also the discussion of
	\citealt{AntoniniMerritt2012}). Whether or not an individual secondary
	BH can make use of this route to coalescence depends on its initial
	eccentricity and on the role of three-body scatterings with stars. In
	addition, the degree of nuclear rotation can affect whether or not
	they circularize or radialize
	\citep[e.g][]{RasskazovMerritt2017,Mirza+2017}.

	If binary lifetimes are sufficiently short that BHs coalesce
        before another BH makes it to the core, then BH merger rates
        and infall timescales of incoming BHs should have an inverse
        correlation.  Given the shorter infall times of the more
        massive BHs, BH merger rates should hence increase as
        $q_{\star}$ increases.  We confirm this relation in Figure
        \ref{fig:galaxy_mergercorrelation}. The plot shows the
        fraction of galaxy mergers of mass ratio $q_{\star}$, for
        which the central BHs are able to coalesce up to
          $z=0$ in our simulations.  A number fraction of 1 means
          two BHs introduced by a galaxy merger always successfully
          merge whereas a fraction of 0 means they fail to merge.
        In the FLC-model, as $q_{\star}$ increases, it is more likely
        that BH mergers take place, and the BH merger rates can be
        directly related to the frequency of major galaxy mergers.
        
  	\begin{figure}
  	\centering
  	\includegraphics[width=8.4cm]{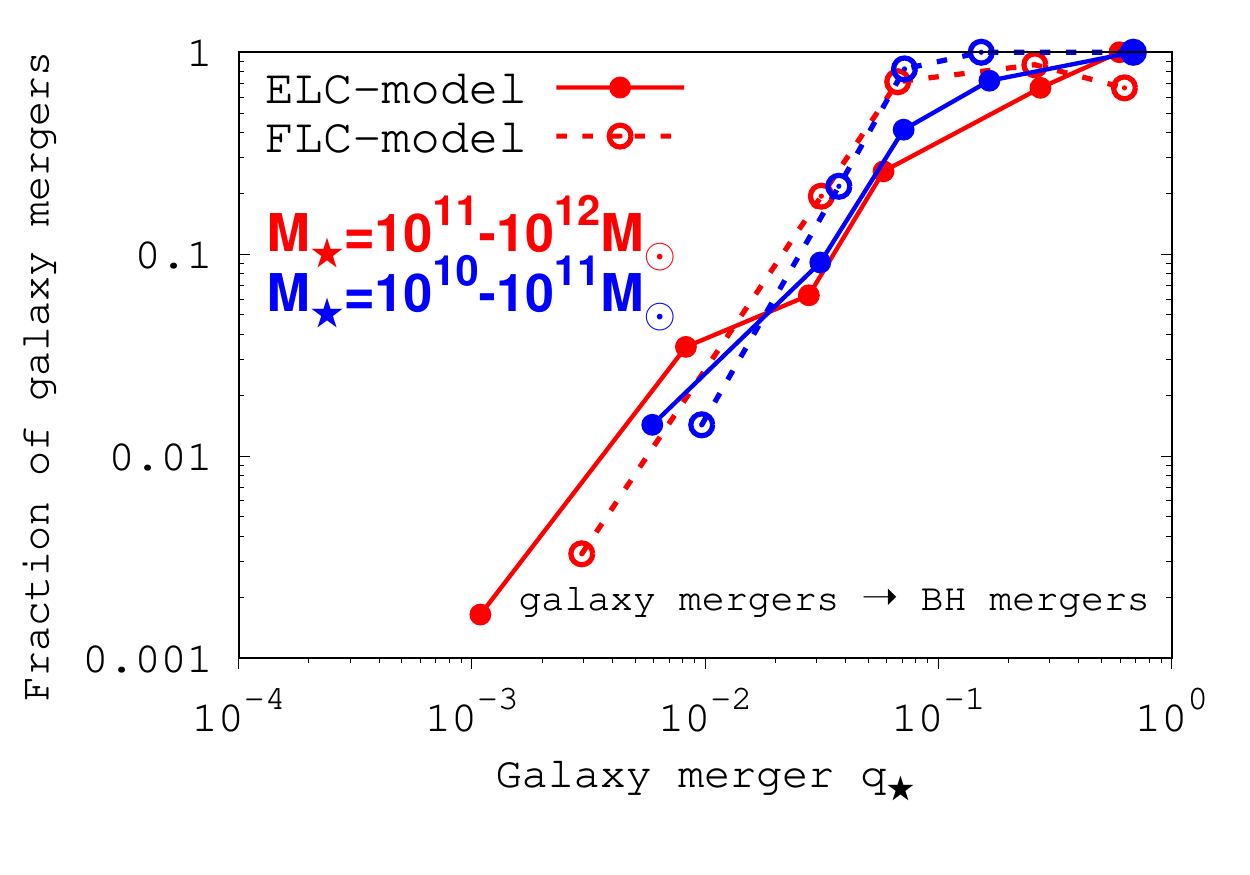}
  	\caption{The fraction of galaxy mergers, as a function of
  		their mass ratio $q_{\star}$, for which the central BHs have
  		merged over our entire merger trees. For
  		example, the number fraction of 1 (0) means that the galaxy
  		merger always (never) leads to coalescence of the central
  		BHs. }
  	\label{fig:galaxy_mergercorrelation}
  \end{figure}

\vspace{0.1in}	
	\item \textit{ELC model}
	
	\label{subsection:ELCmodel}
	On the other hand, in the ELC-model, the central binaries
        typically stall at $r\sim $ a few $10\pc$ at $z=0$ . This separation may be
          somewhat larger than generally assumed. In our simulations, due to dynamical
           friction, the BH binary orbits efficiently decay to near the hardening 
           radii, which are at least on the order of a few tens of pc at low $z$ for the very high mass BHs we consider.
        Under these conditions, unless a $3^{\rm rd}$ BH approaches
        sufficiently close to the core, the central binaries do not
        merge.  This means that in order for the central binaries to
        further decay and finally merge, multiple (at least $N\geq2$)
        major mergers are necessary, so that new BHs can make it to
        the core rapidly and effectively interact with the central
        binaries.  Therefore, there is a longer delay in time from
        binary formation to BH merger. This is clearly different from
        the FLC-model.  As a result, coalescences of BHs
        preferentially take place in the host galaxies experiencing
        more than one major mergers. We find in our
        simulations that 99\% of BH mergers in the ELC-model occur in
        such galaxies (experiencing multiple major mergers) in both
        mass bins of $M_{\star}=10^{10}-10^{11}\Msol$ and
        $M_{\star}=10^{11}-10^{12}\Msol$. Furthermore, we see similar
        correlations between the galaxy merger mass ratios and the
        likelihood of BH mergers within the FLC-model, as shown in
        Figure \ref{fig:galaxy_mergercorrelation}. However, we note
        that the fraction is slightly lower for high $q_{\star}$ than
        in the FLC-model. In the FLC-model, major galaxy mergers
        favorably lead to BH mergers, but because of ejections
        ($\sim1-5\%$ of BHs found at $r>r_{\rm h}$ at $z=0$) via
        multiple BH interactions, this is not always the case in the
        ELC-model.
	
The general picture of multi-body BH interactions in our
  simulations is as follows. When a third BH is orbiting far from the
  core region, its orbit is governed by dynamical friction and the galactic potential. Every time the intruder BH gets sufficiently close to the central binary at the pericenter of its orbit around the galactic potential, it goes through multiple gravitational slingshots with the
  central binary (typically, its apocenter remains outside the core). The intruder BH gains energy at the pericenter via
  the slingshot mechanism, and loses energy outside the core region via dynamical
  friction. In this case, the background potential when the 
  ejected BH is outside the core can additionally provide more chances to
  return back for another slingshot \citep{Ryu2017}. This appears to
  make the intruder BH linger a little bit longer 
  before its apocenter completely falls into the core. These repeated ejections confined in
  $r<r_{\rm h}$ do not always lead to significant decreases in the semimajor axes of the central binaries
   \footnote{For a hard binary (primary mass of
    $m_{\rm 1}=10^{8}\Msol$, $q=0.1$ and $a_{\rm h}\simeq1\pc$) with
    orbital energy $E_{\rm b,hard}$, the energy taken from the
    binary by a light BH of mass $m_{\rm 3}$ approaching with velocity
    $v=\sigma_{\star}$ and subsequently ejected at $v<v_{\rm
      esc}(r=r_{\rm h})$ (the escape velocity at $r=r_{\rm h}$) is
    $|\Delta E_{\rm b, hard}/E_{\rm b,hard}|\simeq0.003-0.3$ for
    $m_{\rm 3}/m_{1}=0.001-0.1$.}, but initially wide binaries can
  benefit from these slingshots, becoming hardened to some
  extent. 
  
  Finally, when the three BHs become bound, they either go
  though chaotic interactions followed by ejections, or form a
  hierarchical triple. Due to the gravitational pull from the third BH,
  the central binaries are usually located off-center when the triples
  form. The central binaries go through this course of interaction,
  similarly described by \citet{HoffmanLoeb2007}, one or even more times
  before they finally merge.  We find that it is less likely for
  ejected BHs to return and manage to merge. Typically ejected BHs are
  the less massive ones, hence they tend to be easily ejected again even
  though they can make it to the core. Additionally, we find that escapes
  of all three BHs are rare (also similarly to \citealt{HoffmanLoeb2007}). Even for that
  case, cores empty of BHs are transient, and are rapidly re-filled with
  other BHs from minor mergers or the ejected BHs themselves when they return. In our simulations, BH binaries
  merge in hierarchical triples and due to strong binary-single BH
  interactions (see also \citealt{Iwasawa+2006}). However, the majority of BH mergers
  occur when they are in hierarchical triples.

	\begin{figure}
		\centering
		\includegraphics[width=8.4cm]{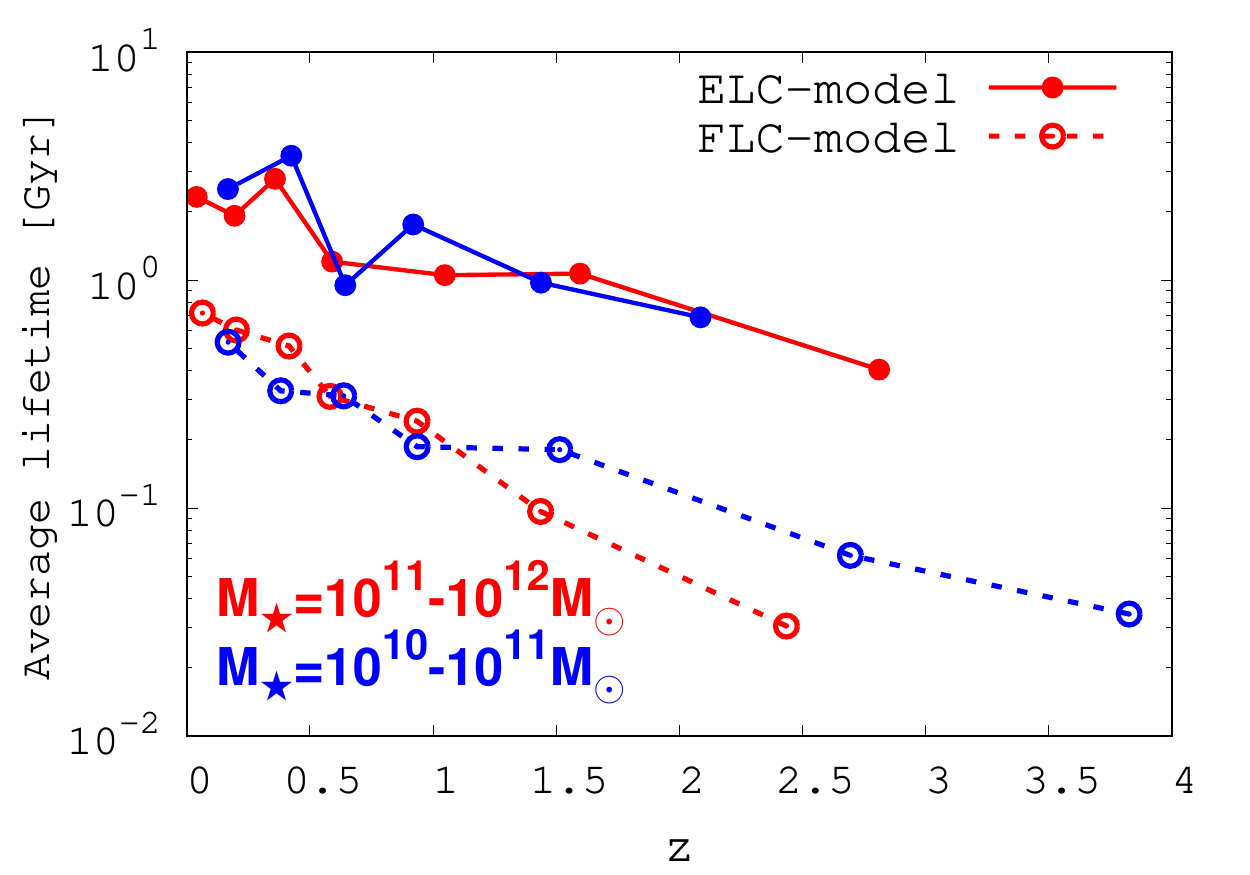}
		\includegraphics[width=8.4cm]{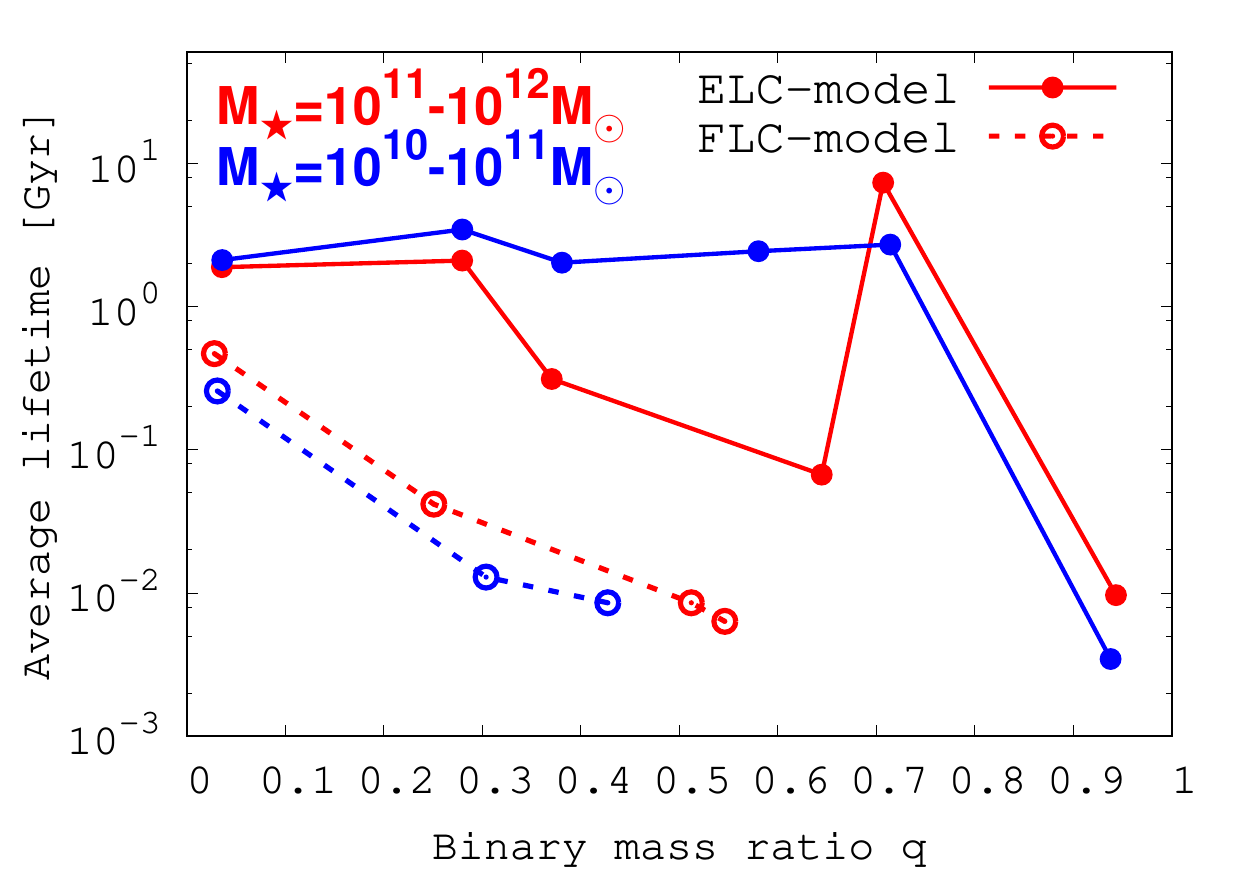}
		\caption{The average lifetimes of merged binaries as a
			function of $z$ (\textit{upper} panel) and the binary mass
			ratio $q$ (\textit{bottom} panel) for the galaxies of
			$M_{\star}=10^{10}-10^{11}\Msol$ (blue lines) and
			$M_{\star}=10^{11}-10^{12}\Msol$ (red lines). We use solid
			(dotted) lines to represent the ELC- (FLC-) model.  We
			define the lifetimes of binaries as the time from binary
			formation to coalescence. }
		\label{fig:lifetime_redshiftevolution}
	\end{figure}

\end{enumerate}

\subsubsection{Merger efficiencies and binary lifetimes}
\label{subsub:binarylifetime}
In order to highlight the differences between the two models, we provide
the average lifetimes of coalescing binaries in Figure
\ref{fig:lifetime_redshiftevolution} as a function of $z$
(\textit{upper} panel) and the binary mass ratio $q$ (\textit{lower}
panel).  We define the lifetime of a binary as the time from binary
formation to coalescence.  In the ELC$-$model, three-body interactions
can cause the ionization of existing binaries. In this case 
we estimate the lifetime as the time between when a binary forms 
 and when it merges, for the subset of binaries that avoid ionization. In
both panels, as expected, the lifetimes of the merged binaries in
the ELC-model ($\geq1\Gyr$) are longer than those in the FLC-model ($\leq
1\Gyr$). 

\begin{figure*}
	\centering
	\includegraphics[width=8.4cm]{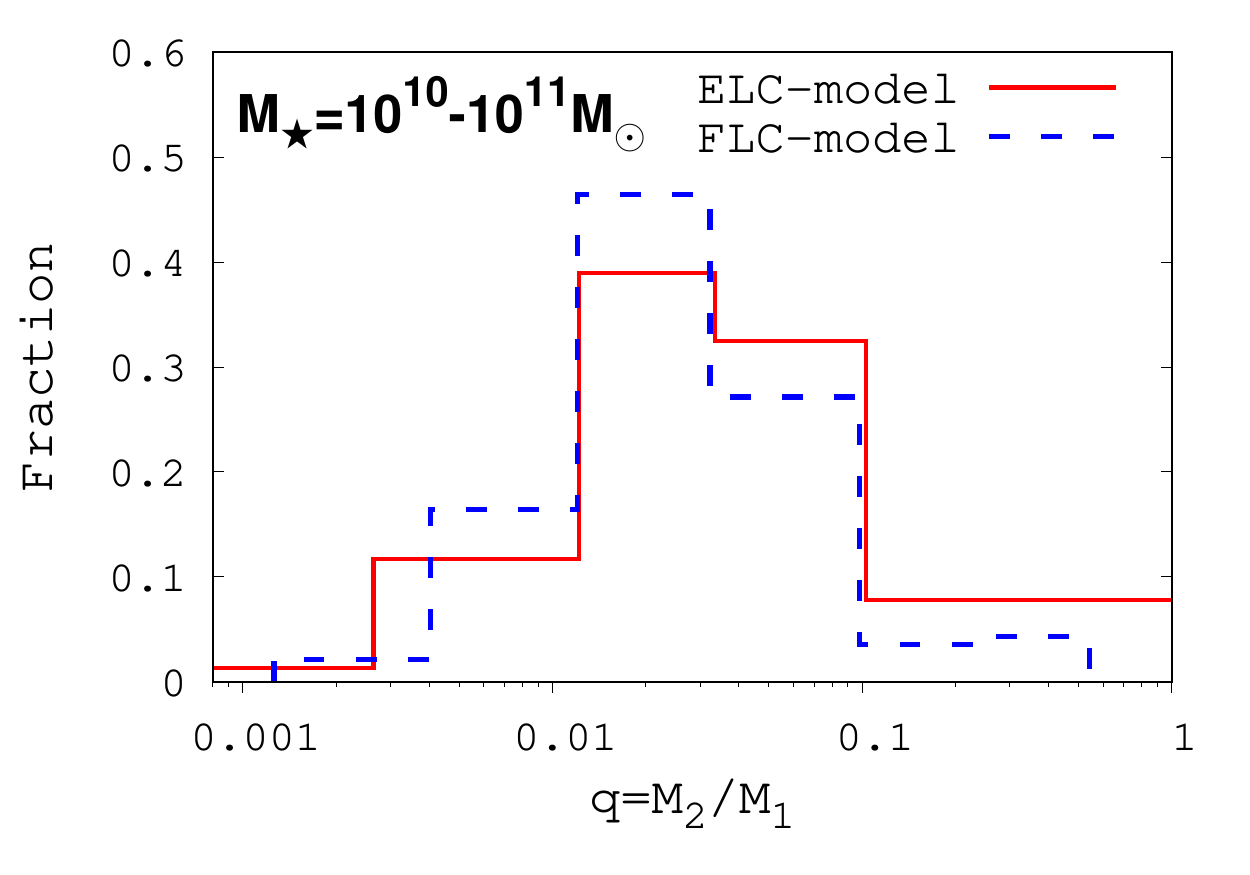}
	\includegraphics[width=8.4cm]{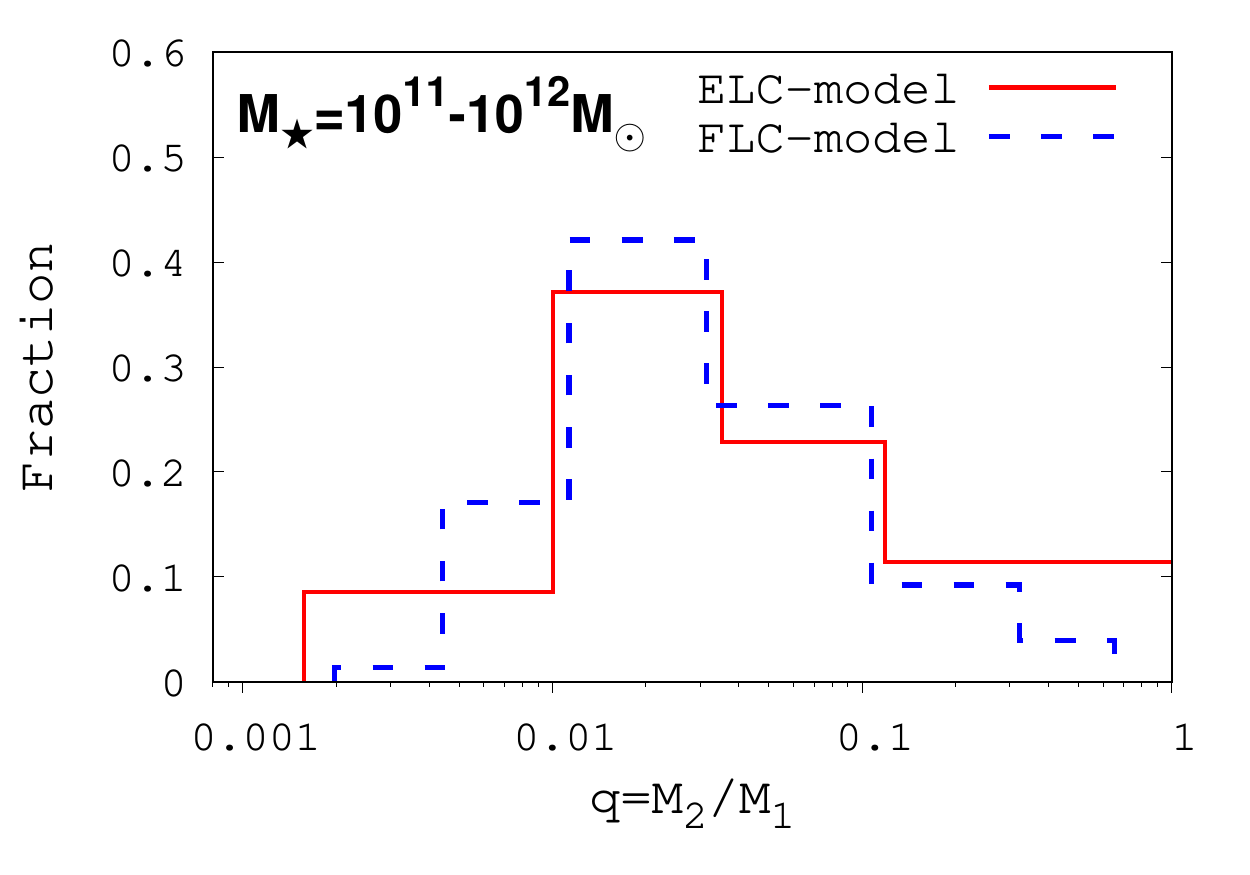}
	\caption{The relative fraction of merged central BH binaries as a function
		$q$ (in logarithmic intervals) in host galaxies of masses
		$M_{\star}=10^{10}-10^{11}\Msol$ (\textit{left} panel) and
		$M_{\star}=10^{11}-10^{12}\Msol$ (\textit{right}
		panel). Red (blue) solid lines refer to the ELC-
		(FLC-) model. It is normalized such that the sum of the fractions is unity.}
	\label{fig:merger_massratio}
\end{figure*}

 \begin{figure*}
	\centering
	\includegraphics[width=8.4cm]{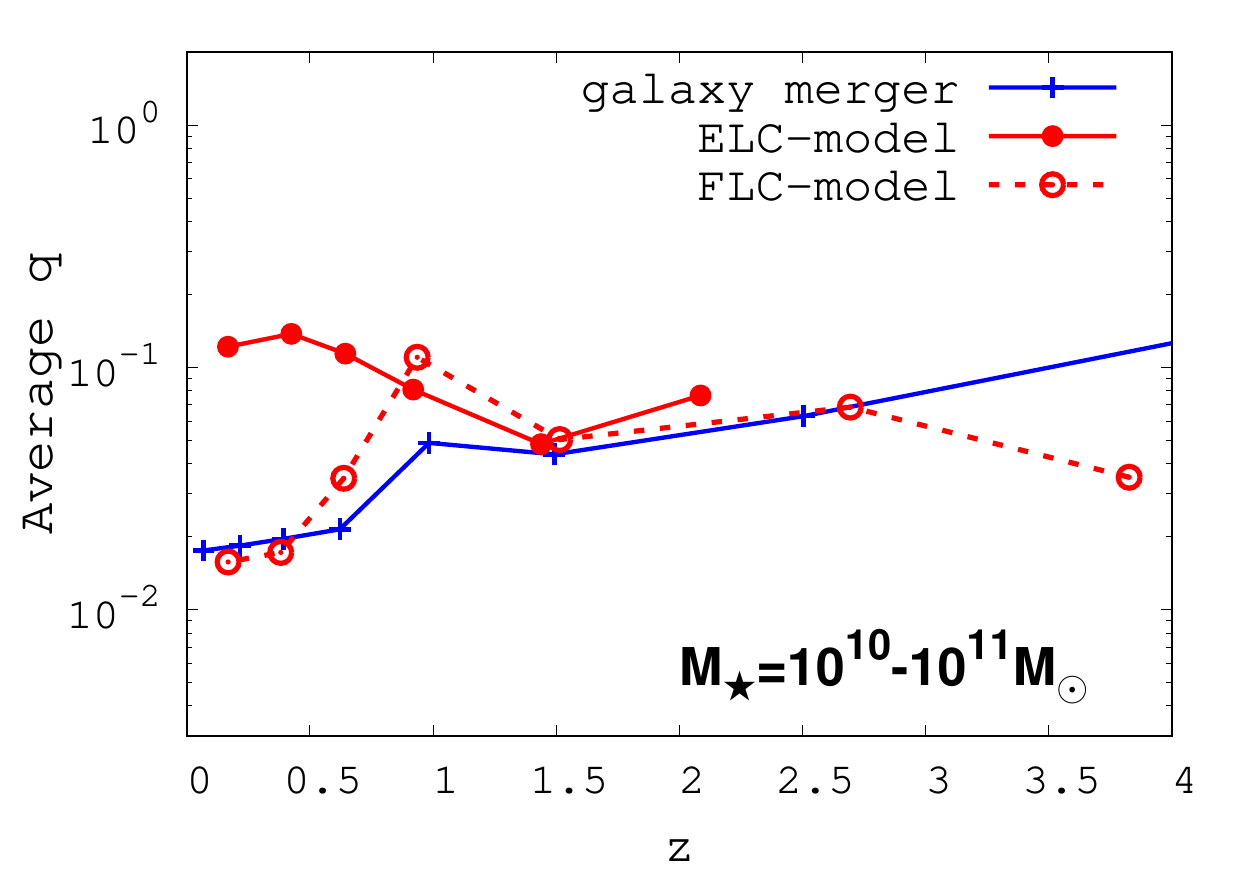}
	\includegraphics[width=8.4cm]{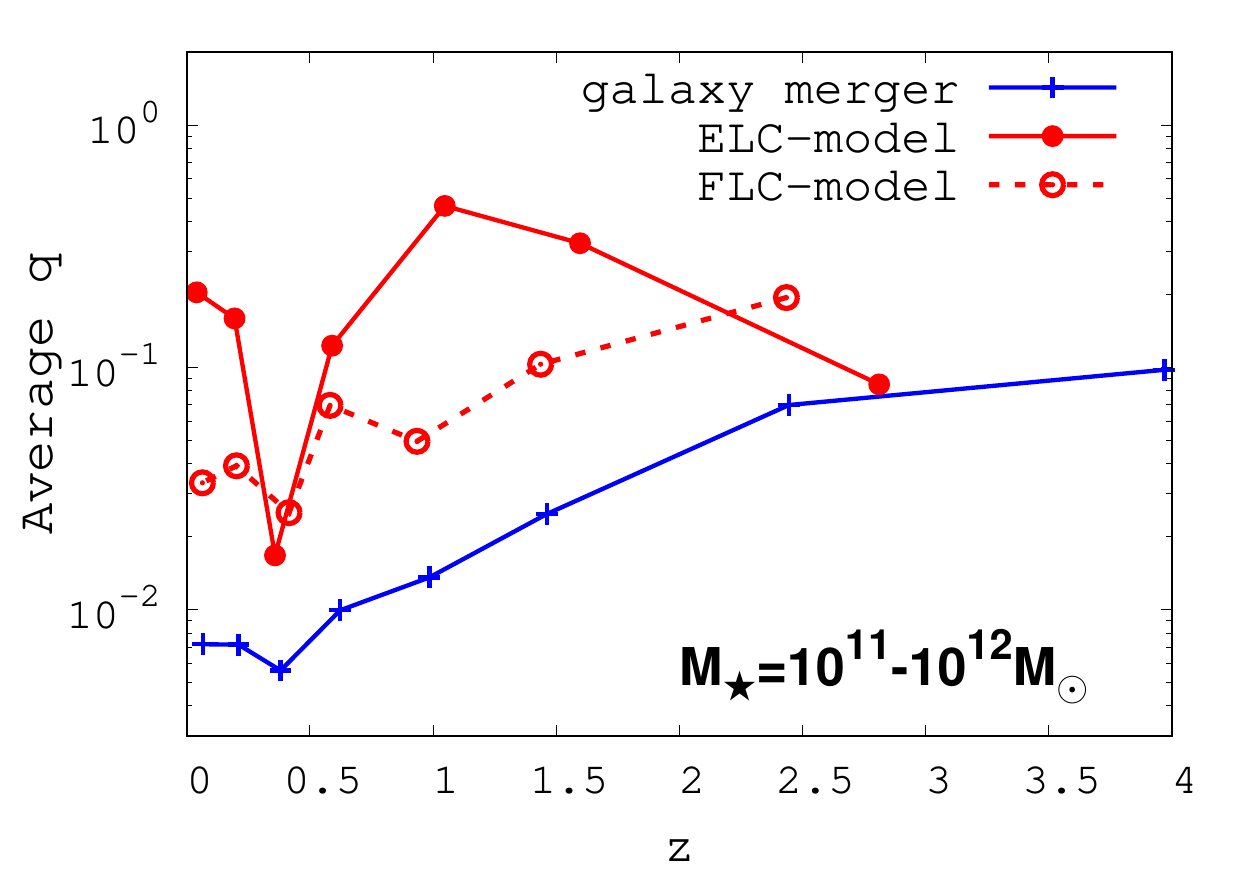}
	\caption{The average $q$ in each Gyr from $z=3.5$ to $z=0$,
		along with the average galaxy merger ratio $q_{\star}$ (same
		lines as in the \textit{middle} panels in Figure
		\ref{fig:galaxymerger_q_count}, but each line separately
		drawn in each panel). }
	\label{fig:BHmerger_massratio_z}
\end{figure*}

The lifetimes for the FLC-model that we find are relatively short
compared to the ones reported by \citet{Kelley+2017}. Besides different models and
different prescriptions for binary decay mechanisms, this difference may be
primarily due to highly eccentric binary orbits and the assumption
of efficient decay due to dynamical friction at all times. In
particular, in the \textit{upper} panel, as $z$ decreases, the
lifetimes become longer for both models.  However, such longer
times may be due to different reasons in each model.  In
the FLC-model, we can understand this as a result of galaxy mergers of
smaller $q$ at lower $z$ (see the \textit{middle} panel in Figure
\ref{fig:galaxymerger_q_count}), hence binaries with lower
$q_{\star}$. Remember that the dynamical friction timescale for a tightly
bound binary is roughly estimated as $t_{\rm df}=E_{\rm b}/(f_{\rm
  df}\cdot v)\propto M_{\rm BH,1}^{1/2}q^{-1}$. This can be also found in
the \textit{bottom} panel, which shows that the lifetimes
rise as $q$ declines. In the ELC-model, on the other hand, the longer lifetimes may be
attributed to mainly two reasons: 1) as galaxy mergers occur with smaller
$q_{\star}$ the central binaries have to wait for a longer time until
new BHs fall into the core (or longer infall times of less massive
BHs); and 2) it is harder for the central binaries to be ionized or to
get hardened via three-body interactions. Interestingly,
differently than in the FLC-model, the dependence on the mass ratio $q$ is
weakened (even flat for $q<0.3$) as the central binaries go through
chaotic interactions with other BHs, followed by ionization and
exchange in binary members.

Because of such differences between the two models, we find different
statistical properties of the merged BH binaries including 
their merger rates and mass ratios.  This is the subject of the next section.

\subsection{Coalescence of BHs - BH merger rate and mass ratio}
\label{sub:BHmergerrates}
In this section we focus on a detailed analysis of the statistical
distributions of BH mergers, such as merger rates, mass ratios and
their evolution as a function of $z$. We provide an overview of BH coalescence events for the FLC- and
ELC- models in Table \ref{table:BHmergercount}.

\subsubsection{Mass ratios and chirp mass of coalescing BH binaries}

\begin{figure*}
	\centering
	\includegraphics[width=8.4cm]{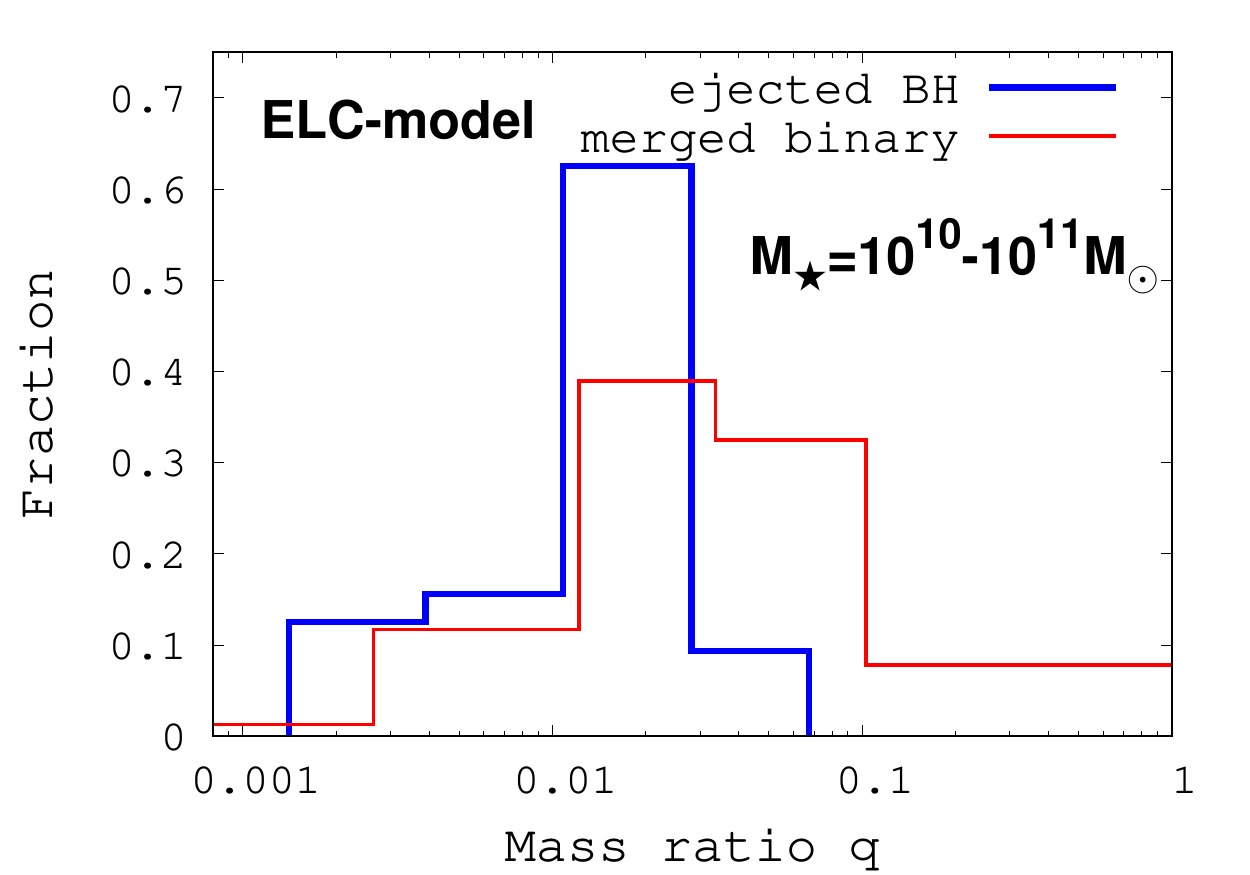}
	\includegraphics[width=8.4cm]{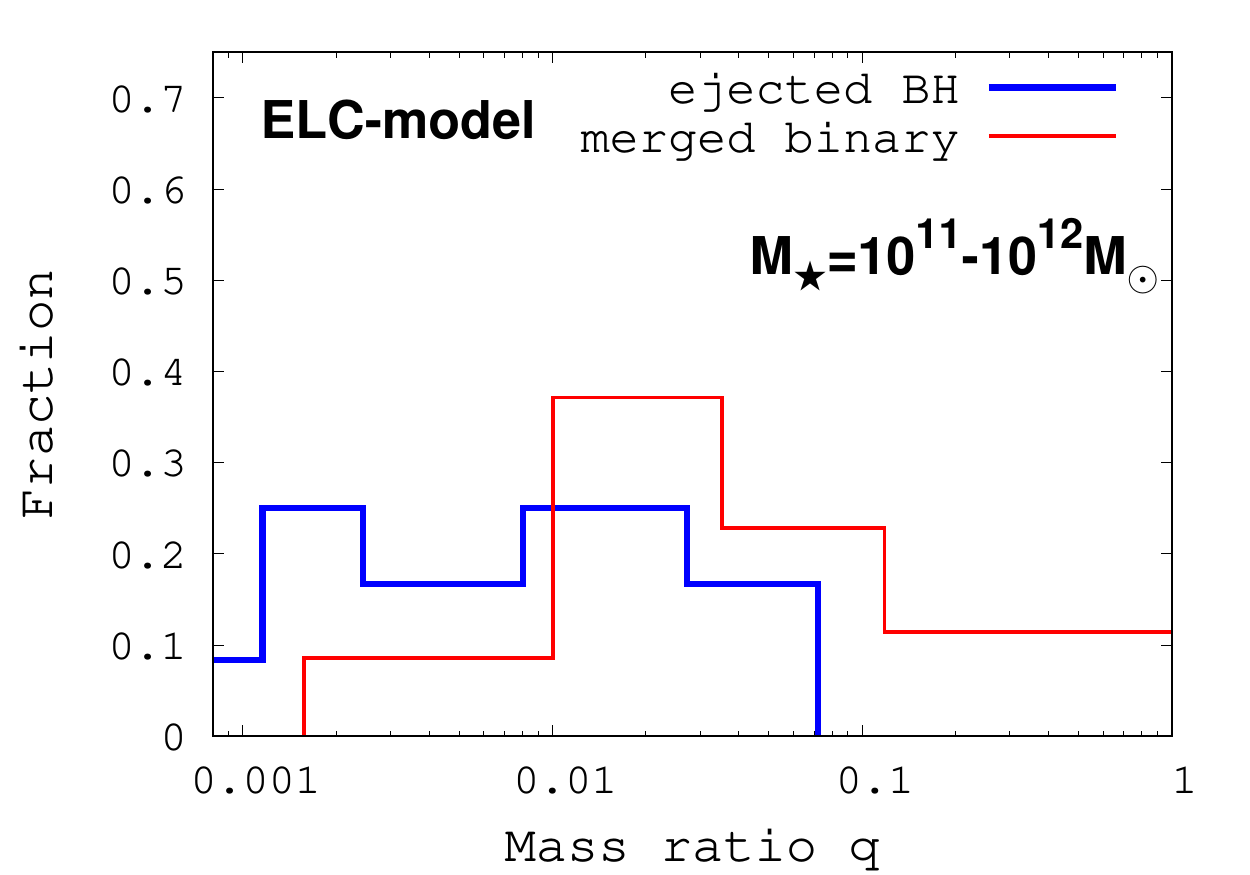}
	\caption{The relative fraction of ejected BHs
		(thick blue solid line) as a function of mass ratio $q$ in
		the ELC-model. This is normalized so that the
		sum of the fractions is unity. Here, $q$ of the
		``ejected BHs'' is defined as the mass ratio between ejected
		BHs and central BHs during 3-body interactions.  As a
		comparison, we also depict the lines (thin red solid)
		corresponding to the merged binaries shown in Figure
		\ref{fig:merger_massratio}.}
	\label{fig:massratio_ELCmodel}
\end{figure*}

\begin{figure}
	\centering
	\includegraphics[width=8.4cm]{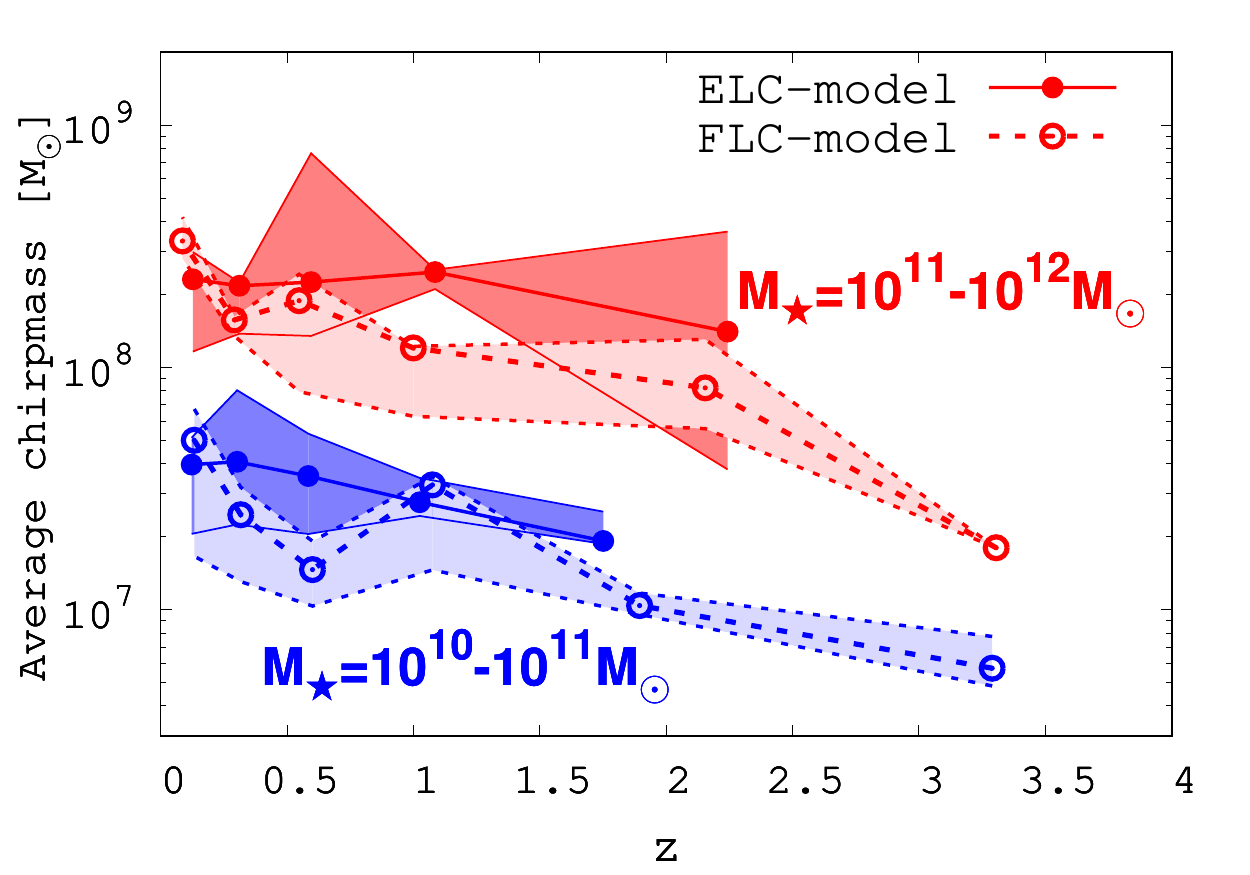}
	\caption{The redshift evolution of the average chirp mass for merged BHs 
		in galaxies of $M_{\star}=10^{10}-10^{11}\Msol$ (blue lines) and 
		$M_{\star}=10^{11}-10^{12}\Msol$ (red lines). The dotted lines represent 
		the FLC-model and the solid lines the ELC-model. The shaded regions indicate 68\% of BH mergers at a given redshift. We use same line types for the average values (lines with circles) and the (slightly thinner) lines running along with the boundaries of the shaded regions.}
	\label{fig:chirmass}
\end{figure}

In Figure \ref{fig:merger_massratio} we present the number fraction of
merged central BH binaries as a function of $q$ (in logarithmic
intervals) in host galaxies of $M_{\star}=10^{10}-10^{11}\Msol$
(\textit{left} panel) and $M_{\star}=10^{11}-10^{12}\Msol$
(\textit{right} panel).  A noticeable difference between the ELC and
the FLC-model is that BH mergers with larger mass ratios are more
common in the ELC-model (see longer high-$q$ tails for the ELC-model in
both galaxies).  The reason for this is likely the nature of
three-body interactions, i.e., less massive objects being easily
ejected, leaving behind more massive
binaries \citep{ValtonenKarttunen2006}. This trend is more pronounced
in the host galaxies of $M_{\star}=10^{11}-10^{12}\Msol$
(\textit{right} panel). Considering more frequent major mergers (see
Table \ref{table:galaxymergercount}) as well as higher $q_{\star}$
(see Figures \ref{fig:galaxymerger_q_count}), the BH merger ratios in
such galaxies for the FLC-model and ELC-model are generally
high. However, for the galaxies of $M_{\star}=10^{11}-10^{12}\Msol$,
the number of host galaxies going through a single major merger and
multiple major mergers are comparable (ratio of $\sim3:4$ in Table
\ref{table:galaxymergercount}). This means that BH mergers in the ELC-model
more ``selectively'' occur in the galaxies experiencing multiple major
mergers. Even though the merger rates are low (see Figure
\ref{fig:mergercount_z}), this can possibly lead to a shift to higher
$q$.

Such enhancement of higher $q$ (or ``selective mergers'' in more massive galaxies) 
for the ELC-model can also be found 
in Figure \ref{fig:BHmerger_massratio_z}. In this figure 
we show the average $q$ for every Gyr from $z=4$ to $z=0$ along with the average 
galaxy merger ratio $q_{\star}$. 
As explained above, typically the mass ratios for the ELC-model are higher than 
for the FLC-model. 
However, comparing with the galaxy merger ratios, the difference becomes noticeable. 
For galaxies of $M_{\star}=10^{10}-10^{11}\Msol$ (\textit{left} panel), the BH merger 
mass ratios $q$ are quite moderately following the line for the galaxy merger mass 
ratio $q_{\star}$. For those of $M_{\star}=10^{11}-10^{12}\Msol$ (\textit{right} panel), 
however, the lines for $q$ are always positioned above that for $q_{\star}$, and $q$ for 
the ELC-model is generally higher than that for the FLC-model.

	As a consequence of
three-body interactions, the chirp mass is higher for BH
mergers in the ELC-model. 
We present these effects in Figures
\ref{fig:massratio_ELCmodel} and \ref{fig:chirmass}. Figure \ref{fig:massratio_ELCmodel}
shows the fraction of ejected BHs as a
function of mass ratio $q$ in the ELC-model. Here, $q$
labeled ``ejected BHs'' refers
to the mass ratio of ejected BHs to the central BHs
during 3-body interactions.  As a comparison, we also
depict the lines corresponding to
the merged binaries shown in Figure
\ref{fig:merger_massratio}.  We can see a higher fraction
of ejected BHs with smaller $q$ for galaxies in both mass bins.  This
implies that less massive BHs are more likely to be ejected,
resulting in more massive binaries retained in the core
regions. Additionally, a comparison between the two
panels shows that the mass ratios of ejected BHs to the central BHs in larger
galaxies (\textit{left} panel) are lower than those in
smaller galaxies (\textit{right} panel).  Therefore,
given the central binary masses required by the
$M-\sigma$ relation (i.e., the average mass of merged
binaries $\sim4.5\times 10^{-3}M_{\star}$ in Table
\ref{table:BHmergercount}) and the larger mass ratios,
the chirp mass for the ELC-model also becomes higher for galaxies in both mass bins as found in Figure \ref{fig:chirmass}. The shaded regions indicate 68\% of BH mergers at a given redshift. The lines for the average values and those demarcating the shaded regions share the same line types, but slightly thinner.

\begin{figure*}
	\centering
	\includegraphics[width=8.4cm]{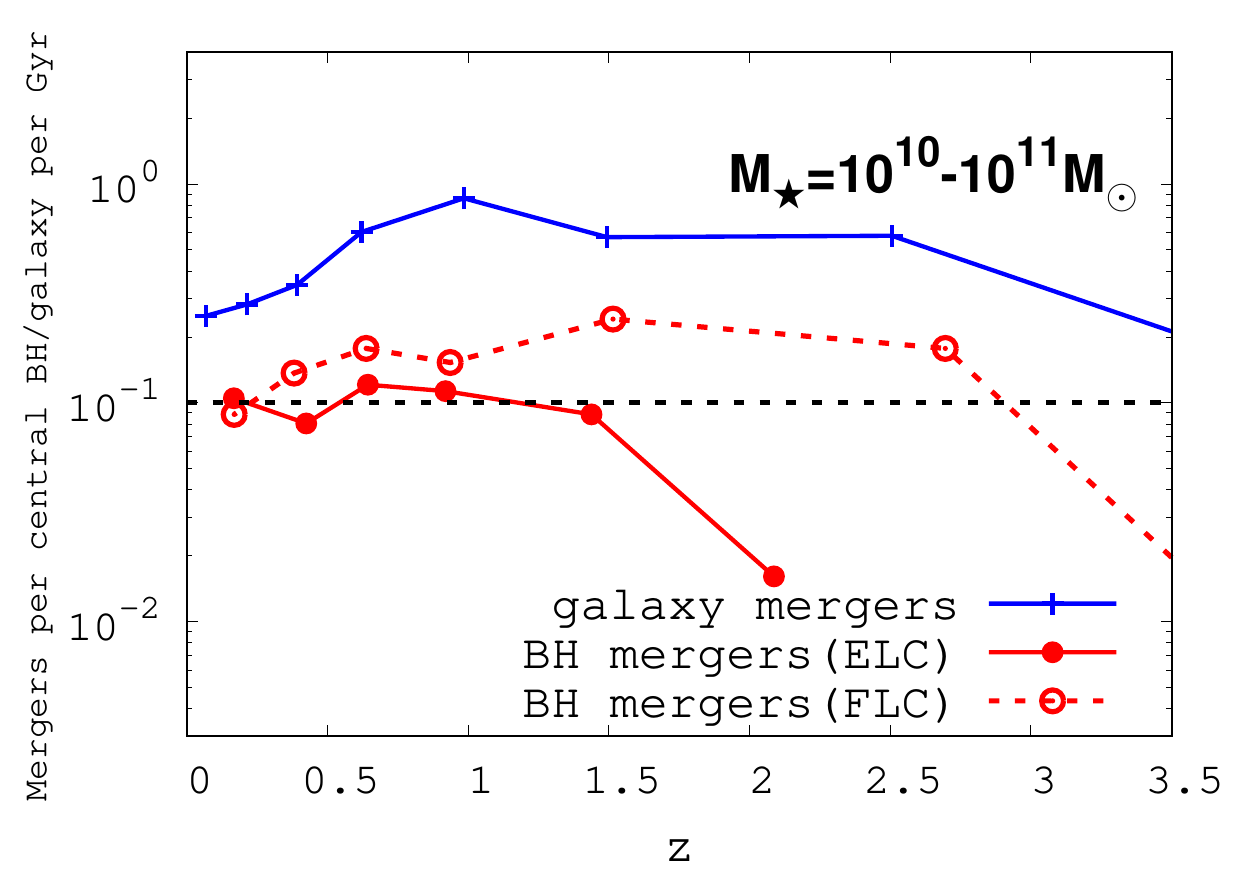}
	\includegraphics[width=8.4cm]{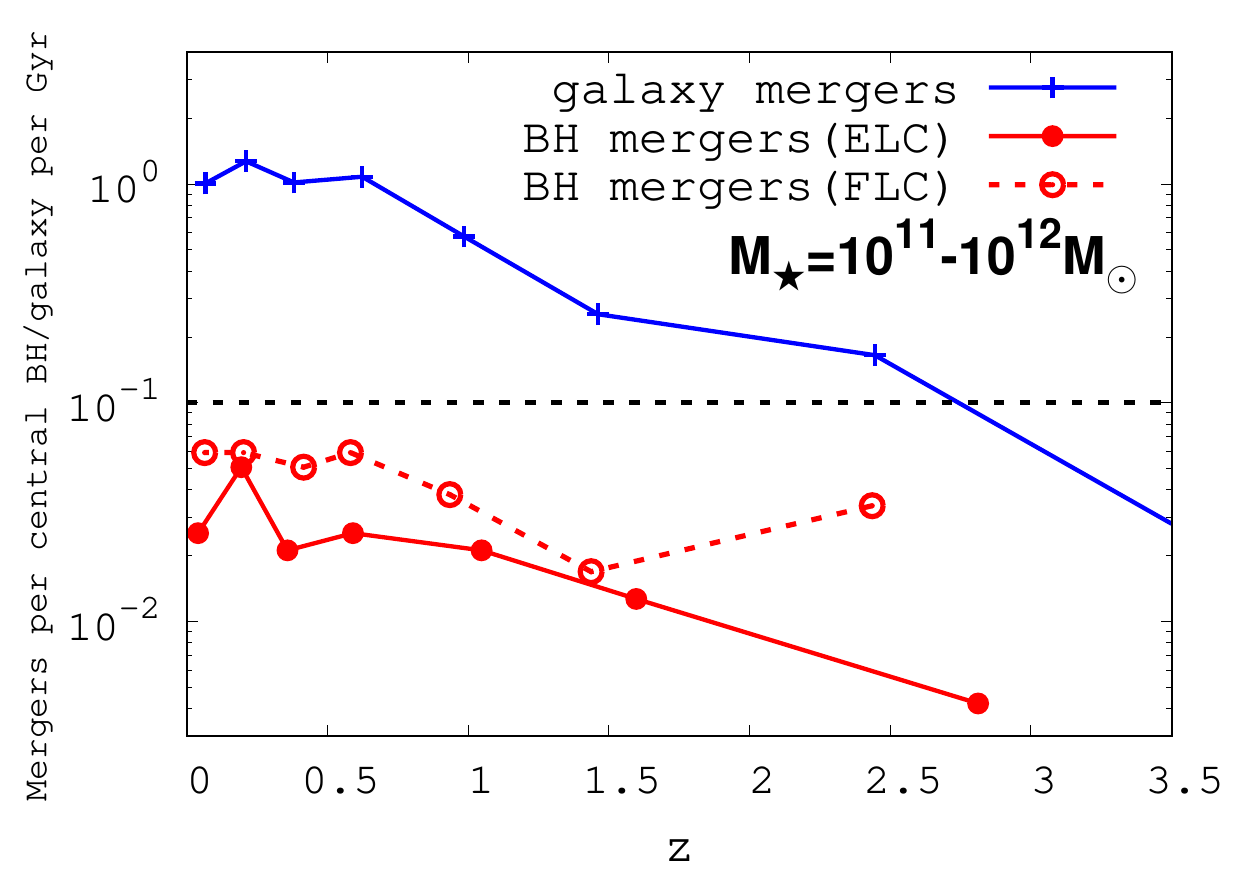}
	\caption{The merger counts per central BH (red lines)/galaxy (blue line) averaged per
          Gyr, or $\Delta N/\Delta t$ for the host galaxies
          of $M_{\star}=10^{10}-10^{11}\Msol$ (\textit{left} panel)
          and $M_{\star}=10^{11}-10^{12}\Msol$ (\textit{right}
          panel). The blue solid line with crosses
            indicates the merger counts for the host galaxies (same
            lines with the thickest lines as in the \textit{bottom} panel
            of Figure \ref{fig:galaxymerger_q_count}). We adopt the
            solid line with solid squares for the ELC-model and the dotted line
            with hollow squares for the FLC-model. For an easy
            comparison, we additionally depict a reference line
            (black dotted line) corresponding to $\Delta N/\Delta
            t=0.1$.}
	\label{fig:mergercount_z}
\end{figure*}

\begin{figure*}
	\centering
	\includegraphics[width=8.4cm]{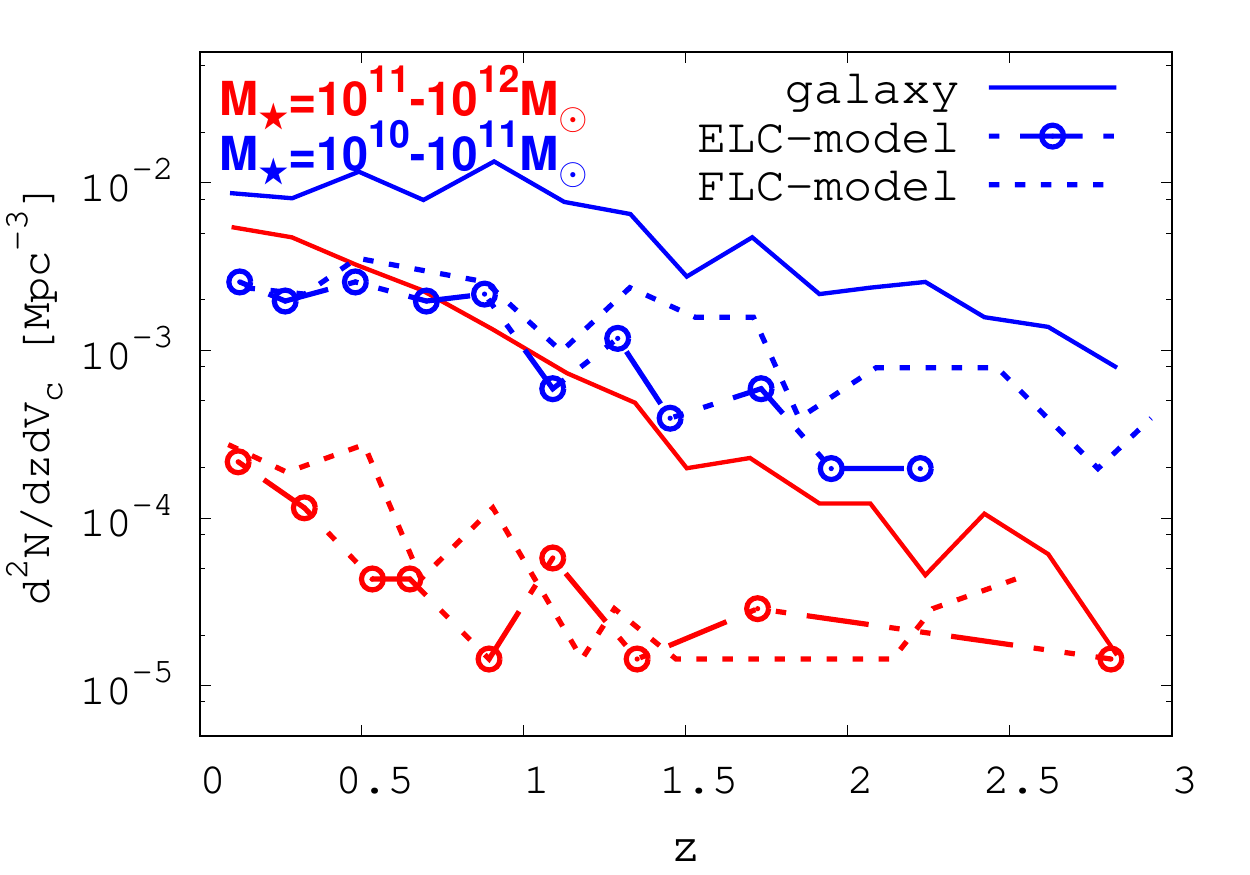}
	\includegraphics[width=8.4cm]{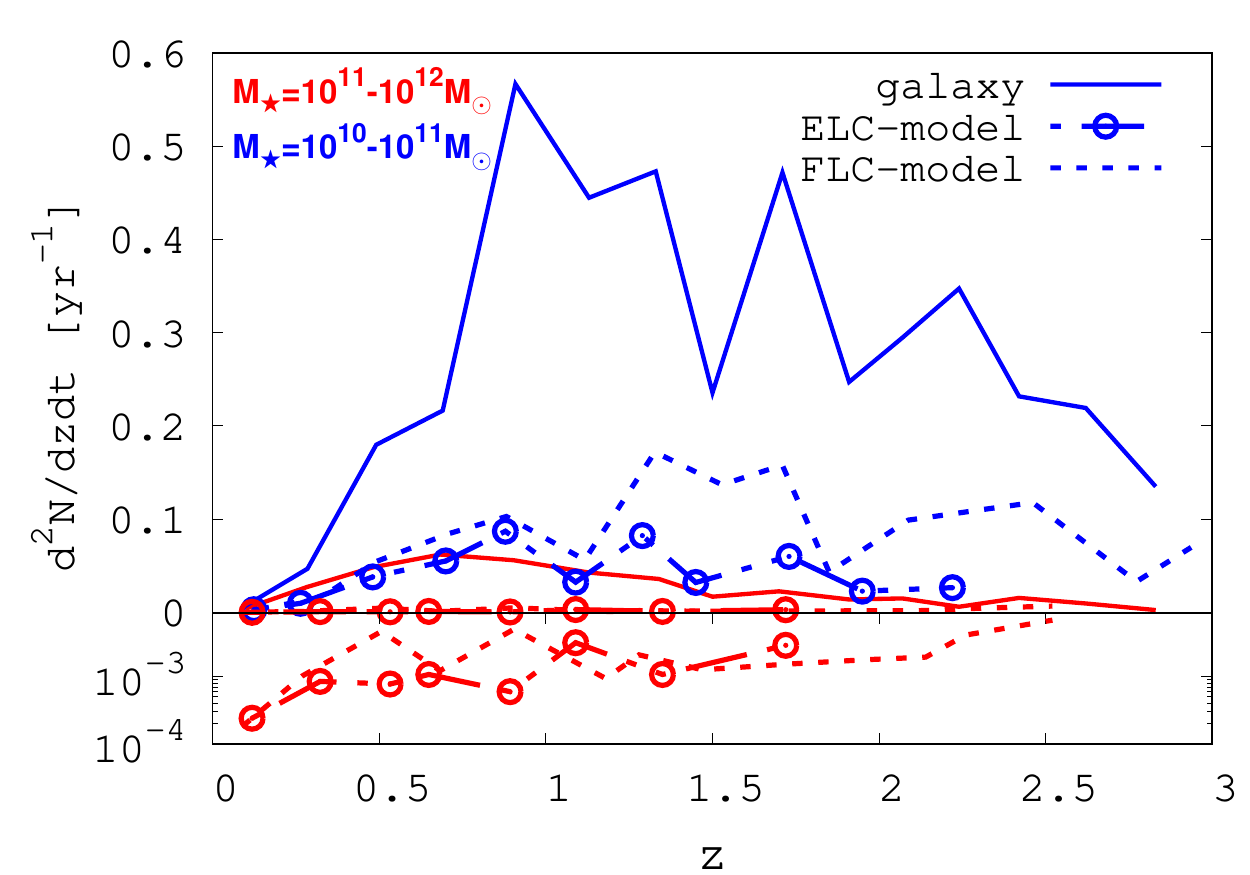}
	\caption{The merger rates of BHs and
		galaxies in two different units.  \textit{Left} panel:
		the number of BH mergers per unit redshift per comoving volume $V_{\rm c}$, or
		$d^{2}N/dzdV_{\rm c}$, for the FLC- (dotted lines) and the
		ELC- models (dot-dashed lines with circle). Here, we
		take for $V_{\rm c}$ the size of the computation box in
		the Milli-Millennium simulation. \textit{Right} panel:
		the number of BH/galaxy mergers per unit time per unit
		redshift, or $d^{2}N/dzdt$. We use equation (4) in
		\citet{Menou+2001} for the unit
		conversion between the merger rates in the two panels.
		The same line colors and types are adopted as in the
		\textit{left} panel. In the bottom panel, for clarity,
		we further draw on a logarithmic scale the lines for the
		BH merger rates in the galaxies of
		$M_{\star}=10^{11}-10^{12}\Msol$.}
	\label{fig:mergerrate_comoving}
\end{figure*}

\begin{figure}
	\centering
	\includegraphics[width=8.4cm]{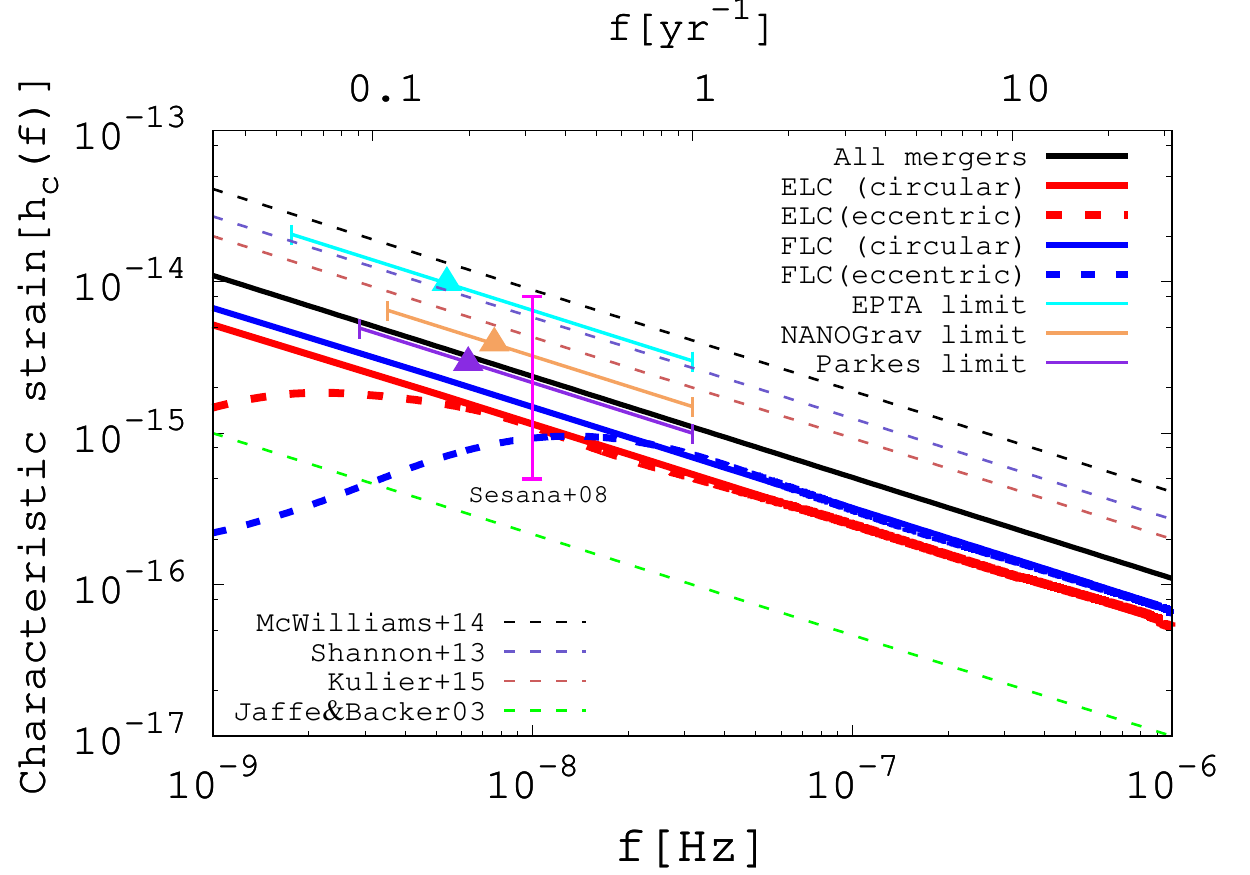}
	\caption{The characteristic strain $h_{\rm c}$ for the
          FLC-model (thick blue solid line) and the ELC-model (thick
          red solid line). The black solid line (labeled "All
          mergers") above both models indicates the strain assuming
          all galaxy mergers lead to BH mergers given the sampled
          galaxy merger trees (see more details in \S
          \ref{sub:semi-analytic}). 
          We additionally indicate the upper limit in each experiment at
          		its peak sensitivity (triangles), and we extrapolate this limit to other
          		frequencies assuming a power law of $f^{-2/3}$ within the frequency range of $1/T<f<1\yr^{-1}$, where $T$ is the total observing time. The dotted lines refer to the PTA
          		estimates from other studies. We estimate $A=0.70\times10^{-15}$ for the
          FLC-model and $A=0.53\times 10^{-15}$ for the ELC-model. The
          curved dotted lines indicate the deviation due to eccentric
          orbits. The line colors are shared with those for the
          circular orbit case (thick blue/red solid lines).}
	\label{fig:characteristic_h}
\end{figure}

\subsubsection{BH merger rate}
\label{subsub:BHmergerrate}
We present in Figure \ref{fig:mergercount_z} and Figure
\ref{fig:mergerrate_comoving} two different realizations of the BH
merger rates as a function of $z$. Figure \ref{fig:mergercount_z}
shows the merger counts per central BH/galaxy averaged over every Gyr,
or $\Delta N/\Delta t$ for the host galaxies of
$M_{\star}=10^{10}-10^{11}\Msol$ (\textit{left} panel) and
$M_{\star}=10^{11}-10^{11}\Msol$ (\textit{right} panel), with a
reference line corresponding to $\Delta N/\Delta t=0.1$. There are a
few noticeable features seen in both panels as follows: (i) the BH
coalescence rates for the FLC-model are higher for galaxies in both
mass bins than those for the ELC-model.  This is also seen in Figure
\ref{fig:mergerrate_comoving}. This is expected given the longer
lifetimes of BH binaries in the ELC-model, possibly leading to
ionizations of binaries as well as ejections of BHs; (ii) the merger
rates are higher for BHs in less massive host galaxies (\textit{left}
panel). Notice that the BH merger rates for more massive galaxies are
always below the reference line; but the differences in the BH merger
rates between the galaxies get smaller as $z$ decreases. Finally, the
rates tend to converge to $10^{-2}\Gyr^{-1}<\Delta N/\Delta
t<10^{-1}\Gyr^{-1}$ at $z\simeq0$. The rate at $z=0$ is consistent
with what has been assumed as a present-day merger rate for a single
object in \citet{JaffeBacker2003}. (iii) Comparing the BH merger rates
with the galaxy merger rates, the BH coalescence rates are smaller
than the galaxy merger rates by a factor of $3-20$ depending on the
model and redshift. As shown in Figure
\ref{fig:galaxy_mergercorrelation}, every galaxy merger with a small
mass ratio does not always lead to a BH merger. BHs, which either
never fall into the core or are ejected, are left orbiting outside the
core regions. If one only considers major mergers ($q_{\star}>0.1$),
then as indicated in Figure \ref{fig:galaxy_mergercorrelation}, the
differences should be smaller. However, such differences should be
considered for studies including both minor and major mergers.

 In Figure \ref{fig:mergerrate_comoving}, we also
     show the merger rates of BHs and galaxies in two different
     units. In the \textit{left} panel, we show the number of
 BH/galaxy mergers per unit redshift per comoving
 volume $V_{\rm c}$, or $d^{2}N/dzdV_{\rm c}$. For this, we take for
 $V_{\rm c}$ the size of the computation box in the Milli-Millennium
 simulation ($V_{\rm c}\simeq6.28\times10^{5}\Mpc^{3}$). It is clear 
 that the merger rates are rising
 towards lower $z<1.5$ (as those for galaxies) except for the rate of
 the galaxies of $M_{\star}=10^{10}-10^{11}\Msol$ for the FLC-model,
 which remain roughly flat.  The counts for all models reach up to
 $d^2N/dzdV_{\rm c}\sim2\times10^{-4}\Mpc^{-3}$ for more massive galaxies 
and $d^2N/dzdV_{\rm c}\sim3\times10^{-3}\Mpc^{-3}$ for less massive galaxies at $z\simeq0$.
 This is attributed to the tendency for a larger number of BHs to
 accumulate in the core region at $z\sim0$.  Even smaller BHs (with
 longer decay times) can have enough time to decay to the core
 regions, increasing the chances of BH mergers in both models.
 Additionally, given the high merger rates for lower-mass galaxies,
 and especially the higher mass ratios in the ELC-model, we can expect
 that the contribution of BH mergers in lower-mass galaxies to the
 GWB is not negligible (see Figure
 \ref{fig:characteristic_h_differentmass}).
 		
 In the \textit{right} panel, the number of BHs/galaxy mergers per
 unit time per unit redshift, or $d^2N/dzdt$, is
 presented. This represents the detectable merger rate
   that originates from a comoving shell in redshift (corresponding to
   the comoving volume in the \textit{left} panel)}. For the conversion
 between the merger rates in the \textit{left} in the \textit{right}
 panels, we use equation (4) in \citet{Menou+2001}. The same line
 colors and line types are used as in the \textit{left} panel. Also
 note that, for a clearer view, we further draw on a logarithmic scale
 the lines for the BH merger rates in the galaxies of
 $M_{\star}=10^{11}-10^{12}\Msol$ (bottom box).  The BH merger event
 rate is $0.1-0.2$ per yr at $z\simeq1-2$ and $10^{-4}-10^{-2}$ per yr
 for the galaxies of $M_{\star}=10^{10}-10^{11}\Msol$ and
 $M_{\star}=10^{11}-10^{12}\Msol$, respectively. 
 
In the next section, we use the BH merger rates in our models to 
estimate the amplitude and spectrum of the stochastic GWB.

\begin{figure*}
	\centering
	\includegraphics[width=8.4cm]{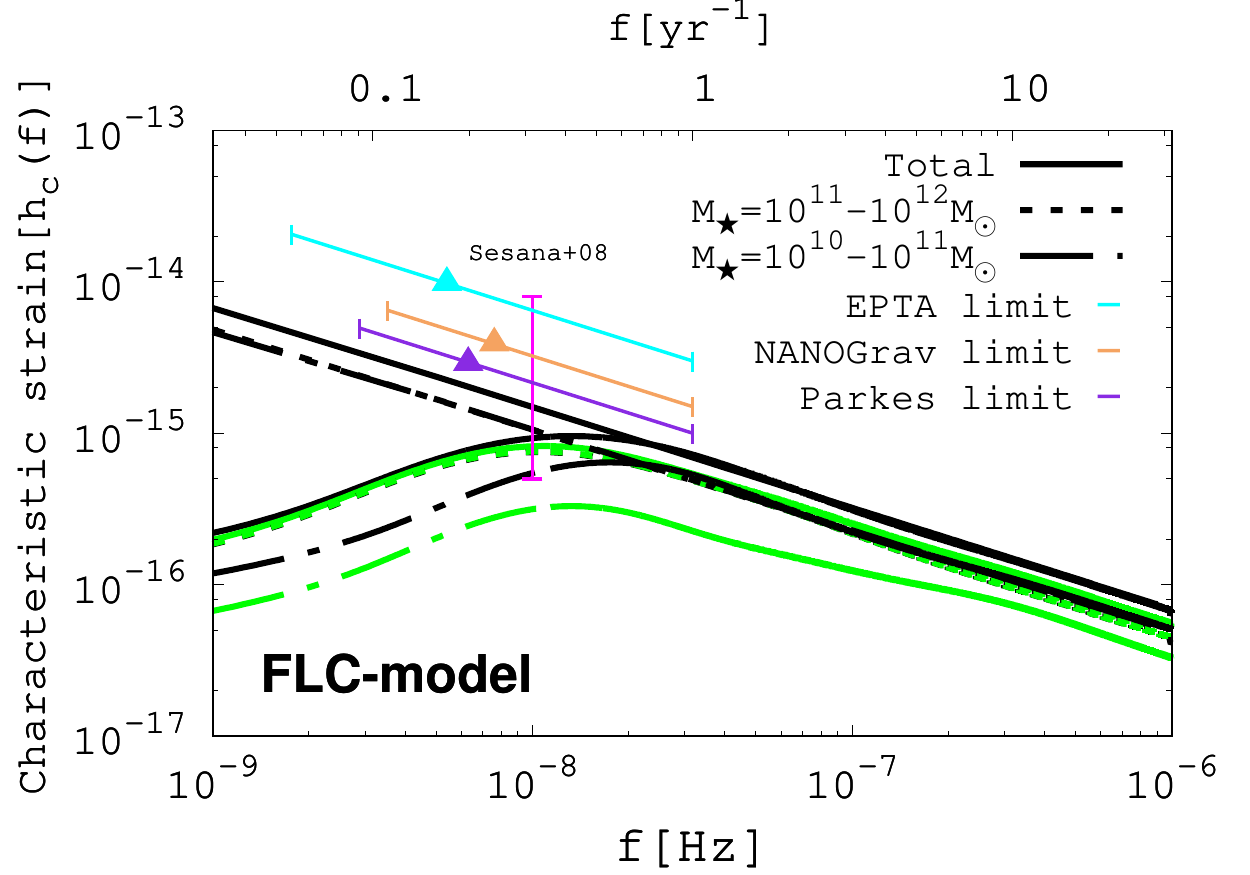}
	\includegraphics[width=8.4cm]{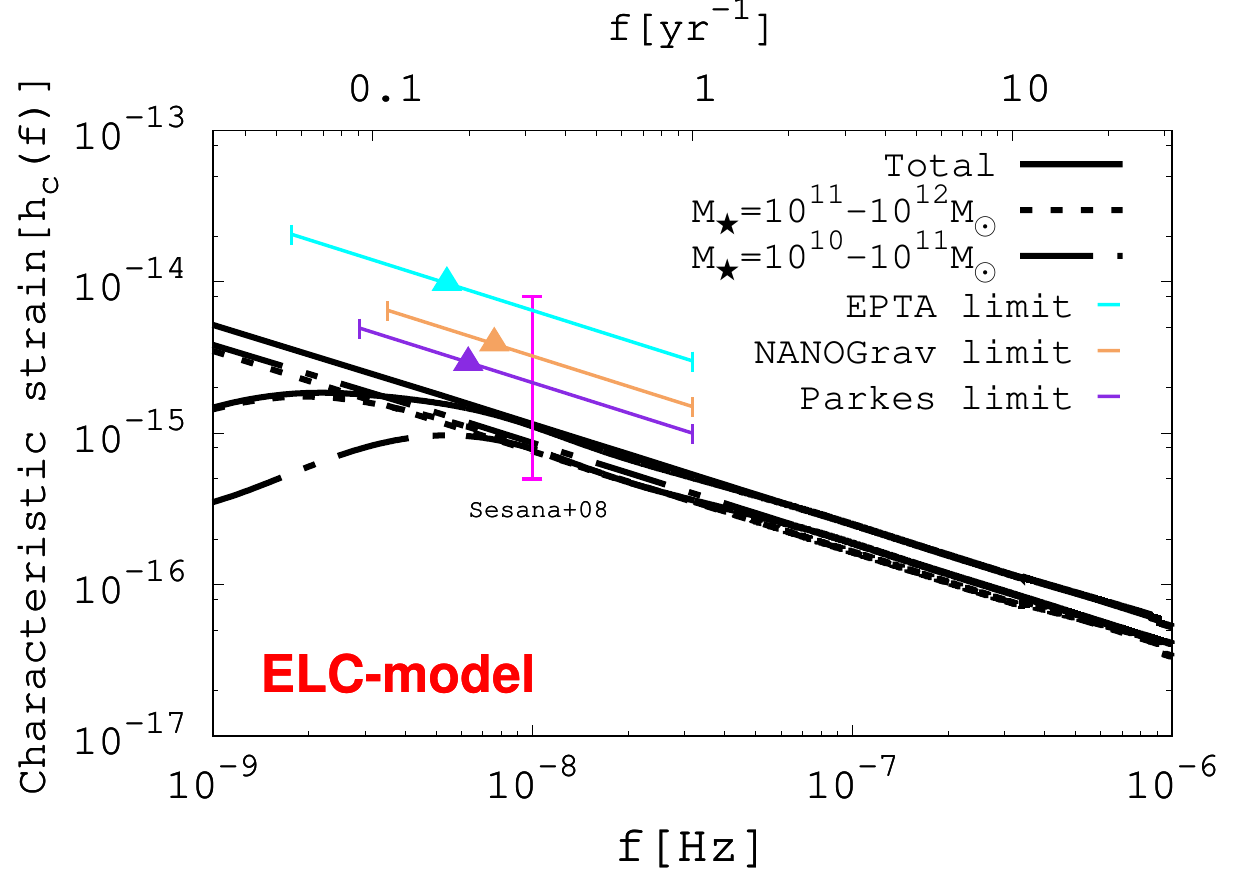}
	\caption{The spectra of the characteristic strain from the
          galaxies of each mass range (dotted line for
          $M_{\star}=10^{11}-10^{12}\Msol$ and dot-dashed line for
          $M_{\star}=10^{10}-10^{11}\Msol$) as well as the total
          estimate (solid line) for the FLC-model (\textit{left}
          panel) and the ELC-model (\textit{right} panel).  The
          straight lines represent the strain assuming circular
          orbits, while the curved lines show the modification when
          the orbital eccentricities are taken into account. As a
          reference, we also indicate the upper limits for the strain with thin solid lines. The triangles show the upper limit in each experiment at
          its peak sensitivity. In
            the \textit{left} panel, the green lines indicate the
            spectra made with only SMBH binaries fulfilling the bypass
            condition (Equation \ref{eq:e_rGW}). As a consequence of
          three-body interactions, the chirp mass is higher for BH
          mergers in the ELC-model (see Figure \ref{fig:chirmass}). As
          a result, despite the lower BH merger rates, BH mergers in
          the smaller galaxies of $M_{\star}=10^{10}-10^{11}\Msol$
          almost equally contribute to the GWB as those in more
          massive galaxies. }
	\label{fig:characteristic_h_differentmass}
\end{figure*}

\begin{figure}
	\centering
	\includegraphics[width=8.2cm]{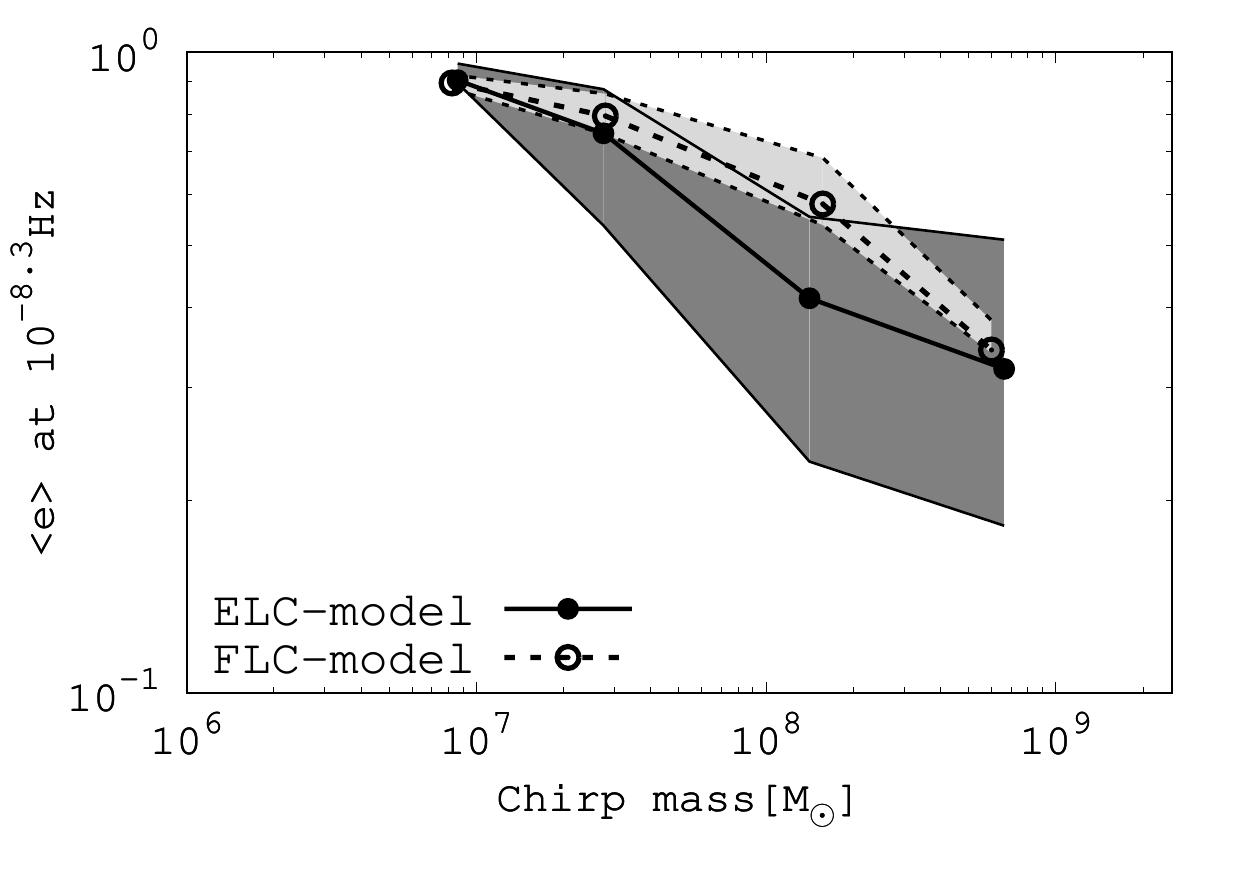}
	\includegraphics[width=8.2cm]{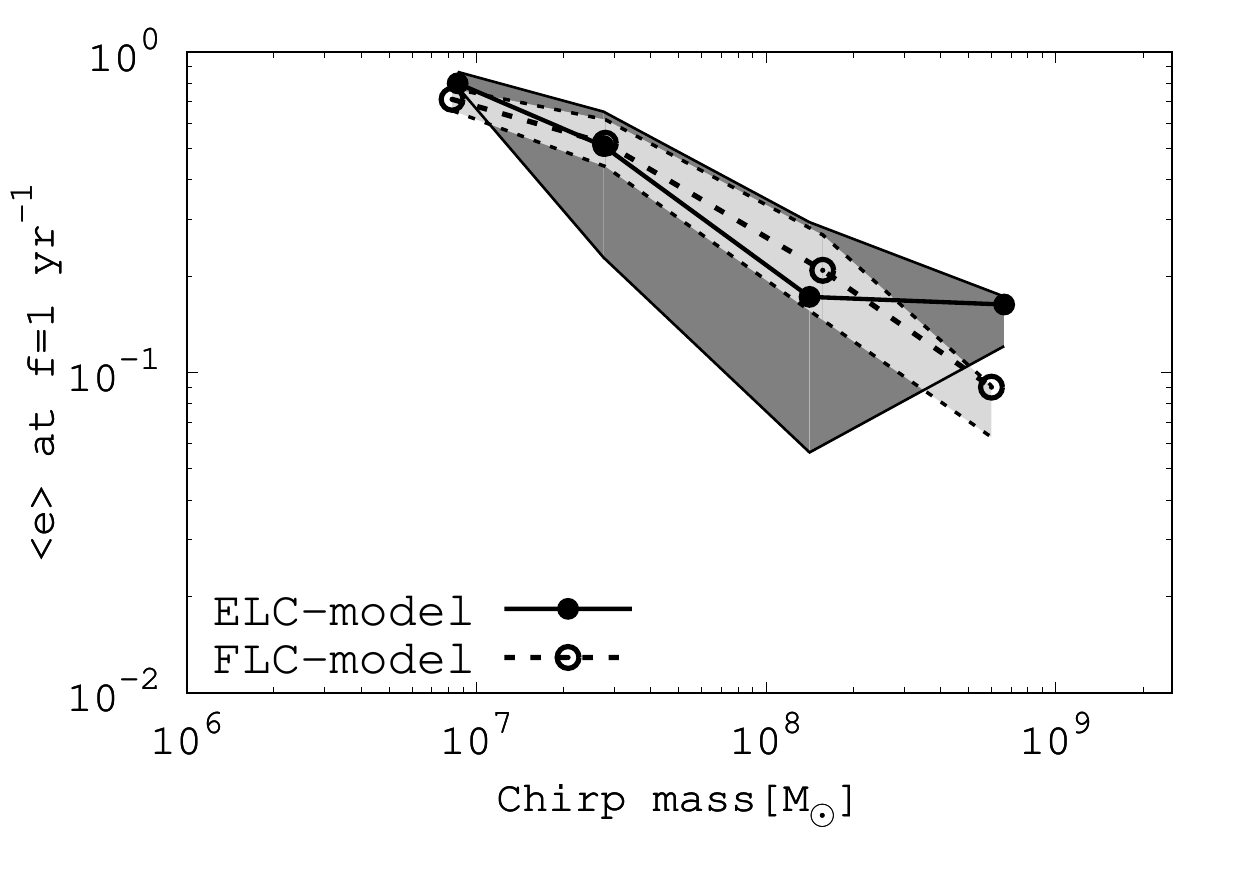}
	\includegraphics[width=8.2cm]{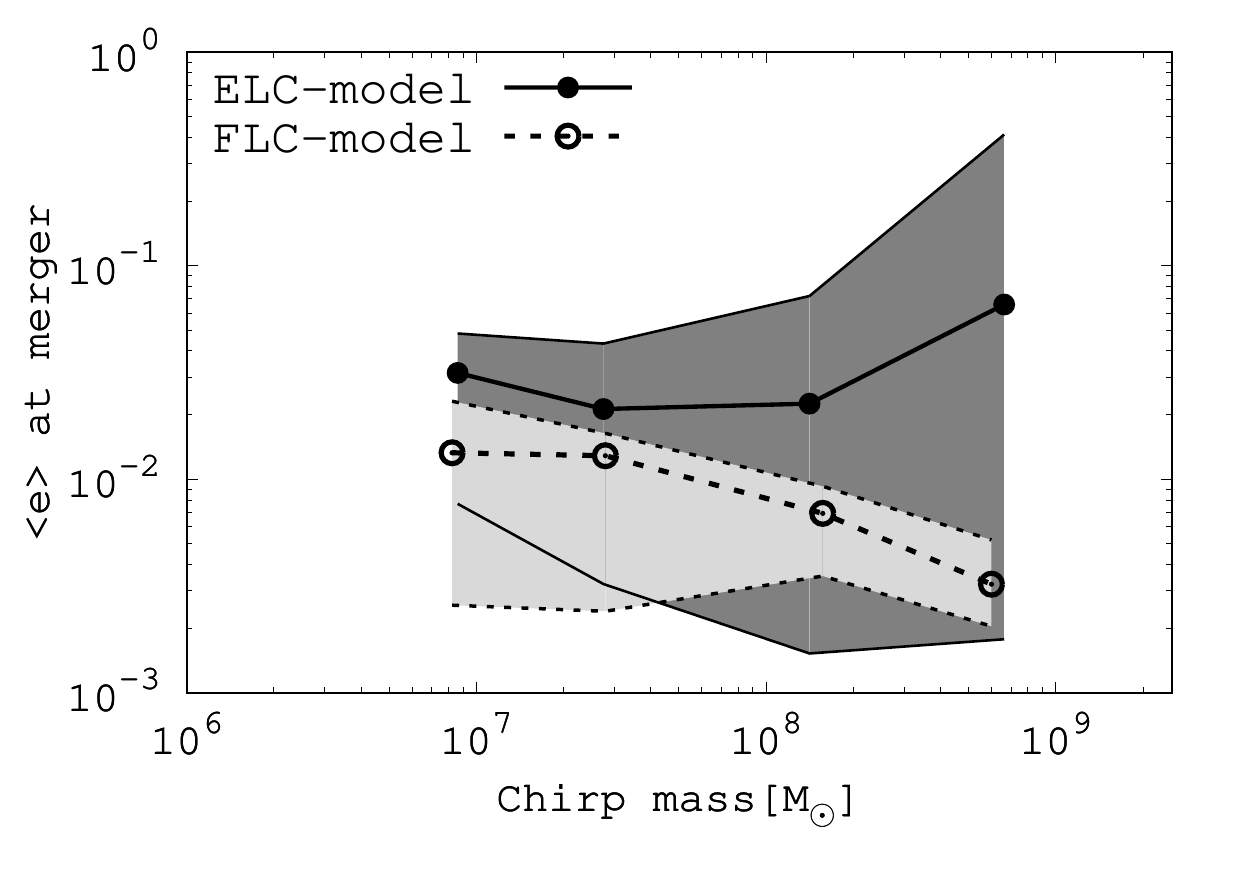}
	\caption{$M_{\rm chirp}^{5/6}$-weighted (corresponding to
          weighting by GW intensity) average $e$ as a function of
          $M_{\rm chirp}$ for the FLC-model (dotted line with hollow
          circles) and the ELC-model (solid line with solid circles)
          at three characteristic frequencies, i.e. $f=10^{-8.3}~{\rm
            Hz}$ (near peak sensitivity, see Figure
          \ref{fig:A_sensitivity}), $f=1 \yr^{-1}~{\rm
            Hz}=10^{-7.5}~{\rm Hz}$ and $f$ at merger.  We
          analytically estimate the eccentricities at which GW
          emission becomes more efficient 
          \citep{PetersMathews1963}. The shaded regions indicate 68\%
          of BH mergers at a given chirp mass. }
	\label{fig:ecc_atmerger_year}
\end{figure}

\begin{figure}
	\centering
        \includegraphics[width=8.5cm]{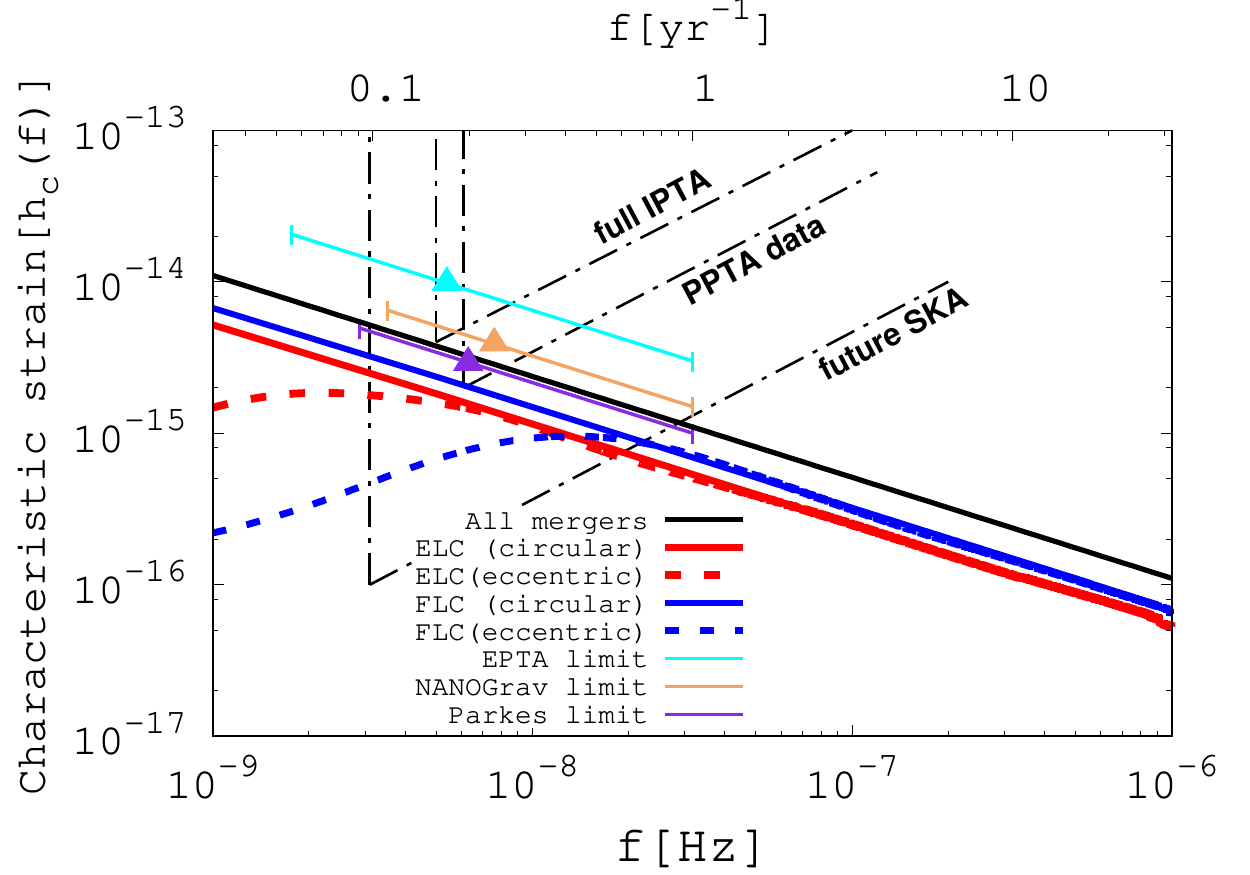}
	\caption{Our estimates for $h_{\rm c}$ are compared with the
          current/future-expected sensitivities (wedge-shaped
          dot-dashed lines).  The upper-most wedge indicates the
          sensitivity set by the full IPTA \citep[$h_{\rm c}\sim
            10^{-16}-10^{-17}$ at $f=5\times10^{-9} {\rm ~Hz}$,
          ][]{Janssen+2015}.  The predictions of our models cannot yet
          be tested with the current instruments.  The other two lines
          refer to the sensitivity set by complete Parkes PTA (PPTA,
          20 pulsars for $5\yr$) data set (labeled ``PPTA data'') and
          that achievable with the planned Square Kilometer Array
          (SKA) assuming monitoring of 20 pulsars over 10$\yr$
          (labeled ``future SKA'').  The sensitivity
            provided by the PPTA dataset may not be sufficient to
            reach the strain inferred from our models.  In the future,
            we expect that the planned SKA will be able to impose
            constraints over wider frequency ranges.}
	\label{fig:A_sensitivity}
\end{figure}

\section{Discussion}
\label{discussion}
\subsection{Stochastic GW background - Pulsar timing array estimate}
\label{sub:PTA}

Over observing times of a few years to a few months,
binary supermassive BHs are one of the most promising
astrophysical sources of GWs in the nHz frequency band accessible to
pulsar timing arrays (PTAs).  In this section, based on the merger
rates inferred from our two models, we estimate the characteristic
strain $h_{\rm c}(f)$ first assuming circular orbits,
and then including the effects of high orbital
    eccentricity.

\subsubsection{GW from circular orbits}
The characteristic strain $h_{\rm c}(f)$ of the GW signal from a circular binary can 
be calculated as follows \citep{Phinney2001,Sesana+2008},
\begin{equation}
h_{\rm c}^{2}(f)=\frac{4G}{\uppi f^{2}c^{2}}\int_{0}^{\infty}dz\int_{0}^{\infty}d\mathcal{M}\frac{d^{2}n}{dzd\mathcal{M}}\frac{1}{1+z}\frac{dE_{\rm gw}(\mathcal M)}{d\ln f_{\rm r}}\,,
\label{eq:characteristic_strain}
\end{equation} 
where $f$ is the observed frequency and $\mathcal{M}$ is the chirp
mass, defined as $\mathcal{M}=(M_{\rm BH,1}M_{\rm BH,2})^{3/5}/(M_{\rm
  BH,1}+M_{\rm BH,2})^{1/5}$.  Here $f$ is related to the rest-frame
frequency $f_{\rm r}$ and the Keplerian orbital frequency $f_{\rm
  orb}$ such that $f(1+z)=f_{\rm r}=2f_{\rm orb}$. $E_{\rm gw}$ is the
energy emitted in GWs. $d^{2}n/dzd\mathcal{M}$ represents the
differential merger rate density (i.e., the number of BH mergers per
comoving volume) of SMBH binaries per unit redshift per unit chirp
mass. It is easily shown that the strain scales as $f^{-2/3}$
\citep{Phinney2001} and is usually described in terms of $A$
\citep{Jenet+2006},
\begin{equation}
h_{\rm c}(f)=A\left(\frac{f}{\yr^{-1}}\right)^{-2/3}\,.
\label{eq:hc}
\end{equation}
In particular, for a finite number of sources in a comoving volume $V_{\rm c}$ with 
the rest frame frequency in the range of $f_{\rm min}<f_{\rm r}<f_{\rm max}$, 
Equation \ref{eq:characteristic_strain} can be re-written as follows,
\begin{equation}
h_{\rm c}^{2}(f)=\frac{4\uppi^{-1/3}}{3c^{2}}f^{-4/3}\sum_{\substack{i\\
f_{\rm min}<f_{\rm r}<f_{\rm max}}}\frac{1}{(1+z_{i})^{1/3}}\frac{(G\mathcal{M}_{i})^{5/3}}{V_{\rm c}}\frac{N_{\rm galaxy,total}}{N_{\rm galaxy}}
\end{equation}
where $i$ represents each GW source (BH merger event) in the galaxies
of both mass ranges. Assuming that our galaxy sample of size $N_{\rm
  galaxy}$ is representative of the properties of the entire set of
galaxies in the Millennium simulation of number $N_{\rm galaxy,total}$, we
normalize our estimate of the strain with a factor of $N_{\rm
  galaxy,total}$/$N_{\rm galaxy}$. The variable $f_{\rm min}$
represents twice the Keplerian orbital frequency calculated with the
values of the binary mass and the semimajor axis at the time when GWs
become dominant to shrink the orbit (i.e., when the merger condition
\ref{item:mergercondition1} is fulfilled). For $f_{\rm max}$, we
assume the frequency at the innermost circular orbit, or $f_{\rm
  max}=[2/(1+z)]f_{\rm orb}(r=3r_{\rm sch})$
\citep{Hughes2002,Ravi+2012,McWilliams+2014}, which is written as,
\begin{equation}
f_{\rm max}=2.2\times10^{-5}\left(\frac{M_{\rm BH,1}}{10^{8}\Msol}\right)^{-1}
\left(1+\frac{M_{\rm BH,2}}{M_{\rm BH,1}}\right)^{1/2}{\rm Hz}\,.
\end{equation}

Assuming circular orbits and given the amplitude scaling as $f^{-2/3}$
(see Eq.\ref{eq:hc}), from our merger rates we find that
$A=0.70\times10^{-15}$ for the FLC-model and $A=0.53\times 10^{-15}$
for the ELC-model. We show our estimates for the characteristic strain
$h_{\rm c}$ for the FLC-model (denoted by ``FLC (circular)'') and the
ELC-model (denoted by ``ELC (circular)'') in Figure
\ref{fig:characteristic_h}. The black solid line (labeled "All
mergers") above the two lines for the FLC-model and the ELC-model
corresponds to the strain assuming all galaxy mergers lead to BH
mergers given the sampled galaxy merger trees (see more details in \S
\ref{sub:semi-analytic}). We additionally depict the GW spectra
inferred in other published studies
\citep{JaffeBacker2003,Sesana+2008,KocsisSesana2011,Shannon+2013,
  McWilliams+2014,Kulier+2015} and observational upper
limits set by EPTA \citep[$A=3.0\times10^{-15}$,
][]{Lentati+2015}, NANOGrav \citep[$A=1.5\times10^{-15}$,
][]{Arzoumanian+2016} and Parkes \citep[$A=1.0\times10^{-15}$,
][]{Shannon+2015}. For the latter, we indicate the upper limit in each experiment at
		its peak sensitivity (triangles), and we extrapolate this limit to other
		frequencies assuming a power law of $f^{-2/3}$.  The frequency range
		shown in each case is $1/T<f<1\yr^{-1}$, where $T$ is the total observing time. In spite of different dominant mechanisms for
orbital decay in the FLC- and ELC- model, the values are
comparable. We believe the reasons are as follows:
\begin{enumerate}

	\item\label{A:reason2} The merger rates for the ELC-model are
          lower at $0.5<z<2$ than those for the FLC-model (see Figure
          \ref{fig:mergercount_z}). The resulting decrease in the GWB,
          however, is relatively minor, because it is the mergers
          involving the lowest-mass BHs that are missing. The mergers
          which dominate the GWB, involving more massive BHs, are
          still occurring in the ELC-model.

	\item \label{A:reason1} In the ELC-model,
            we find that binaries have longer lifetimes (see \S
            \ref{subsub:binarylifetime}) due to the time taken for
            multiple SMBHs to accumulate in the cores as the host
            galaxies go through successive mergers. This can cause an
            overall delay of the BH mergers compared to nearly prompt
            mergers in the FLC-model. This results in sparse mergers
            at higher $z$ and, more importantly, copious GW emissions
            at lower $z$. Moreover, individual GW emissions are more
            powerful because the delay of mergers causes BHs to
            accrete more mass before they undergo mergers. This would
            compensate for the decrease in the GWB due to the loss of
            some BH mergers, as described in \ref{A:reason2} above. 
		
	\item In the ELC-model, BH mergers in the
	smaller galaxies of $M_{\star}=10^{10}-10^{11}\Msol$ 
	contribute more to the GWB than those in more massive
	galaxies. Those contributions are even higher than those for the FLC-model because of the more frequent mergers with larger chirp mass at lower $z$ in the ELC-model. Figure
	\ref{fig:characteristic_h_differentmass} shows how much BH mergers in
	the galaxies of each mass range contribute to the total estimates for
	the FLC-model (\textit{left} panel) and the ELC-model (\textit{right}
	panel). The dotted line for $M_{\star}=10^{11}-10^{12}\Msol$
	and the dot-dashed line refers to $h_{\rm c}$ for
	$M_{\star}=10^{10}-10^{11}\Msol$ assuming circular orbits.  As shown
	in the \textit{right} panel, the strain is higher in the
	ELC-model from BH mergers in smaller galaxies.

	Also note that, as will be explained in \S
	\ref{subsubsec:eccentricity}, the curved lines 
	show the effect on the strain of binary eccentricities.
	
\end{enumerate}

We also find that the amplitude of the characteristic strain is
dominated by BH mergers at low redshift $z<2$ (see also
\citealt{WyitheLoeb2003}). In the FLC-model, 86\% of the BH
coalescences occur at $z<2$ with an average chirp mass of
$\mathcal{M}=1.0\times10^{8}\Msol$, while in the ELC-model the
fraction of mergers at $z<2$ is 98\% with
$\mathcal{M}=1.6\times10^{8}\Msol$. If we impose
    more stringent constraints on $z$, in the FLC-model the fraction
    decreases to 65\% with $\mathcal{M}=1.2\times10^{8}\Msol$ at $z<1$
    and to 35\% with $\mathcal{M}=1.4\times10^{8}\Msol$ for $z<0.5$.  In
    the ELC-model the fraction becomes 79\% with
    $\mathcal{M}=1.8\times10^{8}\Msol$ and 49\% with
    $\mathcal{M}=2.1\times10^{8}\Msol$. However, still the majority
of SMBH binaries effectively emit GWs at $z<1$. The increase in the
chirp mass especially for the ELC-model can be seen in the redshift evolution
of the average chirp mass shown in Figure
\ref{fig:chirmass}. Here, we separately show the results for the galaxies of each mass range,
but the average chirp mass (including the fraction of BH mergers) given above is
estimated based on all merger events for galaxies in both mass bins.

\subsubsection{GWs from eccentric orbits} 
\label{subsubsec:eccentricity}

 An eccentric orbit emits GWs at all integer harmonics of the orbital
 frequency \citep{PetersMathews1963,Peters1964}. Especially for very
 eccentric orbits, the GW radiation power is greater at higher
 harmonics. Since the evolution of a binary orbit strongly depends on
 the evolution of the eccentricity, this may change the shape of the
 spectrum. In fact, larger contributions from higher harmonics
 effectively suppress power at lower frequencies, leading to a low
 frequency flattening or even a turnover in the spectrum
 \citep{EnokiNagashima2007,Sesana2010,Sesana2015a}. Therefore, it is necessary to
 take into account such effects of the eccentricity for more realistic
 estimates of the GWB. 
 
We find that the binary orbits are very eccentric when GW emission
becomes more efficient in our simulations (see Figure
\ref{fig:lifetime_eccentricity_q}). In Figure
\ref{fig:ecc_atmerger_year}, we show $M_{\rm chirp}^{5/6}$-weighted
average $e$ as a function of $M_{\rm chirp}$ for the FLC- and
ELC-model at three characteristic frequencies: $f=10^{-8.3}~{\rm Hz}$
(near peak sensitivity), $f=1 \yr^{-1}=10^{-7.5}~{\rm Hz}$ and $f$ at
merger. Given that these are eccentricities at which GW emission plays
a dominant role in causing the decay of the binary orbits, we
analytically estimate the eccentricities at those frequencies
\citep{PetersMathews1963}. The shaded regions indicate 68\% of BH
mergers at a given chirp mass. Generally speaking, the eccentricities
for the ELC-model tend to have large scatter compared to those for the
FLC-model. The eccentricities are still quite high at
$f=10^{-8.3}~{\rm Hz}$ and $f=1 \yr^{-1}$.
 
 To account for such high eccentricities, we have to consider
 harmonics up to $n_{\rm max}\simeq10(1-e)^{-3/2}$ (for $e=0.99$,
 $n_{\rm max}\simeq10000$), which means that a direct summation of the
 contributions of each harmonic is computationally expensive. We
 instead emply the fitting formula (16) given in \citet{Chen+2017},
 which has been shown to successfully reproduce the spectrum within a
 maximum error of 1.5\% in log amplitude (i.e. 3.5\% in amplitude) for
 a reference case ($e=0.9$). The thick dotted lines in Figure
 \ref{fig:characteristic_h} and Figure \ref{fig:A_sensitivity} show
 the spectra when high eccentricities are taken into account. As
 expected, the spectra at lower frequency of $f<1\yr^{-1}$ are
 flattened and turn over towards lower $f$. The strain for both models
 predicted under the assumption of circular orbit is hardly
 distinguishable. However, clear deviations between the two models can
 be seen when the different eccentricity evolutions are taken into
 account in the computation. Eccentric spectra start differing from
 their circular counterparts at frequencies of $f\sim10^{-7.5}~{\rm
   Hz}\sim1\yr^{-1}$ in both models, and display maxima in the region
 around $f<10^{-8}~{\rm Hz}$. Such turnovers of the spectra are
 consistent with the spectra predicted for $e=0.9$ in
 \citet{EnokiNagashima2007}. The overall shapes of the
   spectra are also similar with what is found for the case of initially
   very eccentric binaries in a dense stellar environment (see Figure
   3 in \citealt{Sesana2015a}). The spectra from BH mergers in
 galaxies of both mass ranges are comparable at frequencies of
 $f>10^{-8}~{\rm Hz}$; however at $f<10^{-8}~{\rm Hz}$, the signals
 from more massive galaxies are clearly larger.

We also have checked how the spectra are altered when we
  exclude SMBH binaries not fulfilling the bypass condition (Equation
  \ref{eq:e_rGW}). This exclusion rules out 3 binaries from each galaxy mass
  bin. 
Interestingly, we find that $A=0.54\times10^{-15}$ at
  $f=1\yr^{-1}$, which is even closer to that for the
  ELC-model. The green lines in the \textit{left} panel of Figure
  \ref{fig:characteristic_h_differentmass} show the modified spectra
  as a result of the exclusion.  The turnovers are now less pronouced
  and shifted to lower frequencies of $f<10^{-8}$ Hz.

 As PTA observation periods span decades, the frequency range of
 $f\sim10^{-9}-10^{-8}{\rm Hz}$ is most sensitive to GWs.  In Figure
 \ref{fig:A_sensitivity}, the current and future-expected sensitivities
 and observational upper limits are compared with our estimates.  The
 uppermost edge indicates the sensitivity by the full IPTA \citep[$h_{\rm
     c}\sim 10^{-16}-10^{-17}$ at $f=5\times10^{-9} {\rm ~Hz}$,
 ][]{Janssen+2015}.  The other two lines refer to the sensitivity set
 by the complete Parkes PTA (PPTA) data set with 20 pulsars for $5\yr$
 and that achievable with the planned Square Kilometer Array (SKA)
 with 20 pulsars over 10$\yr$ \footnote{The expected
   level to be reached by the SKA is lower, $h_{\rm c}\sim
   10^{-16}-10^{-17}$ at a reference frequency of $\yr^{-1}$
   \citep{Janssen+2015}.}.  {\it Our models predict amplitudes below
   the observational upper limits. The current PPTA dataset may not be
   sufficient to confirm/rule out our models; however in the future,
   the planned SKA will be able to give constraints on our models over
   wider frequency ranges.}

\subsection{Semi-analytic analysis on the estimate of $A$ - Comparison with previous works}
\label{sub:semi-analytic}
In this work, using few-body simulations in analytic background potentials, we follow the dynamical
evolution of multiple SMBH systems and estimate the BH coalescence
rates in the host galaxies undergoing multiple mergers with a wide
range of mass ratios.  Using the computed merger rates, we next estimated
 the stochastic GWB. 
For a more thorough understanding of our results, it is hence
important to compare our results with previous works. 

As an informative comparison, given our sampled merger trees, we analytically estimate $A$ following the assumptions about BH mergers in \citet{McWilliams+2014} and \citet{Kulier+2015}. They assume that, 

\begin{enumerate}
	\item\label{assumption1} {\it every bound pair of BHs efficiently solves the final parsec problem on its own}; 
	\item\label{assumption2} {\it BH binaries are always in circular orbits.} 
\end{enumerate}

\begin{figure}
	\centering
	\includegraphics[width=8.2cm]{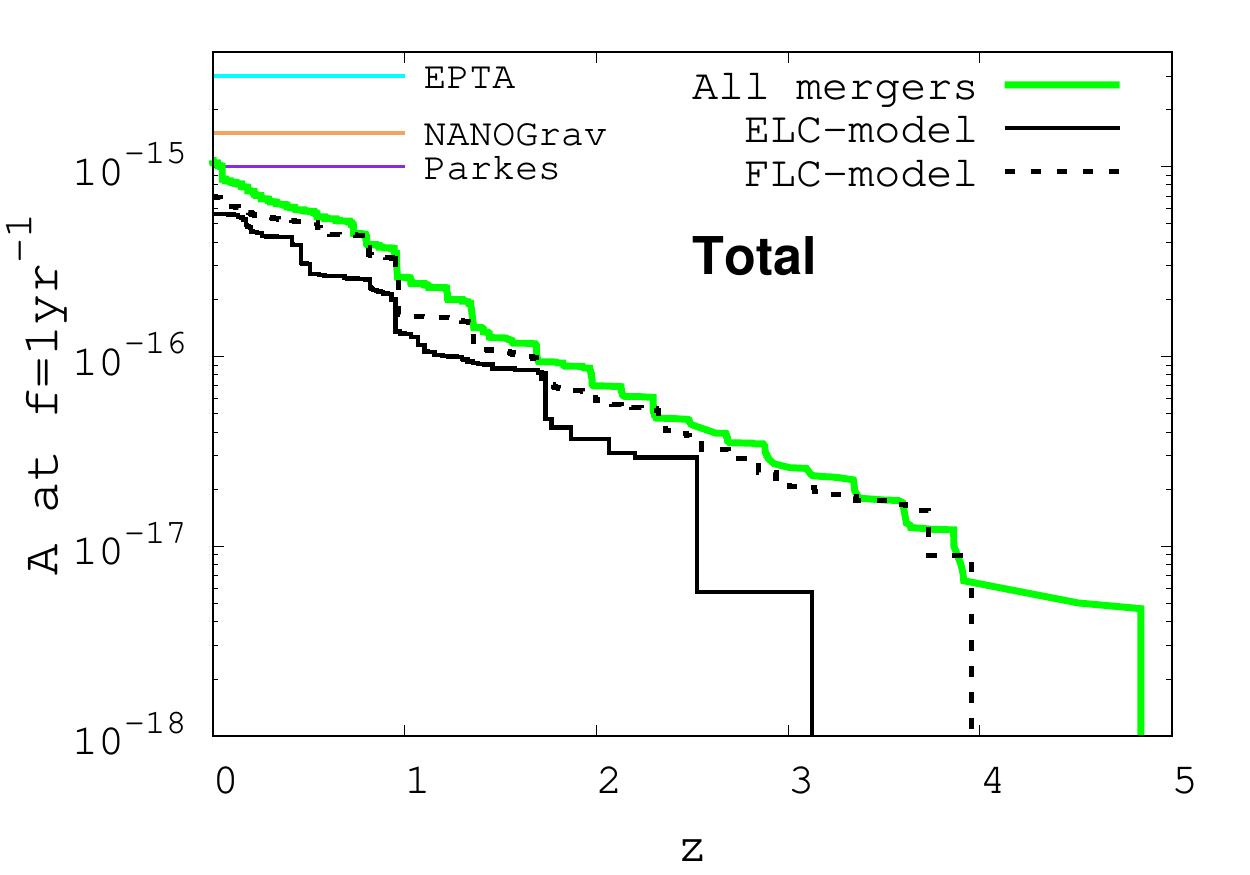}
	\includegraphics[width=8.2cm]{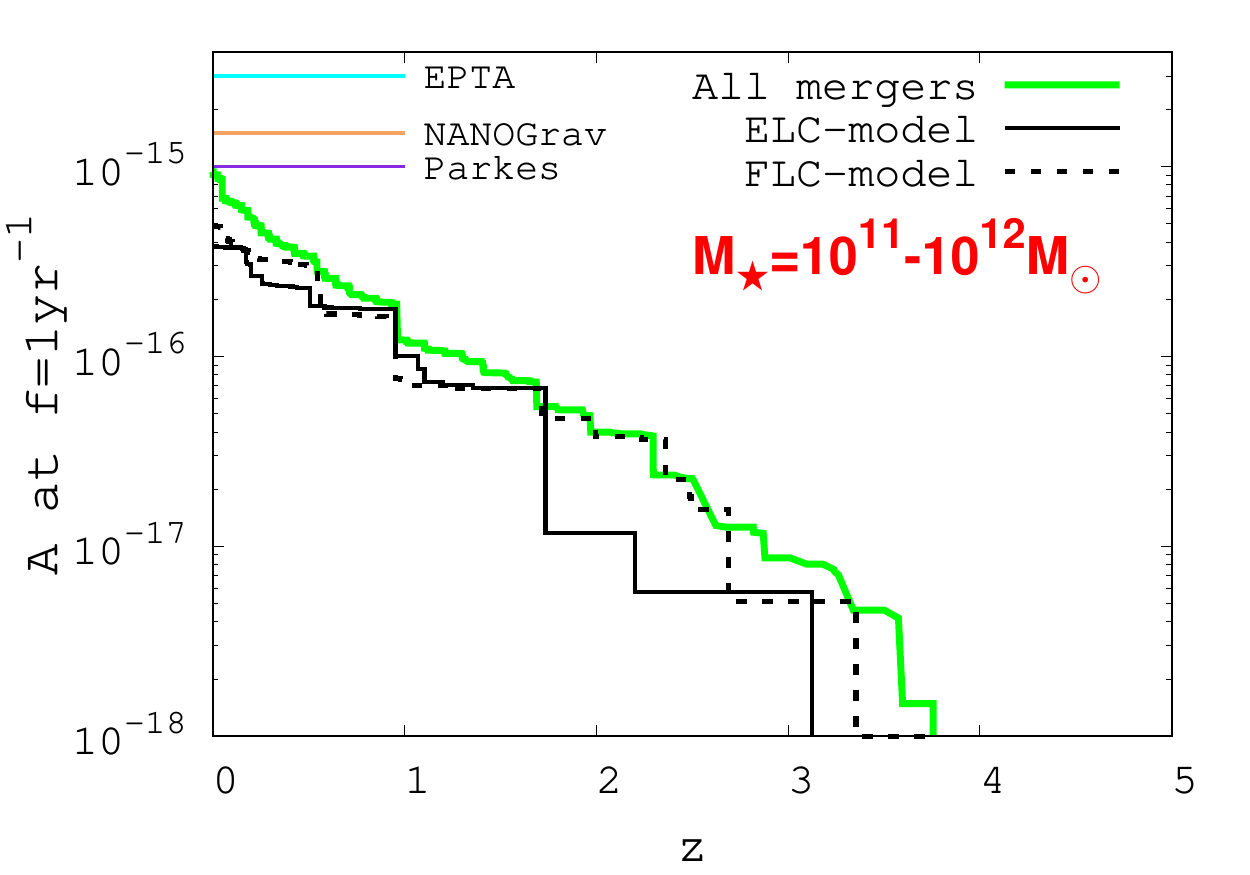}
	\includegraphics[width=8.2cm]{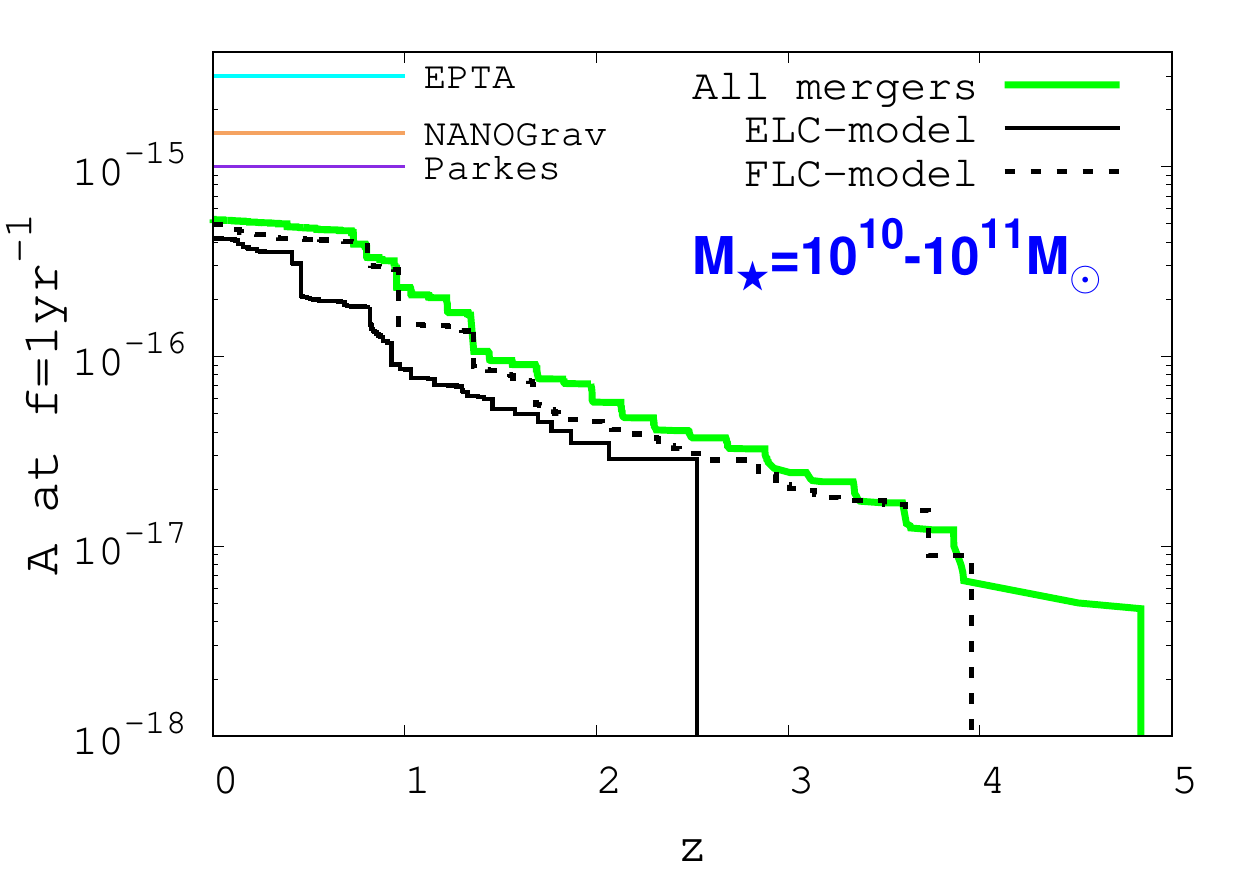}
	\caption{The redshift evolution of the GWB amplitude $A$ for
		the optimistic case (labeled ``all mergers'', thick green
		solid line), the FLC-model (dotted black line) and the
		ELC-model (thin solid black line). The \textit{top} panel
		shows the total $A$ and we separately depict the evolution
		of $A$ contributed by more massive galaxies (\textit{middle}
		panel) and less massive galaxies (\textit{bottom}
		panel).  As a reference, 
		the observational upper limits for EPTA, NANOGrav and Parkes are
		indicated.}
	\label{fig:hc_redshift}
\end{figure}

While these calculations broke new ground in estimating BH merger
rates in a cosmological context, their assumption \ref{assumption1}
is optimistic, and their predicted GW emission
rates should be considered upper limits. As a result, the GWB
predicted by \citet{McWilliams+2014} and \citet{Kulier+2015} is higher
than the one given by our detailed computations.  By comparing $A$ for
the optimistic case to $A$ for the FLC and the ELC models, we may
be able to understand how much each assumption affects $A$. For this
estimate, we additionally assume that BH mergers occur after a dynamical
friction timescale (equation 3 in \citealt{Kulier+2015}) since galaxy
merger events. This leaves 17 out of our total 1744 mergers incomplete
by $z=0$. We take the total mass of merging binaries to be the maximum
value between the BH mass required by the $M-\sigma$ relation at the BH merger
redshift and the sum of the masses of two merging BHs. We find the
total $A=1.10\times10^{-15}$ for the optimistic case, which is larger
by a factor of $1.5-2$ than those for the FLC-model and the
ELC-model. The total $A$ decreases as we impose $q_{\star}$ cutoffs:
Assuming only galaxy mergers of $q_{\star}>0.01$ lead to BH mergers,
$A=1.00\times10^{-15}$. For $q_{\star}>0.1$, $A=0.77\times10^{-15}$,
which is fairly close to $A$ for the FLC-model. As shown in Figures
\ref{fig:galaxy_mergercorrelation} and \ref{fig:mergercount_z} (galaxy
merger rates higher than BH merger rates, see (iii) in \S
\ref{subsub:BHmergerrate}), we can confirm again that not all galaxy
mergers lead to BH mergers in both the FLC and the ELC models,
resulting in smaller $A$. In Figure \ref{fig:hc_redshift}, we show the
evolution of $A$ with redshift for the optimistic case (labeled
``all mergers''), the FLC-  and the ELC-model. In addition to the total $A$
(\textit{upper} panel), we separately show the evolution of $A$ for
more massive galaxies (\textit{middle} panel) and less massive
galaxies (\textit{bottom} panel). In all three panels, we can see that
the amplitudes for the FLC-model and the ELC-model remain below those
for the optimistic case. Due to nearly prompt BH mergers for the
optimistic case and delayed mergers in the ELC-model, the first GW
signals for the optimistic case appear earliest, followed by those for
the FLC-model and the FLC-model at last. The first mergers in the
ELC-model are delayed by $\Delta z\simeq0.3-1.5$ with respect to those
in the FLC-model.

If we relax the assumption of circular orbits, as seen in Figures
\ref{fig:characteristic_h} and
\ref{fig:characteristic_h_differentmass}, the GWB further decreases,
especially at low frequencies. In our two models, the effects of the
eccentricities at $f=1 \yr^{-1}$ are not significant, but the
difference could exceed an order of magnitude at lower frequencies
depending on whether merged binary orbits are assumed to be circular
or eccentric.

A suppression of the GW signal at higher frequencies can be caused by
the presence of a circumbinary disk. In another recent study,
\citet{Kelley+2017}, using the galaxy population in the Illustris
simulation, coevolve massive BHs to predict the GWB. They take into
account various possible environmental mechanisms in their
calculations including dynamical friction, stellar 'loss cone'
scattering and tidal-viscous drag from a circumbinary disc. Similarly
to our models, they explore different degrees of loss cone
filling. Their fiducial model predicts an amplitude within the range
of $0.3\times10^{-15}<A<0.4\times10^{-15}$ (with the upper limit of
$A\simeq0.7\times10^{-15}$). This is smaller than our values roughly
by a factor of $1-2$. We believe that this may be caused by
different strategies to populate SMBHs
  \citep{Sesana+2016}. Furthermore, tidal torques from the gaseous
  circumbinary disk could also come into play. This was 
  studied before by \citet{KocsisSesana2011} with BH merger rates
from the Millennium simulation and adopting the models for gas-driven
inspirals of \citet{Haiman+2009}. Typically, the presence of
circumbinary discs drives very rapid inspirals of binaries via
migration, leading to a significant suppression of the signal at
frequencies $f>10^{-8}~{\rm Hz}$ compared to mergers in a gas-poor
environment.

Generally speaking, adopting the scaling relations to populate SMBHs
in the galaxies, our estimates for $A$ are generally in good agreement
with other studies (see Figure \ref{fig:characteristic_h}), especially
with models constructed on the Millennium simulation
\citep[e.g][]{Sesana+2009}.  However, noting the discussion in
\citet{Sesana+2016} and \citet{RasskazovMerritt2017}, we
  emphasize that our results could also be affected by the use of
  different observational relations. \citet{Sesana+2016} employ
  different SMBH-galaxy scaling relations and accretion prescriptions
  to populate and grow the SMBHs, and they study the impact of
  selection bias in determining SMBH masses on the PTA
  measurements. In another study, \citet{RasskazovMerritt2017}, taking
  into account the effects of rotating and aspherical nuclei on the
  orbital evolution of SMBH binaries, compute the GWB and study the
  dependence on the $M_{\rm BH}-M_{\star}$ relation. Even though they tackle this
  problem within different frameworks, both studies suggest that
  the GWB amplitude has been overestimated and may decrease by a 
  factor of a few if different galaxy scaling relations are
  used \footnote{\citet{Taylor+2016} discuss the similar
    issue of the overestimate of SMBH merger rates from an observational
    perspective.}.

After the original submission of this manuscript, we became aware of a similar recent study by \citet{Bonetti+2017}. They employ a semi-analytic model of galaxy evolution and model SMBH mergers and their GW signals, by incorporating three-body PN effects to study the role of triple and quadruple interactions between SMBHs (adopted from \citealt{Bonetti+2016}). Their inferred 
		merger rates are fairly consistent with those shown in \S\ref{sub:BHmergerrates}, and the physical picture of the mergers we find in our work (see \S\ref{sub:dynamicalfeatures}) 
		is also  similar to that discussed in \citet{Bonetti+2017}.

\subsection{Caveats}
Our results were obtained in models with
observationally and numerically-motivated assumptions, but they are
subject to several caveats. In this section, we discuss the major
limitations of our models.

\begin{enumerate}
	\item {\it Dynamical friction}. In the FLC-model, we assume
          dynamical friction operates very efficiently to decay the
          orbits of BH binaries at all times.  This regime may
          underestimate the true hardening rate in the presence of a
          full loss cone, since inside the influence radius stellar
          scattering hardens SMBH binary orbits by a factor $\sim 1/q$
          faster than the hardening rate from a naive application of
          the dynamical friction formula \citep{Merritt2013}. 
          Furthermore, our merging galaxy model assumes a flat core in
          the inner region, not accounting for the dynamical changes
          in shape of the inner stellar potential as SMBH binaries in
          it mutually evolve. The shape of the stellar potential
          around SMBH binaries is correlated with the hardening rates
          of the binaries \citep{DosopolouAntonini2017}, hence their
          fate and the BH merger rates in the FLC-model. 
          
\item {\it Three-body interactions}
    
We have not self-consistently included the 3-body PN terms (see Equation 5 in \citealt{LoustoNakano2008}) during the 3-body
  interactions \citep[e.g][]{Bonetti+2016}.
 However, as explained in \S\ref{sub:dynamicalfeatures}, 
  most of the important 3-body interactions are in the
  hierarchical regime, with only 2 of the 3 bodies sufficiently close
  for PN terms to be needed. Therefore we expect that our results
  are robust, but we will explore this in future work.      
   
	\item {\it Assumptions on SMBH mass growth}
	
	There are several factors related to the assumptions on SMBH
        mass growth and sampled galaxies which may affect the GWB.
        First of all, given the requirement for the total mass of the
        central BHs, very loud signals from a few massive binary
        mergers of $q\simeq1$ at $z\simeq0$ can cause remarkably large
        jumps in the GWB. As explained in \S
        \ref{subsubsec:DM_stars_BHmass}, this could occur when
        binaries, with the central BHs and other initially small BHs
        which rapidly grow in the cores while the central BHs are
        temporarily ejected, coalesce.  Such temporary ejection of the
        central BHs can occur as a result of GW recoil kicks.  Given
        our galaxy samples, we do not find that $A$ at $z=0$ is
        dominated by a few of these loud GW emission events. But it is
        still possible at lower $z$, especially more likely in the
        ELC-model with its more frequent ejections\footnote{In the
          FLC-model, we find that the BH merger mass ratios decrease
          and the BH masses grow as $z$ decreases, following the trend
          in the galaxy merger histories. Hence GW kick-driven ejections,
          with subsequent rapid growth of small BHs, are more likely at
          higher $z$. However, their contributions may not be
          significant to $A$ at $z=0$ and chances of such giant binary
          formations and mergers at small $z$ may be low.}. If we
        explore a large number of galaxy merger trees, the statistical
        errors from finite sample size\footnote{The Poisson error of
        $\sqrt{N_{\rm BH~merger}}/N_{\rm BH~merger}\simeq 0.1$, where
        $N_{\rm BH~merger}$ is the number of BH mergers for the
        current sample size.} will decrease, but chances of large
        signals from a few individual sources may
        increase. Furthermore, in this study we do not consider large
        kicks driven by nonlinear spin-orbit interactions
        \citep{LoustoZlochower2011,Lousto+2012,LoustoZlochower2013}
        with the maximum recoil velocities larger than typical escape
        velocities of galaxies. We point out that the frequencies of
        ejection and merger events would vary when such large kicks
        are taken into account.

	 \item 	{\it Galaxy samples}
	 
We have not included the contributions from smaller galaxies of
$M_{\star}<10^{10}\Msol$ (or $M_{\rm DM}<5\times10^{11}\Msol$). We
find that BH mergers from less massive galaxies contribute more to the
overall GWB as a result of higher BH merger rates as shown in Figure
\ref{fig:mergerrate_comoving}. Therefore, it is also possible that BH
coalescences in galaxies with $M_{\star}<10^{10}\Msol$ can noticeably
increase the predicted GWB. However, it is uncertain whether the BH
merger rates increase further for galaxies of $M_{\star}<10^{10}\Msol$
and, more importantly, the offset between the increase in the BH
merger rates and the decrease in the chirp mass lead to a significant
rise in the GWB. For such low mass galaxies, the relationship between
stellar mass $M_\star$ and halo mass $M_{\rm DM}$ is more complicated
than the single power law we have
assumed\citep[e.g][]{Behroozi+2010,Moster+2013}, and our model would
need to be modified accordingly. In addition, we assume that
each galaxy always harbors a SMBH as long as the BH mass is larger
than the minimum mass. However, the occupation fraction in the low
mass galaxies is more likely to be affected by the assumptions on
BH seed formation and initial occupation fraction at high redshift
\citep{Menou+2001}. Therefore, considering that less massive halos
tend to possess relatively small number of early progenitors as well
as even small mass progenitors, the contributions to the GWB from low
mass galaxies of $M_{\star}<10^{10}\Msol$ may not be significant
\citep{Sesana2013}. But these estimates are approximate, so more
systematic studies are necessary for better understanding of the
contributions of the BH mergers in dwarf galaxies.
	
	Another caveat is that, given that we follow up to $10-12$
        galaxy mergers, for host galaxies experiencing a large number
        of mergers at low redshift, we may miss some galaxy mergers in
        their histories, hence possibly leading to an
        underestimate of the GWB.

		Last, in our merging galaxy model, we assume
                one central BH per satellite galaxy at galaxy
                mergers. Multiple BHs in satellite galaxies are
                definitely possible. For those cases, more frequent
                multi-BH interactions and ejections will take place,
                which possibly influence the BH merger rates, ejection
                rates as well as the chances of such giant binary
                mergers explained above.

\end{enumerate}

\section{Summary}
\label{summary}

In this work, using few-body simulations in analytic background
potentials, we have examined the evolution of SMBH binaries and higher
multiples, from their formation to coalescence, as the host galaxies
go through mergers with mass ratios of $10^{-4} < q_{\star} < 1$. For
host galaxies of mass $M_{\star}=10^{10}-10^{12}\Msol$ at $z=0$
extracted from the Millennium simulation, we followed their merger
trees by assuming a SMBH in each of the host/satellite galaxies, with
the BH mass determined by standard scaling relations. We have explored
two limiting scenarios for the decay of the binary orbits,
approximating full and empty loss cone regimes. In the full loss cone
model (FLC-model), we assume dynamical friction efficiently shrinks
the orbits until binaries merge, whereas in the empty loss cone model
(ELC-model), we assume that dynamical friction is no longer able to
operate and cause orbital decay when the binaries become hard. The
ELC-model allows us to examine multi-body BH interactions in a
cosmological context, and test their utility as a ``solution of last
resort'' to the final parsec problem in large galaxies where other
solutions may fail.  The FLC-model, while highly
    idealized, serves as a valuable comparison case, and as a testing
    ground for an underexplored regime: inspirals where $e$ is excited
    to very high values by dynamical friction in a flat stellar core
    characteristic of the highest mass galaxies.  Based on the
inferred merger rates from our simulations, we estimate the stochastic
GWB in the two models.  We summarize our findings as follows:

\begin{enumerate}
	\item \textit{Dynamical features of SMBH binaries and multiple
          systems}: we find a few clear differences in the evolution
          of SMBH binaries between the FLC-model and the ELC-model.
          For the FLC-model, dynamical friction tends to increase the
          binary eccentricity.  When energy loss to GWs becomes
          dominant, the binary eccentricities are almost unity
          ($e>0.99$). The evolution of the orbital eccentricity of
          SMBH binaries in various stellar distributions will be
          explored in a future paper (Stone et al. 2017, in
          prep). However, our FLC results can be understood in the
          context of past work, which finds eccentricity excitation
          due to dynamical friction in flat density profiles,
          particularly in Keplerian potentials
          \citep{DosopolouAntonini2017}. A critical
              assumption behind the eccentricity excitation seen in
              the FLC model is the existence of a flat stellar density
              core.  While this assumption is reasonable for the very
              high-mass galaxies considered in this paper, it would
              not apply to lower-mass galaxy mergers.  Subsequently,
          SMBH binaries merge on a short time scale and the lifetimes
          of coalescing binaries are less than $1\Gyr$. We also find
          that the BH merger events are strongly coupled with major
          mergers ($q_{\star}>0.1$) of the host galaxies (Figure
          \ref{fig:galaxy_mergercorrelation}).  For the ELC-model,
          however, there is a time delay before the central SMBH
          binaries merge because they must wait for other BHs to come
          close and effectively interact with them. This results in
          longer binary lifetimes ($\gtrsim1\Gyr$). This is a clear
          difference from the FLC-model.  As a result, coalescences of
          SMBHs in the ELC-model preferentially occur in the host
          galaxies experiencing multiple major mergers.
	
	\item \textit{BH merger rates}: we find that SMBH binaries do
          merge in both models, but with typically higher coalescence
          rates in the FLC-model than in the ELC-model. {\it There is no
          ``final parsec'' problem in either scenario}. Even though the
          BH coalescence rate for the ELC-model is lower, the BH
          mergers in the ELC-model strongly indicate that, as galaxies
          go through a series of mergers and binaries stall due to
          inefficient decay mechanisms (e.g., empty loss cone), a
          cluster of multiple SMBHs is naturally produced in the core
          regions, and these BHs can eventually merge via multi-body
          interactions.
	
	\item \textit{Mass ratio of coalesced BH binaries}: another noticeable feature of the ELC-model is that the mass ratios, and hence the chirp masses, of coalescing SMBHs tend to be higher. As they go through chaotic three-body interactions,
	the less massive BHs will typically be ejected, leaving behind a
	binary of the more massive BHs.

	\item \textit{Stochastic GW background}: using the inferred BH
          coalescence rates, we estimate the strain amplitudes
          $A=0.70\times10^{-15}$ and $A=0.53\times10^{-15}$ for the
          FLC-model and the ELC-model, respectively. In spite of the
          lower BH merger rates for the ELC-model, we find that the
          amplitudes are quite similar. This is because (a)
          mergers of BH binaries, especially
              with large chirp masses, still occur in the ELC-model.
              Only those with lower mass ratios, which make minor
              contributions to the GWB, are missing; (b) in the
              ELC-model, BH coalescence events preferentially take
              place at a later time with larger chirp mass, as BHs
              have more time to grow, given the delayed mergers. In
              other words, louder GW emissions more abundantly occur
              at lower redshift. This would counterbalance the
              decrease in the GWB due to the loss of some BH mergers,
              as described in \ref{A:reason2} above;  (c) due to
          the larger mass ratios of the merged binaries, the
          contributions of the less massive binary mergers (i.e.,
          coalesced BHs in less massive galaxies) to the GWB are
          relatively high in the ELC-model compared to the FLC one.
          Our inferred strain is consistent with current observational
          limits and a factor of roughly two below the rates predicted
          by the simple model in which every galaxy merger leads to a
          BH merger.
	 
	 \item \textit{Effect of high eccentricities on GW spectra}:
            Given the high eccentricities of the merged SMBH binaries,
           our models predict significant suppression of GW power at
           lower frequencies.  This causes a low frequency flattening
           as well as a turnover in the stochastic background spectrum as shown in Figures
           \ref{fig:characteristic_h} and
           \ref{fig:characteristic_h_differentmass}, which will be
           observationally important for comparison to future data.

\end{enumerate}

By adopting a dynamical approach to study the coalescence of SMBH
binaries, our work shows clear distinctions between two limiting
regimes of loss cone physics. In particular, different expectations
for chirp masses, mass ratio distributions, and flattening of the GW
spectra due to high eccentricities can all be observationally
relevant. Furthermore, multi-body interactions between SMBHs are a
natural consequence of galaxy mergers, and are clearly a plausible
channel for driving BH coalescences.  Our predictions show that
ongoing PTA searches can potentially discriminate between different
models of black hole binary orbital evolution.

\vspace{0.5cm}

\section*{Acknowledgments}
We are grateful to Alberto Sesana, Chiara Mingarelli for their valuable and constructive feedbacks. We also thank the anonymous referee for comments and suggestions that helped us to improve the paper. ZH acknowledges support by a Simons Fellowship in Theoretical 
Physics (ZH) and by NASA grant NNX15AB19G. NCS received financial support from NASA through Einstein Postdoctoral Fellowship Award Number PF5-160145.
Results in this paper 
were obtained using the
high-performance LIred computing system at the Institute for Advanced
Computational Science at Stony Brook University, which was obtained
through the Empire State Development grant NYS \#28451.








\appendix
\section{Scaling relations}
\label{appendix:scalingrelation}

We provide in Table \ref{table:scalingrelation}
    the scaling relations between the relevant variables in our model
    in terms of $M_{\rm BH}$ (as well as $M_{\rm DM}$), derived with
    the four scaling relations \ref{eq:scaling1}-\ref{eq:scaling4}. We
    show derivations for some of the exponents in the table which are
    less immediate.
\begin{table*}
	\centering
	\setlength\extrarowheight{5pt}
	\begin{tabulary}{1\linewidth}{c c}
		\hline
		\hline
		\multirow{2}{*}{total stellar mass $M_{\star}$} & $\left(\frac{M_{\star}}{10^{11}\Msol}\right)=2.00\left(\frac{M_{\rm  DM}}{10^{13}\Msol}\right)$ \\
		& $\left(\frac{M_{\star}}{10^{11}\Msol}\right)=1.85 \left(\frac{M_{\rm BH}}{10^{9}\Msol}\right)^{0.86}$\\
		\multirow{2}{*}{stellar dispersion $\_{\star}$} & $\left(\frac{\sigma}{200\km s^{-1}}\right)=1.34\left(\frac{M_{\rm DM}}{10^{13}\Msol}\right)^{0.26}$ \\
		& $\left(\frac{\sigma}{200\km s^{-1}}\right)=1.31\left(\frac{M_{\rm BH}}{10^{9}\Msol}\right)^{0.23}$\\
		\multirow{2}{*}{half-mass radius $r_{\rm h}$}& $\left(\frac{r_{\rm h}}{\rm kpc}\right)=5.34\left(\frac{M_{\rm DM}}{10^{13}\Msol}\right)^{0.47}$ \\
		& $\left(\frac{r_{\rm h}}{\rm kpc}\right)=5.14\left(\frac{M_{\rm BH}}{10^{9}\Msol}\right)^{0.41}$ \\
		\multirow{2}{*}{core radius $r_{\rm c}$}& $\left(\frac{r_{\rm c}}{\rm pc}\right)=0.92\left(\frac{M_{\rm DM}}{10^{11}\Msol}\right)^{0.99}$ or $\left(\frac{r_{\rm c}}{\rm kpc}\right)=0.089\left(\frac{M_{\rm DM}}{10^{13}\Msol}\right)^{0.99}$ \\
		& $\left(\frac{r_{\rm c}}{\rm pc}\right)=1.56\left(\frac{M_{\rm BH}}{10^{7}\Msol}\right)^{0.86}$ or $\left(\frac{r_{\rm c}}{\rm kpc}\right)=0.082\left(\frac{M_{\rm BH}}{10^{9}\Msol}\right)^{0.86}$\\
		\multirow{2}{*}{core density $\rho_{\rm c}$}& $\left(\frac{\rho_{\rm c}}{\Msol\pc^{-3}}\right)=240\left(\frac{M_{\rm DM}}{10^{13}\Msol}\right)^{-1.46}$\\
		& $\left(\frac{\rho_{\rm c}}{\Msol\pc^{-3}}\right)=270\left(\frac{M_{\rm BH}}{10^{9}\Msol}\right)^{-1.26}$ \\
		\multirow{2}{*}{core mass $M_{\rm c}$} & 	$\left(\frac{M_{\rm c}}{10^{9}\Msol}\right)=0.453 \left(\frac{M_{\rm DM}}{10^{13}\Msol}\right)^{1.52}=0.158 \left(\frac{M_{\star}}{10^{11}\Msol}\right)^{1.52}$ \\
		&$\left(\frac{M_{\rm c}}{10^{9}\Msol}\right)=0.403 \left(\frac{M_{\rm BH}}{10^{9}\Msol}\right)^{1.31}$ \\
		\hline
		\hline
	\end{tabulary}
	\caption{The variables relevant for the model galaxy in this work in terms of BH mass $M_{\rm BH}$ and DM halo mass $M_{\rm DM}$, derived using the scaling relations \ref{eq:scaling1} to \ref{eq:scaling4}.}
	\label{table:scalingrelation}
\end{table*}

\begin{enumerate}
	\item  \textit{Relations of $\sigma$}:
	
	Using the scaling relation \ref{eq:scaling3} and the relation between $M_{\rm BH}$ and $M_{\rm DM}$ (same as the scaling relation \ref{eq:scaling2}),
	\begin{align}
	\sigma\sim M_{\rm BH}^{1/4.38}\sim M_{\rm DM}^{1.16/4.38}=M_{\rm DM}^{0.26}.
	\label{eq:derive1}
	\end{align}
	
	\item \textit{Relations of $r_{\rm h}$}:
	
	Combining the relation $r_{\rm h}\sim M_{\star}/\sigma^{2}$ and the scaling relation \ref{eq:scaling1} and \ref{eq:scaling3} (or Equation \ref{eq:derive1} derived above), 
	
		\begin{align}
	r_{\rm h}\sim M_{\star}\sigma^{-2}\sim M_{\rm DM}\left(M_{\rm DM}^{-0.26}\right)^{2}\sim M_{\rm DM}^{0.47}\sim M_{\star}^{0.47}.
	\label{eq:derive2}
	\end{align}
And the relation $M_{\rm DM}\sim M_{\rm BH}^{0.86}$ gives,
		\begin{align}
r_{\rm h}\sim M_{\rm DM}^{0.47} \sim M_{\rm BH}^{0.41}.
\label{eq:derive3}
\end{align}
	
	\item \textit{Relations of $r_{\rm c}$}:
	
	From the scaling relation \ref{eq:scaling4} we find 
	\begin{align}
	r_{\rm c}\sim M_{\rm BH}^{0.86}\sim M_{\rm DM}^{0.86\times1.16}\sim M_{\rm DM}^{0.99}\sim M_{\star}^{0.99}.
	\label{eq:derive4}
	\end{align}

	\item \textit{Relations of $\rho_{\rm c}$}:
	
	Given Equation (5) in \citet{StoneOstriker2015} and Equations \ref{eq:derive3} and \ref{eq:derive4}, for $r_{\rm h}\gg r_{\rm c}$,
	\begin{align}
	\rho_{\rm c}\sim M_{\star} r_{\rm h}^{-1} \left(r_{\rm c}\right)^{-2}\sim M_{\rm DM} M_{\rm DM}^{-0.41}\left(M_{\rm DM}^{-0.99}\right)^{2}\sim M_{\rm DM}^{-1.46}\sim M_{\star}^{-1.46}.
	\label{eq:derive5}
	\end{align}
	The scaling relation \ref{eq:scaling2} gives 
		
		\begin{align}
	\rho_{\rm c}\sim M_{\rm DM}^{-1.46}\sim M_{\rm BH}^{-1.26}.
	\label{eq:derive6}
	\end{align}

	\item \textit{Relations of $M_{\rm c}$}:
	
	Starting with Equation (6) in \citet{StoneOstriker2015}, 
	\begin{align}
	M_{\rm c}\sim r_{\rm c} \left(r_{\rm h}\right)^{-1}M_{\star},
		\label{eq:derive7}
	\end{align}
	and inserting Equation \ref{eq:derive2} and \ref{eq:derive4}
        in to Equation \ref{eq:derive7} above, we arrive at the expressions
	\begin{align}
	M_{\rm c}\sim r_{\rm c} \left(r_{\rm h}\right)^{-1}M_{\star}\sim M_{\rm DM}^{0.99}M_{\rm DM}^{-0.47}M_{\rm DM}\sim M_{\rm DM}^{1.52}\sim M_{\star}^{1.52},
	\label{eq:derive8}
	\end{align}
	and
	\begin{align}
	M_{\rm c}\sim M_{\rm BH}^{1.31}.
	\label{eq:derive9}
	\end{align}

	\item \textit{$r_{\rm c} - r_{\rm h}$ relation}:
	
	\begin{equation}
	\frac{r_{\rm h}}{r_{\rm c}}=63  \left(\frac{M_{\rm BH}}{10^{9}\Msol}\right)^{-0.45}=60  \left(\frac{M_{\rm DM}}{10^{13}\Msol}\right)^{-0.52},
	\end{equation}
	or
	\begin{align}
	\left(\frac{r_{\rm h}}{\rm kpc}\right)&=16.9\left(\frac{r_{\rm c}}{\rm kpc}\right)^{0.48} \rightarrow \left(\frac{r_{\rm h}}{\rm kpc}\right)=0.61\left(\frac{r_{\rm c}}{\rm pc}\right)^{0.48},\\
	\left(\frac{r_{\rm c}}{\rm kpc}\right)&=0.0026\left(\frac{r_{\rm h}}{\rm kpc}\right)^{2.10} \rightarrow \left(\frac{r_{\rm c}}{\rm pc}\right)=2.6\left(\frac{r_{\rm h}}{\rm kpc}\right)^{2.10}.
	\end{align}

\end{enumerate}

\section{Gravitational wave recoils and remnant masses}
\label{appendix:GWkicks}

For the recoil kick in the simulations, we adopt 
the fitting formula provided by \citet{Lousto+2010}:

\begin{align}
\textit{\textbf{v}}_{\rm recoil}(q,\boldsymbol{\alpha})&=v_{\rm m}\hat{e}_{1}+v_{\perp}(\cos\xi~\hat{e}_{1}+\sin\xi~\hat{e}_{2})+v_{\parallel}\hat{n}_{\parallel},\\
v_{\rm m}&=A\frac{\eta^{2}(1-q)}{1+q}[1+B\eta]\nonumber\\
v_{\perp}&=H\frac{\eta^{2}}{1+q}(1+B_{\rm H}\eta)(\alpha_{2}^{\parallel}-q\alpha_{1}^{\parallel})\nonumber\\
v_{\parallel}&=K\frac{\eta^{2}}{1+q}(1+B_{\rm K}\eta)(\alpha_{2}^{\perp}-q\alpha_{1}^{\perp})\cos(\Theta_{\Delta}-\Theta_{0})\,,\nonumber\nonumber\\
\end{align} 
where $q$ is the mass ratio of two BHs in binaries, 
$M_{\rm BH, 1}/M_{\rm BH, 2}(<1)$, $\eta=q/(1+q)^{2}$ and 
$\boldsymbol{\alpha}_{i}=\textbf{\textit{S}}_{i}/M_{\rm BH, i}^{2}$ is the 
intrinsic spin of BH $i$ and the indices $\perp$ and $\parallel$ refer to 
perpendicular and parallel to the orbital angular momentum, respectively.
$\hat{e}_{1}$ and $\hat{e}_{2}$ are orthogonal unit vectors in the orbital 
plane, $\xi$ measures the angle between the unequal mass and spin
contribution to the recoil velocity in the orbital plane. $\Theta_{\Delta}-\Theta_{0}$ 
is the angle difference between the in-plane component and the infall 
direction at merger. Adopting their findings, we take $A=1.2\times10^{4}\km\s^{-1}$, 
$B=-0.93$, $H=6.9\times10^{3}\km\s^{-1}$, $B_{H,K}=0$, 
$K=6.0\times10^{4}\km\s^{-1}$ and $\xi=145^{\circ}$. 
Following \citet{SchnittmanBuonanno2007}, 
we randomly assign spin magnitudes to both BHs of the binary from 
a uniform distribution in the range of $0.0\leq \boldsymbol{\alpha}_{1,2}\leq0.9$. 
We take $\Theta_{0}=0$, while $\Theta_{\Delta}$ is also arbitrarily drawn from a uniform distribution.

Using the same parameters drawn for the recoil velocities, we estimate 
remnant masses using Eq.~(4) up to the leading order and Eq.~(5) in 
\citet{Lousto+2010}. For two BHs of $M_{\rm BH, 1}$ and $M_{\rm BH, 2}$ 
of a binary, the remnant mass $M_{\rm remnant}$ is expressed as follows,
\begin{align}
\frac{\Delta M_{\rm BH}}{M_{\rm BH, 1}+M_{\rm BH, 2}}&=\eta \tilde{E}_{ISCO},\\
\tilde{E}_{ISCO}&=(1-\frac{\sqrt8}{3})+0.103803\eta\nonumber\\
&+\frac{1}{36\sqrt{3}(1+q)^{2}}\left[q(1+2q)\alpha_{1}^{\parallel}+(2+q)\alpha_{2}^{\perp}\right]\nonumber\\
&-\frac{5}{324\sqrt{2}(1+q)^{2}}\Big[\boldsymbol{\alpha}_{2}^{2}-3 (\alpha_{2}^{\parallel})^{2}-2q(\boldsymbol{\alpha}_{1}\cdot\boldsymbol{\alpha}_{2}-3\alpha_{1}^{\parallel}\alpha_{2}^{\parallel})\nonumber\\
&+q^{2}(\boldsymbol{\alpha}_{1}^{2}-3(\alpha_{1}^{\parallel})^{2})\Big]\\
M_{\rm remnant}&=M_{\rm BH, 1}+M_{\rm BH, 2}-\Delta M_{\rm BH}
\end{align}

In our simulations, given the frequent merger mass ratio of
$\simeq10^{-2}$, the mass loss $-\Delta M_{\rm BH}$ corresponds to
$\Delta M_{\rm BH}\sim10^{-3})[M_{\rm BH, 1}+M_{\rm BH, 2}]$.

\bsp	
\label{lastpage}
\end{document}